\documentclass[]{aa}  

\usepackage{graphicx}
\usepackage{txfonts}
\usepackage{natbib}
\usepackage[colorlinks=true, linktocpage, linkcolor={blue!60!black}, citecolor={blue!60!black}, urlcolor={blue!60!black}]{hyperref}

\usepackage[dvipsnames,svgnames,x11names]{xcolor}

\setcitestyle{citesep={,}}

\usepackage{siunitx}
\newcommand*{\ra}[2][]{%
    \ang[
        angle-symbol-over-decimal,
        math-degree=\textsuperscript{h},
        text-degree=\textsuperscript{h},
        math-arcminute=\textsuperscript{m},
        text-arcminute=\textsuperscript{m},
        math-arcsecond=\textsuperscript{s},
        text-arcsecond=\textsuperscript{s},
        #1]{#2}%
}

\newcommand*{\dec}[2][]{%
    \ang[angle-symbol-over-decimal,#1]{#2}%
}

\begin{document} 

\titlerunning{The Merger in SPT-CL~J2106-5844}
\title{Multiwavelength view of SPT-CL~J2106-5844}
\subtitle{The radio galaxies and the thermal and relativistic plasmas\\in a massive galaxy cluster merger at $z\simeq1.13$}

\author{
     Luca Di Mascolo\inst{1,2,3,4}
\and Tony Mroczkowski\inst{5}
\and Yvette Perrott\inst{6}
\and Lawrence Rudnick\inst{7}
\and M. James Jee\inst{8,9}
\and Kim HyeongHan\inst{8}
\and Eugene Churazov\inst{1,10}
\and Jordan D. Collier\inst{11,12}
\and Jose M. Diego\inst{13}
\and Andrew M. Hopkins\inst{14}
\and Jinhyub Kim\inst{8}
\and B\"arbel S. Koribalski\inst{12,15}
\and Joshua D. Marvil\inst{16}
\and Remco van der Burg\inst{5}
\and Jennifer L. West\inst{17}
}

\authorrunning{L. Di Mascolo et al.}

\institute{
     Max-Planck-Institut f\"{u}r Astrophysik (MPA), Karl-Schwarzschild-Strasse 1, Garching 85741, Germany\\\email{luca.dimascolo@units.it}
\and Astronomy Unit, Department of Physics, University of Trieste, via Tiepolo 11, Trieste 34131, Italy
\and INAF – Osservatorio Astronomico di Trieste, via Tiepolo 11, Trieste 34131, Italy
\and IFPU - Institute for Fundamental Physics of the Universe, Via Beirut 2, 34014 Trieste, Italy
\and European Southern Observatory (ESO), Karl-Schwarzschild-Strasse 2, Garching 85748, Germany
\and School of Chemical and Physical Sciences, Victoria University of Wellington, PO Box 600, Wellington 6140, New Zealand
\and Minnesota Institute for Astrophysics, School of Physics and Astronomy, University of Minnesota, 116 Church Street SE, Minneapolis, MN 55455, USA
\and Department of Astronomy, Yonsei University, 50 Yonsei-ro, Seoul 03722, Korea
\and Department of Physics, University of California, Davis, One Shields Avenue, Davis, CA 95616, USA
\and Space Research Institute, Profsoyuznaya str. 84/32, Moscow, 117997, Russia
\and The Inter-University Institute for Data Intensive Astronomy (IDIA), Department of Astronomy, University of Cape Town, Private Bag X3, Rondebosch, 7701, South Africa
\and School of Science, Western Sydney University, Locked Bag 1797, Penrith, NSW 2751, Australia
\and Instituto de Física de Cantabria (CSIC-UC). Avda. Los Castros s/n. 39005 Santander, Spain
\and Australian Astronomical Optics, Macquarie University, 105 Delhi Rd, North Ryde, NSW 2113, Australia
\and Australia Telescope National Facility, CSIRO Astronomy and Space Science, P. O. Box 76, Epping, NSW 1710, Australia
\and National Radio Astronomy Observatory, P.O. Box O, Socorro, NM 87801, USA
\and Dunlap Institute for Astronomy and Astrophysics University of Toronto, Toronto, ON M5S 3H4, Canada
}

\date{Received 28 December 2020 / Accepted 12 April 2021}

\abstract
{SPT-CL~J2106-5844 is among the most massive galaxy clusters at $z>1$ yet discovered. While initially used in cosmological tests to assess the compatibility with $\Lambda$ Cold Dark Matter cosmology of such a massive virialized object at this redshift, more recent studies indicate SPT-CL~J2106-5844 is undergoing a major merger and is not an isolated system with a singular, well-defined halo.}
{We use sensitive, high spatial resolution measurements from the Atacama Large Millimeter/Submillimeter Array (ALMA) and Atacama Compact Array (ACA) of the thermal Sunyaev-Zeldovich (SZ) effect to reconstruct the pressure distribution of the intracluster medium in this system. These measurements are coupled with radio observations from the pilot survey for the Evolutionary Map of the Universe, using the Australian Square Kilometre Array Pathfinder (ASKAP), and the Australia Telescope Compact Array (ATCA) to search for diffuse nonthermal emission. Further, to better constrain the thermodynamic structure of the cluster, we complement our analysis with reprocessed archival \textit{Chandra} observations.}
{We jointly fit the ALMA and ACA SZ data in $u\varv$-space using a Bayesian forward modeling technique. The ASKAP and low-frequency ATCA data are processed and imaged to specifically highlight any potential diffuse radio emission.}
{In the ALMA and ACA SZ data, we reliably identify at high significance two main gas components associated with the mass clumps inferred from weak lensing. Our statistical test excludes at the $\sim9.9\sigma$ level the possibility of describing the system with a single SZ component. While the components had been more difficult to identify in the X-ray data alone, we find that the bimodal gas distribution is supported by the X-ray hardness distribution. The EMU radio observations reveal a diffuse radio structure $\sim400~\mathrm{kpc}$ in projected extent along the northwest-southeast direction, indicative of strong activity from the active galactic nucleus within the brightest cluster galaxy. Interestingly, a putative optical star-forming filamentary structure detected in the HST image is in an excellent alignment with the radio structure, albeit on a smaller scale.}
{}


\maketitle

%-----------------------------------------------------------------------

\section{Introduction}
The Sunyaev-Zeldovich (SZ) effect (\citealt{Sunyaev1972}; see \citealt{Mroczkowski2019} for a recent review) has been effectively employed over the last decade to survey the high-redshift Universe for massive galaxy clusters. One such cluster, SPT-CL~J2106-5844, was discovered by \citet{Foley2011} during the 2009 observational campaign of the $2500~\mathrm{deg^2}$ SZ survey by the South Pole Telescope \citep[SPT;][]{Carlstrom2011} with an exceptional multiband signal-to-noise ratio of 22.1 \citep[][]{Williamson2011}. The follow-up optical and infrared observations reported in \citealt{Foley2011} found the cluster to be at $z=1.132$ and provided evidence that it hosts a significant galaxy overdensity. Furthermore, the spectroscopic analysis of the cluster members indicated a fairly large value for the velocity dispersion, $\sigma_{\varv} = 1230\substack{+270\\-180}~\mathrm{km s^{-1}}$. Together with the subsequent detection of a very luminous and extended X-ray source by \textit{Chandra}, the strong SZ significance and the high velocity dispersion of the cluster members hinted at the possibility that SPT-CL~J2106-5844 is an exceptionally massive cluster for its redshift. By combining the X-ray, SZ, and velocity dispersion data, \citealt{Foley2011} estimated the cluster mass to be $M_{200}=(1.27\pm0.21)\cdot10^{15}~\mathrm{M_{\odot}}$, implying at the time it was the most massive\footnote{In context, the SPTpol Extended Cluster Survey \citep{Bleem2020} recently identified a stronger SZ decrement from an even higher redshift ($z=1.22$), more massive cluster, SPT-CL~J0329-2330. Like SPT-CL~J2106-5844, which is confirmed at a significance of $19\sigma$ in recent SZ survey data \citep{Hilton2020} from the Atacama Cosmology Telescope \citep[AdvACT;][]{Henderson2016}, SPT-CL~J0329-2330 was also confirmed by AdvACT \citep[again, in][]{Hilton2020} to be more massive than SPT-CL~J2106-5844. We note that \cite{Hilton2020} report a slightly lower SZ flux for SPT-CL~J2106-5844 than that reported by \cite{Bleem2015}, but the values are consistent, as are the relative mass rankings. We also highlight MOO~J1142+1527 \citep{Gonzalez2015}, which currently occupies third place in the AdvACT catalog (\citealt{Hilton2020}; see also \citealt{Dicker2020} and \citealt{Ruppin2020} for further comparison). All told, SPT-CL~J2106-5844 remains one of the three most massive $z>1$ clusters currently known when treated as a singular overdensity.} singular, virialized object known at $z>1$ (see also \citealt{Amodeo2016}, \citealt{Schrabback2018}, \citealt{Bulbul2019}, and \citealt{Kim2019} for further independent mass estimates for this system; we note that the assumption made in several previous studies that this system can be described by a single mass component could present a critical source of bias, given the clear evidence for two subclusters we present in this work).

\textit{Chandra} X-ray observations provide further information about the thermodynamic state of the intracluster medium (ICM) within SPT-CL~J2106-5844. The X-ray surface brightness distribution is found to be irregular, indicating that SPT-CL~J2106-5844 may be undergoing a merger. The slight skewness of the velocity distribution of the cluster members measured by \citet{Foley2011} provides additional, marginal evidence for the possible existence of intracluster substructures, supporting the interpretation that the cluster is in the midst of a major merger. The same interpretation was later put forth in several works on or including SPT-CL~J2106-5844 (see, e.g., \citealt{McDonald2013}, \citealt{Bartalucci2017}), based on a number of independent proxies for the dynamical state of the cluster. The reconstruction of the intracluster temperature distribution further indicated that SPT-CL~J2106-5844 harbors a remnant cool core that seems to have survived the potential merger event \citep{Foley2011}. Successive X-ray measurements by \textit{XMM-Newton} confirmed the disturbed morphology of the cluster, pointing to the potential presence of substructures within the ICM \citep{Bartalucci2017}. An extended intracluster feature was independently identified in the \textit{Chandra} observation of SPT-CL~J2106-5844 by \citet{Birzan2017}, who interpreted it as either due to X-ray cavities, in contrast with previous analyses of the same data \citep{Hlavacek2015}, or due to sloshing.

Additional support in confirming this system as a major merger was subsequently provided in a weak-lensing study by \cite{Kim2019}. The reconstructed mass distribution is revealed to be composed of a main structure, co-spatial with the SZ and X-ray signal from SPT-CL~J2106-5844, and a less massive western extension. The main mass component is bimodal mainly in the north-south direction, with the two subcomponents distributed in the same direction as the asymmetry in the X-ray surface brightness. The northern peak approximately corresponds to the central X-ray cusp and is coincident with the position of the brightest cluster galaxy (BCG), while the other, southern peak corresponds to the most prominent region in the galaxy number density distribution. Yet the dynamics of the ICM remained largely unknown.

In this paper, we strengthen the case that SPT-CL~J2106-5844 is undergoing a major merger, incorporating new evidence that includes archival subarcminute-resolution SZ observations using the 12-meter Atacama Large Millimeter/Submillimeter Array \citep[ALMA;][]{Wootten2009} jointly with the 7-meter Atacama Compact Array \citep[ACA, also known as the Morita Array;][]{Iguchi2009}, constraints on the radio emission measured using data from the Evolutionary Map of the Universe \citep[EMU;][]{Norris2011} and the Australia Telescope Compact Array \citep[ATCA;][]{Wilson2011}, and a reanalysis of the {\it Chandra} X-ray data. We provide an overview of the measurements used for our analyses and the data reduction details in Sect.~\ref{sec:data}. In Sect.~\ref{sec:res}, we describe our methodology and results, which reveal the extended and bimodal distribution of the ICM associated with the main mass components in SPT-CL~J2106-5844, as well as a high level of activity from active galactic nuclei (AGNs) within the cluster. The radio observations provide evidence for an extended radio complex, with an overall size of $\sim400~\mathrm{kpc}$ and dominated by a large radio galaxy associated with the BCG. The implications of our observations and an exploration of their potential interpretations are detailed in Sect.~\ref{sec:discuss}. We summarize the present work as well as provide suggestions for future observations in Sect.~\ref{sec:conc}.

Throughout this work, we assume a spatially flat $\Lambda$ Cold Dark Matter cosmological model, with $\Omega_{\mathrm{M}}=0.30$, $\Omega_{\Lambda}=0.70$, and $H_0=70.0~\mathrm{km\,s^{-1}\,Mpc^{-1}}$. Here $1\arcsec$ corresponds to $7.62~\mathrm{kpc}$ at the redshift of SPT-CL~J2106-5844 ($z=1.132$).

%-----------------------------------------------------------------------

\section{Observations and data reduction}\label{sec:data}

\subsection{Atacama Large Millimeter/Submillimeter Array}\label{sec:data:alma}
SPT-CL~J2106-5844 was observed by ACA as part of Cycle 2 operations during October 2016 (project code: 2016.1.01175.S, PI: S. Burkutean). A follow-up Cycle 3 observation using the main 12-meter array (hereafter referred to as ``ALMA'') was carried out between June and July 2018 with the aim of targeting potential contamination due to compact sources within the cluster field (project code: 2017.1.01649.S, PI: S. Burkutean). The integration times amount to 5 minutes and 24 minutes on source in Bands 3 and 4 with ALMA, and 1.7 hours on source with each band with ACA. Both the ACA and ALMA observations were designed to provide wideband measurements of SPT-CL~J2106-5844 over the frequency ranges of $84-100~\mathrm{GHz}$ (Band 3) and $137-153~\mathrm{GHz}$ (Band 4), split over four $2~\mathrm{GHz}$-wide spectral windows per band. The array configurations for the observations were chosen to provide similar coverage of the $u\varv$ (i.e., Fourier) plane, with ACA and ALMA sampling the ranges of $u\varv$ distances $2.1-20.0~\mathrm{k\lambda}$ and $3.9-105.5~\mathrm{k\lambda}$, respectively. For reference, these ranges correspond to angular scales of $2-96\arcsec$ (i.e., $15-731~\mathrm{kpc}$ at the cluster redshift).

In this study, we employ the calibrated measurement sets provided by the European ALMA Regional Centre \citep{Hatziminaoglou2015}. We measure root-mean-square (RMS) noise levels of $107.8~\mathrm{\mu Jy}$ and $99.4~\mathrm{\mu Jy}$ ($27.9~\mathrm{\mu Jy}$ and $15.2~\mathrm{\mu Jy}$) in the Band 3 and 4 ACA (ALMA) data, respectively. We include the nominal $5\%$ uncertainty in our error budget for the flux measurement, as reported in the Cycle 2 and Cycle 3 ALMA Technical Handbooks, in the analysis of the ALMA and ACA data.

\subsection{\textit{Chandra}}\label{sec:data:chandra}
\begin{figure}
    \centering
    \includegraphics[clip,trim=12.15cm 1.80cm 0.25cm 2.68cm,width=\columnwidth]{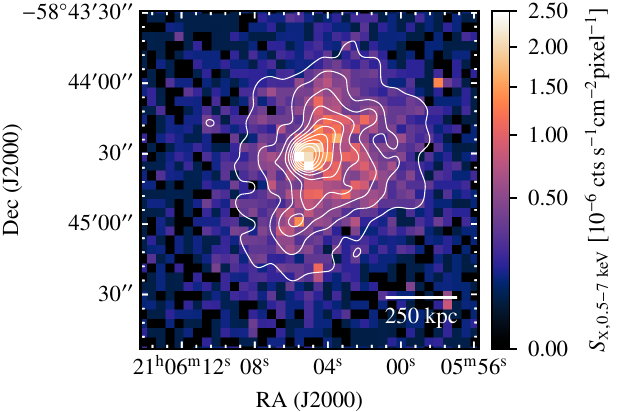}
    \caption{Merged, exposure-corrected 0.5-7~keV map of the {\it Chandra} observations of SPT-CL~J2106-5844. The image is binned into $8\times8$ pixels from the native {\it Chandra} pixel size of 0.492\arcsec. The asymmetric X-ray brightness distribution hints at the disturbed dynamical state of SPT-CL~J2106-5844 \citep{Foley2011,Kim2019}. Overlaid are contours from a smoothed version of the X-ray surface brightness image.}
    \label{fig:xsurf}
\end{figure}

We consider two archival \textit{Chandra} observations of SPT-CL~J2106-5844 of $24.72~\mathrm{ks}$ (ObsID: 12180, PI: S. Murray), and $48.13~\mathrm{ks}$ in ACIS-I VFAINT mode (ObsID: 12189, PI: G. Garmire).
The data were reprocessed using the {\it Chandra} Interactive Analysis of Observations (\texttt{CIAO}) software version 4.12 with \texttt{CALDB} version 4.9.2.1, following a standard reduction procedure (see, e.g., \citealt{Sayers2019}). We employed the \texttt{chandra\_repro} script to produce level 2 event files.
The \texttt{CIAO} task \texttt{merge\_obs} was then used to merge all the available observations and obtain the broadband (0.5-7~keV) exposure-corrected image shown in Fig.~\ref{fig:xsurf}. Finally, \texttt{wavdetect} was applied to the combined map to detect point sources, and we visually inspected the results to identify spurious detections. 

The low number of source counts in the available \textit{Chandra} data did not allow for deriving significant constraints on the temperature distribution of the ICM within SPT-CL~J2106-5844. Instead, we infer a global X-ray temperature of $9.5\substack{+1.8\\-1.6}~\mathrm{keV}$ using approximately 800 source counts after background subtraction and masking of all point sources identified using \texttt{wavdetect}. The inferred temperature is consistent at $1\sigma$ with previous estimates (e.g., \citealt{Foley2011}, \citealt{Bartalucci2017}). We note that the X-ray core (also identified as a point source by \texttt{wavdetect}) includes the radio-loud AGN and thus was excluded in our spectral analysis.

\begin{figure}
    \centering
    \includegraphics[clip,trim=12.15cm 2.68cm 0.25cm 2.68cm,width=\columnwidth]{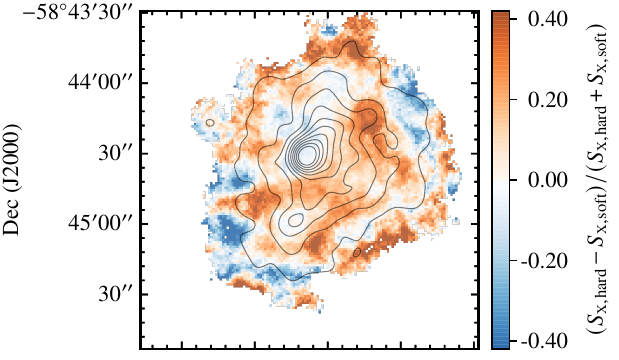}
    \includegraphics[clip,trim=12.15cm 1.80cm 0.25cm 2.68cm,width=\columnwidth]{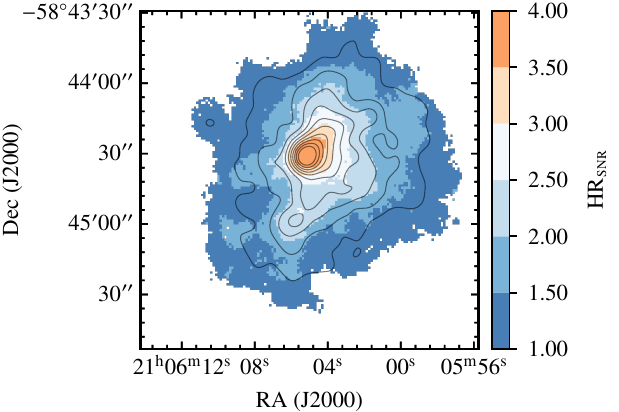}
    \caption{Hardness ratio map (top) and corresponding significance (bottom) based on the surface brightness image at the energy bands 0.3-1.5~keV and 1.5-7.5~keV. Prior to computing the hardness ratio map, the images in each of the energy bands are filtered with a top-hat kernel with radius of $7.5\arcsec$. Further, in order to suppress regions at low signal-to-noise, we mask all the pixels in the hardness ratio map whose corresponding surface brightness is lower than 2\% of the amplitude of the X-ray peak. For reference, we overlay on the hardness ratio map the same contours shown in Fig.~\ref{fig:xsurf} from the smoothed X-ray surface brightness.}
    \label{fig:xhard}
\end{figure}

Shown in Fig.~\ref{fig:xhard} is the X-ray hardness ratio map derived from the exposure-corrected images in the 0.3--1.5~keV and 1.5--7.5~keV range. These ranges were chosen to include the line-dominated and soft emission from the cluster below a rest frame energy of 3.2~keV.  We compute the map as the ratio of the hard (1.5--7.5~keV) minus the soft (0.3--1.5~keV) image, divided by the sum of the two. We estimate the significance of the observed features by bootstrapping over noise realizations for the two energy bands. The inferred uncertainties for the hardness ratio map were then employed to compute the significance map shown in the bottom panel of Fig.~\ref{fig:xhard} as detailed in the Appendix C2 of \citet{Strickland2002}. We note that, aside from the  bright core region, it is not possible to identify any structure with a significance higher than $2\sigma$. Although the inferred hardness ratio map is helpful in providing additional pieces of information on the physical state of the ICM, it should be interpreted with the usual caveats that many apparent structures have low significance.

\subsection{Evolutionary Map of the Universe Pilot Survey}\label{sec:data:askap}
\begin{figure}
    \centering
    \includegraphics[clip,trim=12.15cm 2.68cm 0.25cm 2.68cm,width=\columnwidth]{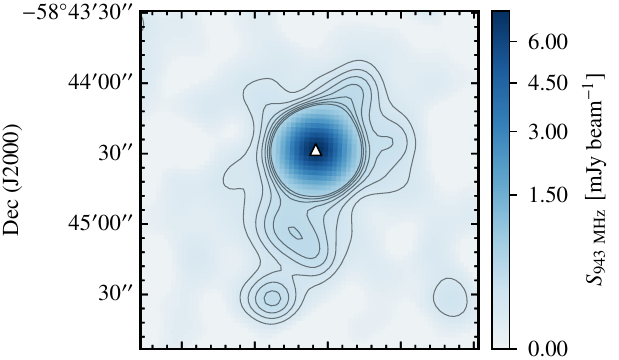}
    \includegraphics[clip,trim=12.15cm 2.68cm 0.25cm 2.68cm,width=\columnwidth]{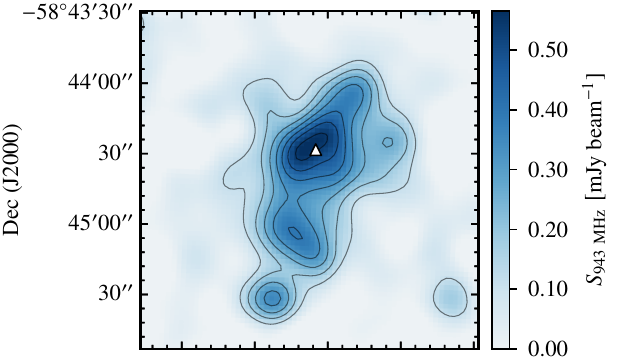}
    \includegraphics[clip,trim=12.15cm 1.80cm 0.25cm 2.68cm,width=\columnwidth]{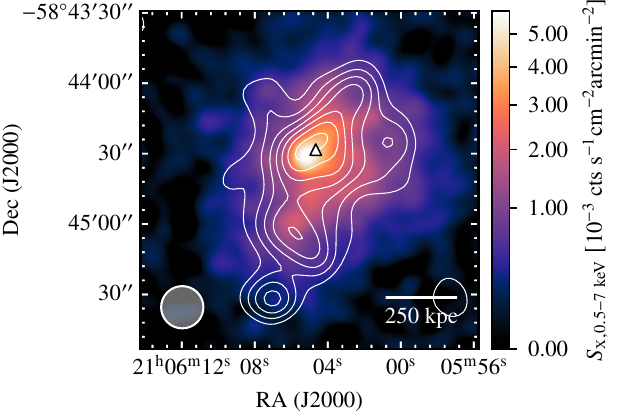}
    \caption{EMU observation of SPT-CL~J2106-5844 before (upper) and after (middle) subtracting the radio emission from the BCG, and comparison of the latter with the smoothed X-ray surface brightness (lower). The bright compact emission associated with the BCG (triangle marker; \citealt{Song2012}) is surrounded by an extended radio structure elongated in the same direction of the X-ray signal. In all panels, the EMU contours start at the $3\sigma$ level and increase by steps of $2\sigma$ ($\sigma=36~\mathrm{\mu Jy~beam^{-1}}$; see Sect.~\ref{sec:data:askap}). The synthesized beam is depicted in the bottom left corner of the lower panel. We note that the $18\arcsec$ resolution is a consequence of the smoothing applied to make the resolution uniform across the entire EMU Pilot Survey.}
    \label{fig:emu}
\end{figure}
EMU\footnote{\url{http://www.emu-survey.org/}} is a wide-field radio-continuum survey project \citep{Norris2011} using the Australian Square Kilometre Array Pathfinder \citep[ASKAP;][]{Johnston2007} for mapping the entire southern sky, up to $+30^\circ$ north, with a continuum sensitivity of $10-20~\mathrm{\mu Jy~beam^{-1}}$ over 288~MHz bandwidth centered at 943~MHz. In this work, we employ observations from the first EMU Pilot Survey (Norris et al.\ in prep), which covers 270 deg$^2$ to an RMS noise level of $25-30~\mathrm{\mu Jy}$ at an average resolution of 10--15\arcsec. The observations include baselines as short as 23 meters, and the maximum observable angular scales are $\gtrsim 48 \arcmin$ (or $\gtrsim 0.8^\circ$, much larger than the field of interest for the present work) depending on the elevation of the field at the time of observation. The EMU Pilot Survey fields were within the SPT 2500 deg$^2$ survey footprint, and include observations of SPT-CL~J2106-5844.

The ASKAP data used in this work are from Scheduling Block 9410, observed for 10 hours on 24 July 2019. They were calibrated with a 2 hour observation of PKS~B1934-638 (Scheduling block 9409) using the flux density model of \citet{Reynolds1994}. The image reduction was performed with the \texttt{ASKAPsoft}\footnote{\url{https://www.atnf.csiro.au/computing/software/askapsoft/sdp/docs/current/index.html}} pipeline, utilizing a two Taylor-term (T0 and T1) multi-scale CLEAN algorithm \citep{Rau2011}. The resulting 36 single-beam images were convolved to a resolution of $18\arcsec$ full-width-half-maximum and combined with a linear mosaic in order to provide uniform angular resolution across the entire survey.
Spectral indices could then be recovered using the ratio T1/T0 of the convolved zero- and first-order Taylor coefficient maps. Local to SPT-CL~J2106-5844, the RMS noise level in the image is measured to be 36~$\mathrm{\mu Jy~beam^{-1}}$.

The EMU image of SPT-CL~J2106-5844 (Fig.~\ref{fig:emu}) provides clear evidence for the presence of a strong radio emission in the direction of the galaxy cluster. The overall morphology of the signal is observed to consist of an extended radio structure covering roughly $500~\mathrm{kpc}$ around a compact radio core that dominates the flux density and is coincident with the BCG. We estimate the flux density of the compact core by subtracting an unresolved source from the image and determining when there were no significant positive or negative residuals on top of the extended emission. The resulting core-subtracted image best highlighting the diffuse radio structure is shown in the lower panel of Fig.~\ref{fig:emu}. The flux density of the compact source cannot be measured unambiguously because of substructure in the residual surrounding diffuse emission.  We note that the quoted errors in the flux estimated from this subtraction procedure are significantly higher than the random off-source noise RMS level.  We find a flux of $6.7\pm 0.3~\mathrm{mJy}$ at the EMU central frequency of $943~\mathrm{GHz}$ and a spectral index of $-1.2\pm0.1$, indicating the flux is likely dominated by an extended but small-scale structure (unresolved at 18\arcsec), rather than more compact emission from the AGN alone. We note that, unfortunately, due to the faintness of the extended structure, we were only able to reliably determine the spectral index for the compact central source.

We additionally checked for any signatures of polarization in the direction of SPT-CL~J2106-5844 by performing rotation measure synthesis across a 144~MHz bandwidth for the low-band EMU data that were processed for the Polarisation Sky Survey of the Universe's Magnetism\footnote{\url{https://askap.org/possum/}} \citep[POSSUM;][]{Gaensler2010} survey. While polarized flux was visible for a number of sources in the field, we detected no polarized emission from SPT-CL~J2106-5844 at a $3\sigma$ level of $350~\mathrm{\mu Jy~beam^{-1}}$. This results in upper limits for the fractional polarization of $\sim 3\%$ for the core, and $\sim 50\%$ for the brightest parts of the diffuse emission. Although emission from radio halos is expected to be unpolarized, radio galaxy or relic structures are typically polarized at the tens of percent level, but these limits are only indicative and do not provide meaningful constraints on the specific nature of the radio emission.

\subsection{Australia Telescope Compact Array}\label{sec:data:atca}
\begin{figure}
    \centering
    \includegraphics[clip,trim=0 0.05cm 0 0.15cm,width=\columnwidth]{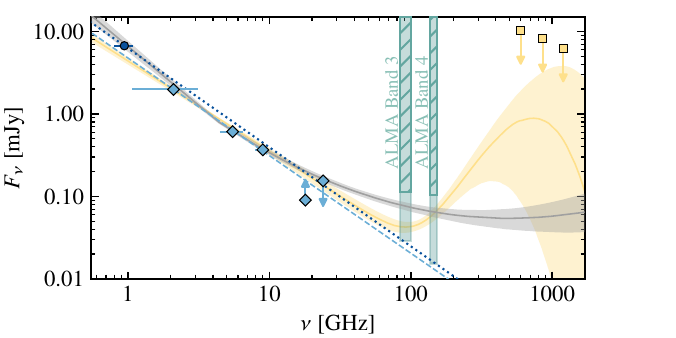}
    \caption{Spectral properties of the BCG. The points denote the flux density of the bright radio core as measured by EMU (dark blue circle) and ATCA (light blue diamonds), while the yellow squares mark the \textit{Herschel}-SPIRE upper limits on the source flux. The dotted and dashed lines correspond to the power-law spectrum inferred from the EMU and the ATCA data, respectively. The solid line is the modified power law derived by fitting the ALMA and ACA observations jointly with the ATCA flux measurements, with respective 68\% credible intervals denoted by the shaded region (Sect.~\ref{sec:res:sz:twosz}). The mint green vertical bands indicate the frequency ranges covered and the sensitivities achieved by the ALMA (solid) and ACA (hatched) Band 3 and Band 4 observations. Here, the bottom edges of all four bands denote the sensitivity of the corresponding measurements. The radio spectrum of the BCG is measured to become shallower when going to higher frequencies. In yellow is the tentative spectrum for the BCG obtained when including a thermal dust spectral component (Sect.~\ref{sec:res:sz:pts}) in addition to a power law model.} 
    \label{fig:indexc}
\end{figure}

To better resolve any potential radio structures and compact sources in the cluster field, we complemented the EMU data with observations from ATCA. For our analyses, we employed measurements in three different frequency bands: 2.1~GHz data from the projects C2585 (PI: R. Kale) and C2837 (PI: M. Johnston-Hollitt), and 5.5 and 9~GHz measurements (project code: C3217, PI: S. Burkutean; C1563, PI: W. Walsh), covering the $u\varv$-ranges $0.1-60~\mathrm{k\lambda}$, $0.4-95~\mathrm{k\lambda}$, and $0.8-150~\mathrm{k\lambda}$, respectively. All bands have a bandwidth of $2048~\mathrm{MHz}$. In this work we mainly focus on the 2.1 and 5.5~GHz data, which yield the most significant detections of the radio emission from SPT-CL~J2106-5844 due to both the spectrum declining with frequency and to resolution effects (Fig.~\ref{fig:indexc}). The 9~GHz data marginally improve our constraints on the spectral properties of the radio sources in the field. We also examined the 18 and 24~GHz ATCA data, observed as part of Project C3217, but do not include them here due to the faintness of the sources at these frequencies. We note for completeness that, at 18~GHz, the flux of the central source is measured to be $90\pm23~\mathrm{\mu Jy}$. However, the emission that we consider to belong to a compact object appears marginally extended in the 18~GHz data, and hence we treated the inferred value as a lower limit on the source flux. On the other hand, the central source is not detected in the 24~GHz ATCA data, placing a $\sim 3\sigma$ constraint of $154~\mathrm{\mu Jy~beam^{-1}}$ on its surface brightness. Despite formally representing an upper limit under the assumption of the source being point-like, we note that the core emission looks resolved in the 18~GHz map and would consequently be affected by an unknown degree of spatial filtering at 24~GHz. As a consequence, the aforementioned estimate is likely a lower limit on the actual flux of the compact source.

We reduce and image all ATCA data using the Common Astronomy Software Application\footnote{\url{https://casa.nrao.edu/}} \citep[\texttt{CASA};][]{McMullin2007} package version 5.3.0. In all cases, we split the 2048~MHz bandwidth into eight spectral windows for calibration.  We use the primary calibrator source 1934-638 to solve for the bandpass in amplitude and phase. We then solve for additional amplitude and phase bandpass solutions using the bright point-like sources observed as interleaved calibrators, bootstrapping the flux densities of these sources. For the 2.1~GHz data, we image using natural weighting and then perform three rounds of self-calibration in phase. The higher-frequency data sets did not contain emission of high enough signal-to-noise ratio to warrant self-calibration. The noise levels of the naturally weighted maps, imaged using multifrequency synthesis over the whole bandwidth, are 40, 20 and 15~$\mathrm{\mu Jy~beam^{-1}}$ for the 2.1, 5 and 9~GHz data sets, respectively.

\begin{figure}
    \centering
    \includegraphics[clip,trim=12.15cm 2.68cm 0.25cm 2.68cm,width=\columnwidth]{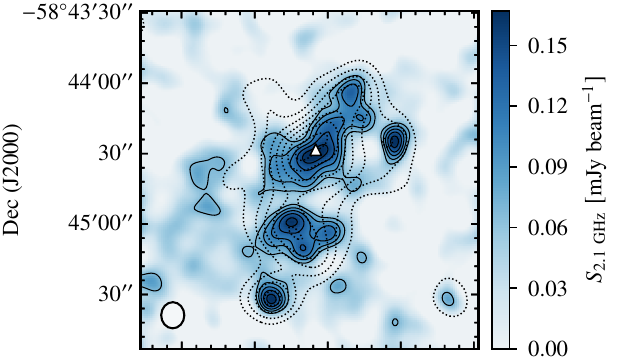}
    \includegraphics[clip,trim=12.15cm 2.68cm 0.25cm 2.68cm,width=\columnwidth]{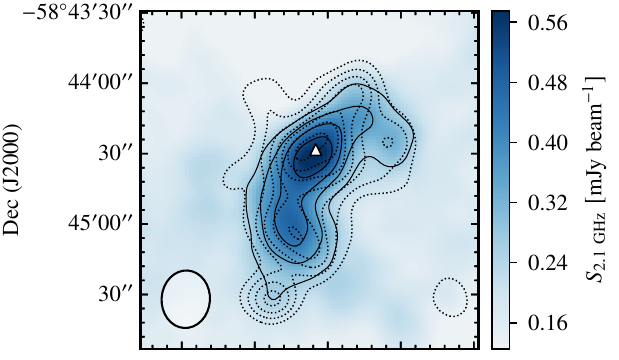}
    \includegraphics[clip,trim=12.15cm 1.80cm 0.25cm 2.68cm,width=\columnwidth]{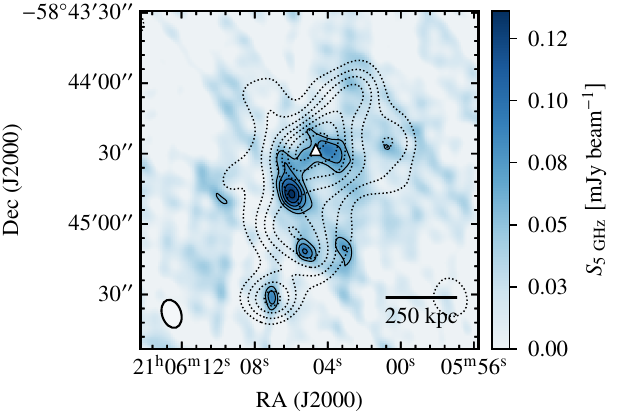}
    \caption{ATCA observations after subtraction of compact source components (see Sect.~\ref{sec:data:atca} for details). The maps correspond to the 2.1~GHz data generated assuming Briggs weighting with robust parameter set to 0.00 (upper; $\sigma=23~\mathrm{\mu Jy~beam^{-1}}$), and to the naturally weighted 2.1~GHz (middle; $\sigma=87~\mathrm{\mu Jy~beam^{-1}}$) and 5~GHz (lower; $\sigma=19~\mathrm{\mu Jy~beam^{-1}}$) images. As in Fig.~\ref{fig:emu}, the triangle marks the BCG position \citep{Song2012}. To better highlight the extended radio features, we applied $u\varv$-tapers of $10$ and $20~\mathrm{k\lambda}$ to the 2.1 and 5~GHz measurements, respectively. For all panels, the contours start from the $3\sigma$ level, with increment every $1\sigma$. The higher resolution offered by ATCA allows us to disentangle the small-scale morphology of the diffuse radio structure detected by EMU (dotted contours).}
    \label{fig:atca}
\end{figure}

To get a more detailed view of the extended radio emission, we generate maps of the ATCA data containing only the diffuse component of the radio signal. Specifically, for each measurement set, we remove any compact sources contaminating the field by performing a first imaging run assuming a uniform weighting, and then employing the reconstructed CLEAN components as input models to be subtracted from the each of the raw interferometric data sets. The resulting images for the 2.1 and 5~GHz measurements are presented in Fig.~\ref{fig:atca}. We do not include the 9~GHz map as no extended emission was observed at a significant level in the compact source-subtracted image.

%-----------------------------------------------------------------------

\section{Results}\label{sec:res}
\subsection{Radio imaging}\label{sec:res:radio}
The radio observations clearly show that SPT-CL~J2106-5844 hosts a complex radio structure. Here, we present the main results obtained through the analysis of the imaging products from the ASKAP and ATCA data introduced in the previous section.

\begin{figure}
    \centering
    \includegraphics[clip,trim=12.15cm 1.80cm 0.25cm 0.00cm,width=\columnwidth]{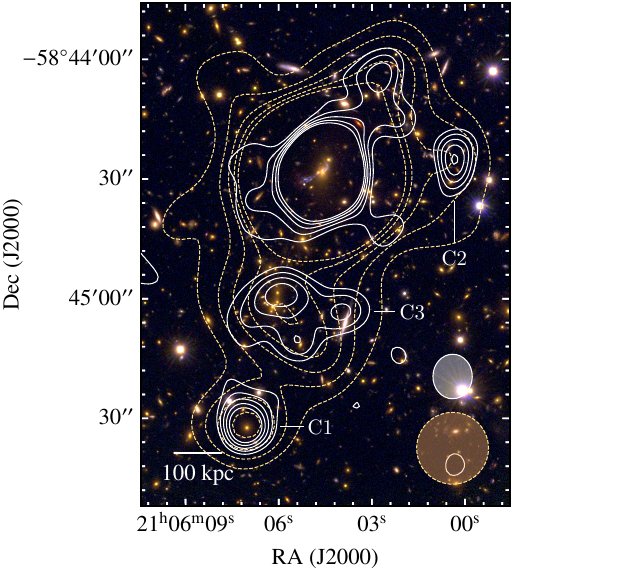}
    \caption{Composite HST image with contours from the EMU image (dashed yellow) from Fig.~\ref{fig:emu} and the 2.1~GHz ATCA map (solid white) generated assuming Briggs weighting and robust parameter at 0.00 to highlight any compact contributions to the observed radio emission. The corresponding synthesized beams are shown in the bottom right corner following the same color convention. The outskirts of the extended structure observed in the EMU image are found to be contaminated by point-like sources. The labeled features correspond to the sources discussed in Sect.~\ref{sec:res:radio:pts} and \ref{sec:res:radio:ext}.}
    \label{fig:compact}
\end{figure}

\subsubsection{Compact sources}\label{sec:res:radio:pts}
The comparison of high-resolution ATCA observations of SPT-CL~J2106-5844 with the distribution of the radio signal in the EMU map provides key information on the population of unresolved sources contaminating the diffuse radio emission. We emphasize that we subtracted from the EMU data only the model corresponding to the central radio source. On the one hand, the EMU resolution does not allow a clean discrimination between other compact sources and any potential co-spatial diffused components. On the other hand, as described above, it was not possible to reliably measure the spatially distributed spectral shape. This prevented us from the possibility of subtracting the signal from point-source model components based on the extrapolation of the individual ATCA fluxes up to the EMU frequency.
As shown in Fig.~\ref{fig:compact}, several compact sources are found in the peripheral regions of the extended radio structure. Among the most prominent features is the southernmost component in the EMU map, though, with no redshift identification, we cannot rule out the possibility that it is not associated with the cluster itself. This feature is found to coincide with a compact emission in the ATCA data (labeled as C1 in Fig.~\ref{fig:compact}), and has an optical counterpart in the Hubble Space Telescope (HST) image. A second source (C2) is identified at the $\sim4\sigma$ level in the uniformly weighted 2.1~GHz ATCA map and found to be surrounded by a combination of diffuse emission and a blend of multiple compact sources. If associated with the central radio galaxy, these may correspond to knot-like features in one arm of the radio jet. On the other hand, the radio emission from C2 may be related to a separate radio galaxy, although with the current resolution it is difficult to determine its optical counterpart.

\subsubsection{Diffuse emission}\label{sec:res:radio:ext}
The high-resolution, diffuse-only ATCA images (Fig.~\ref{fig:atca}) show that the radio emission has a narrow structure elongated in the northwest-southeast direction, and composed of main structure surrounding the BCG and a separate diffuse component (C3) detected midway between the BCG and the compact source C1.
We assessed the possibility that the observed radio morphology is impacted by the underlying SZ decrement. However, extrapolating to $943~\mathrm{MHz}$, the integrated SZ signal within $1\arcmin$ radius expected for SPT-CL~J2106-5844 results in an SZ flux (decrement) $\sim30$ times smaller than the radio power measured from the EMU map. We note that this is just an indicative upper limit, as the SZ effect is more severely affected than the radio structure by the interferometric filtering of the signal components on large scales. While the SZ signal strength is higher at 2.1 and $5~\mathrm{GHz}$ than it is at $943~\mathrm{MHz}$, the lack of baselines sensitive to any arcminute-scale signal ensures that the SZ contamination is negligible in the ATCA data as well.

\begin{figure}
    \centering
    \includegraphics[clip,trim=0 0.05cm 0 0.15cm,width=\columnwidth]{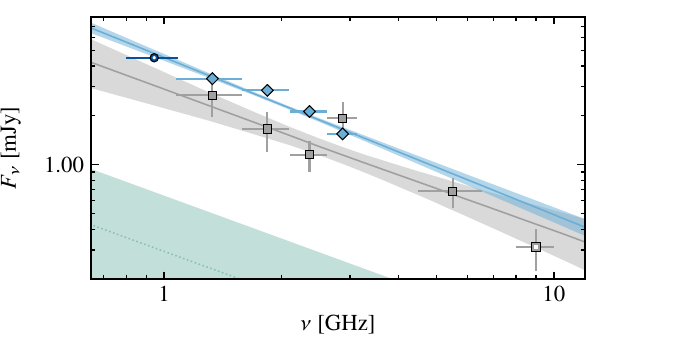}
    \caption{Integrated spectrum of the extended radio structure. The light blue and gray points denote the estimates of the diffuse flux computed by integrating over the regions with significance $>2.5\sigma$ or by fitting with Gaussian profiles the 0.4-$2~\mathrm{k\lambda}$ naturally weighted maps of each spectral band, respectively (see the main text in Sect.~\ref{sec:res:radio:ext} for details). We use the same color scheme for the corresponding best-fit lines and $1\sigma$ confidence intervals. The dark blue point represents the integrated flux measured from the EMU image. All the open markers denote values that were not included in this analysis. For reference, we report in green the expected flux and corresponding uncertainties of a radio halo for a cluster of mass equal to the best-fit value reported in \citet{Kim2019} for SPT-CL~J2106-5844 and following the radio power to mass scaling relation by \citet{Cuciti2021b} under the assumption of a spectral index $\alpha=-0.87$  (Sect.~\ref{sec:res:radio:ext}).}
    \label{fig:indexe}
\end{figure}

In order to gain additional information on the origin of the observed radio emission, we aimed at performing a spectral analysis of the extended radio signal. Unfortunately, the different angular resolutions of the available radio data, combined with the severe sparsity and low significance of the observations at high frequencies, do not allow for building a detailed map of the spectral index distribution of the source. We hence provide in Fig.~\ref{fig:indexe} estimates of the spectral index of the extended emission based on its integrated flux densities. Since the lower resolution of the EMU data does not allow us to disentangle the diffuse emission from any compact signal as shown above in the case of the ATCA observations, we employ only ATCA measurements to compute the integrated spectral index. Given the heterogeneity of the available data, we perform two attempts to get a spectral constraint on the diffuse emission.

First, we split the 2.1~GHz data in four spectral bands and image them separately from the others using natural weighting and a $u\varv$-taper of $10~\mathrm{k\lambda}$. In this case, we further exclude the 5 and 9~GHz data as they do not provide the required $u\varv$-coverage to significantly detect any extended radio structure. The integrated flux for each of the frequency windows (blue points in Fig.~\ref{fig:indexe}) is then obtained by measuring the flux within the region with significance larger than $2.5\sigma$. The spectral fit provides an estimate of the integrated spectral index of $-0.96\substack{+0.06\\-0.06}$. Notably, the extrapolation of the resulting spectrum is consistent with the estimate of the diffuse flux from the EMU image.
 
To probe the spectral behavior of the extended emission over a broader range of frequencies, we repeat the above analysis by imaging the four 2.1~GHz sub-bands and the 5~GHz data after selecting only the baselines between 0.4 and $2~\mathrm{k\lambda}$. Since the diffuse source was only marginally resolved in the resulting maps, we fit its signal in each maps with a Gaussian profile and integrate over the reconstructed model to compute the diffuse flux (gray points in Fig.~\ref{fig:indexe}). We perform a similar analysis on the 9~GHz measurements, but exclude the corresponding flux estimate from the fit. In fact, only a small number of baselines are available in the common $0.4-2~\mathrm{k\lambda}$ range, and we hence consider in the 9~GHz data all the $u\varv$ scales smaller than $3~\mathrm{k\lambda}$. Since the source would be better resolved than for the other data sets, we expect the reconstructed flux to be marginally biased to a lower value. In this case, the integrated spectral index is measured to be $-0.87\substack{+0.22\\-0.23}$.

Despite the clear differences in the normalization of the reconstructed spectra, ascribable to the different method employed to measure the integrated fluxes, both the analyses show that the integrated spectrum seems to follow a well-behaved power law over the entire range of frequencies probed by the ATCA and EMU radio data. The lack of a break in the spectral index at high frequencies indicates that the extended radio emission is a mixture of multiple electron populations with different ages and/or magnetic field strengths (see, e.g., Fig. 5 of \citealt{Rajpurohit2020}). Observations with a higher angular resolution will be key to disentangle the multiple contributions to the observed diffuse structure and to allow an understanding of whether this could be ascribed to emission from a single AGN host or to a complex of distinct radio sources.

\subsection{SZ modeling}\label{sec:res:sz}
The reconstruction of the SZ signature of SPT-CL~J2106-5844 is performed by jointly analyzing the Band 3 and Band 4 ALMA and ACA (hereafter, ALMA+ACA) interferometric data with the same $u\varv$-space Bayesian forward modeling technique employed in \citet{DiMascolo2020}. For details, we refer to the discussion in \citet{DiMascolo2019} and references therein. In brief, the reconstruction method is devised to sample the Fourier transform of a map of the SZ effect from a model galaxy cluster to the $u\varv$-plane coordinates of the sampled visibilities. This avoids issues related to the deconvolution of interferometric data (e.g., heavily correlated image-space noise), as well as taking full advantage of the knowledge of the exact visibility sampling function. 

We compute the average relativistic corrections to the thermal SZ spectrum by means of the approximation by \citet{Itoh2004}. Given the large uncertainties on the temperature distribution, we assume this to be described by an isothermal model and consider a split-normal prior on the average cluster temperature, with mode and standard deviations respectively equal to our spectroscopic X-ray estimate and corresponding uncertainties (Sect.~\ref{sec:data:chandra}).
It is however worth noting that we expect such an approximation to have a marginal impact on our results. In fact, at the Band 3 and Band 4 frequencies, the expected deviation from the nonrelativistic thermal SZ spectrum is comparable to the uncertainties inherent to the absolute flux calibration of the ALMA+ACA data, and subdominant to the statistical error bars.

\subsubsection{Free modeling run}\label{sec:res:sz:onesz}
\begin{figure}
    \centering
    \includegraphics[clip,trim=0 1.25cm 0.75cm 0,width=\columnwidth]{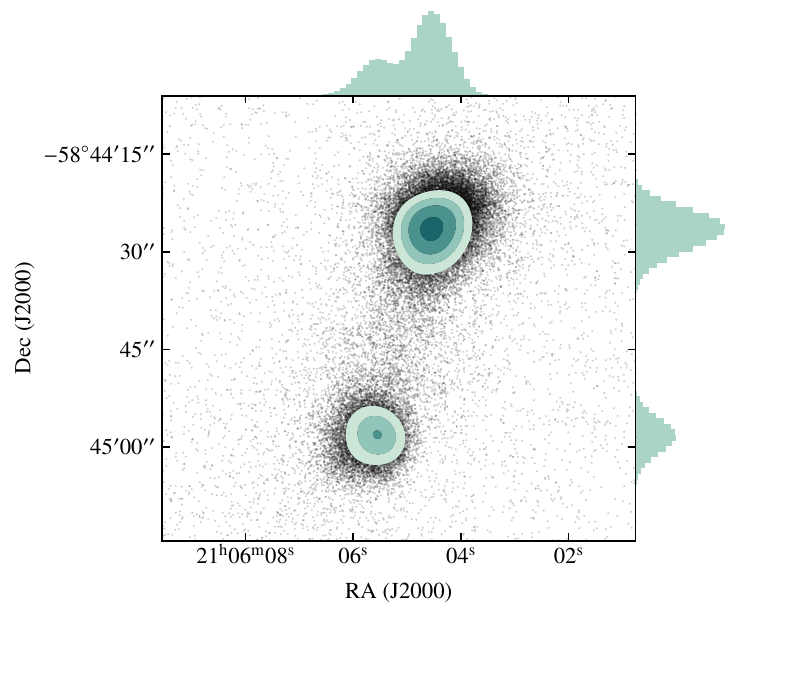}
    \caption{Marginalized posterior probability for the coordinates of the SZ centroid when considering the model with a single gNFW component. The reconstructed posterior distribution is found to track two distinct features in the SZ signal from SPT-CL~J2106-5844. The contours correspond to 0.5, 1, 1.5, and $2\sigma$ significance levels.}
    \label{fig:bimodal}
\end{figure}

The model of the thermal SZ effect from SPT-CL~J2106-5844 is obtained by numerical integration of a generalized Navarro-Frenk-White \citep[gNFW;][]{Nagai2007} pressure profile along the line-of-sight direction (see \citealt{Mroczkowski2009}, \citealt{Arnaud2010}, or \citealt{DiMascolo2020} for specifics). We perform a first modeling run assuming the universal parametrization by \citet{Arnaud2010} but allow the model centroid, mass parameter, and slope at small radii to vary. The centroid coordinates are assumed to have a uniform prior extending over the field of view of ACA, while we consider wide uninformative priors for both the mass and slope parameters.

As discussed in \citet{DiMascolo2020}, we further perform a free search across the entire ALMA+ACA field of view for compact sources that may contaminate the SZ signal. Specifically, we use point source models to describe the contamination from unresolved compact sources with flux characterized by a simple power-law spectral dependence. Uninformative priors are assumed for the position, amplitude, and spectral index. We do not consider any model component to describe the extended radio source, which should not significantly contaminate the measured SZ signal. Extrapolating the extended radio flux on arcminute scales up to the ALMA+ACA frequencies by assuming a power law spectrum with average spectral index $\alpha=-1$ (Sect.~\ref{sec:res:radio:ext}), we obtain values of $16~\mathrm{\mu Jy}$ and $9~\mathrm{\mu Jy}$ in Band 3 and Band 4, respectively. These are roughly an order of magnitude lower than the noise RMS of the ACA data, most sensitive to the SZ effect, as well as around 10,000 times lower than the integrated SZ signal expected for a cluster with the same mass and redshift as SPT-CL~J2106-5844.

Analogous to the cases reported in \citet{DiMascolo2020} for the two Massive and Distant Clusters of WISE Survey (MaDCoWS; \citealt{Gonzalez2019}) galaxy clusters MOO~J0917$-$0700 and MOO~J2146$-$0320, which were targeted in the Verification with the ACA -- Localization and Cluster Analysis (VACA LoCA) program, the marginalized posterior probability of the SZ centroid coordinates is found to be bimodal (Fig.~\ref{fig:bimodal}), which strengthens the hypothesis that SPT-CL~J2106-5844 is also a merger with two distinct pressure peaks. Furthermore, the overall posterior distribution is similar to the asymmetric distributions seen in the mass and galaxy number density maps \citep{Kim2019}. We tested against possible systematics to assess the reliability of this result, and checked that the two SZ components are identified independent of any prior assumptions on the gNFW parameters or point-like component, or of the specific subsets of observations employed in the modeling.

No evidence for thermal SZ effect is found in the direction of the western mass extension reported by \citet{Kim2019}. This is not surprising because the projected mass density of this western substructure is at least a factor of two lower than that of the main component and hence is expected to contribute only a small SZ effect. This combines with the first null of the ACA primary beam to fall right on the position of the western feature, further attenuating its SZ signal.

\subsubsection{Unresolved source}\label{sec:res:sz:pts}
The fitting algorithm further identifies an unresolved (point-like) positive source in the direction of SPT-CL~J2106-5844, whose position is coincident with the bright compact component dominating the low-frequency radio emission from the cluster and consistent with the coordinates reported by \citet{Song2012} for the BCG. 
However, the low significance of the source in the ALMA+ACA observations limits our ability to constrain its spectral index using the ALMA+ACA data alone (see Fig.~\ref{fig:indexc}), leaving its marginalized posterior dominated by the inherent degeneracy with the flux normalization parameter.

To better constrain the spectral signature of the central unresolved source, we repeat the fit by including the flux measurements from the ATCA data at 2.1, 5.5, and 9~GHz. We do not consider here any information from the EMU observations as, at the EMU frequency, the extended radio structure may significantly contaminate the flux measured for the central source. Given the broad range of frequencies covered by the ATCA and ALMA+ACA observations, we extend the spectral model of the unresolved source to a modified power law, hence to fit for any potential spectral curvature at high frequency. We measure a positive curvature of the spectrum toward high frequencies (see Fig.~\ref{fig:indexc}), possibly suggesting the presence of a dust component within the cluster central galaxy (see \citealt{Fogarty2019} or \citealt{Romero2020} for examples of clusters with a dusty BCG).

In order to characterize the level of dust content of the central radio galaxy, we perform another modeling run including constraints on the submillimeter flux in the direction of the BCG obtained from the \textit{Herschel Space Observatory} \citep{Pilbratt2010}. We use Level 2 SPIRE \citep{Griffin2010} observations at 250, 350, and $500~\mathrm{\mu m}$ from the \textit{Herschel} Lensing Survey \citep[ObsID: 1342245542;][]{Egami2010}. It was not possible to distinctly detect any emission from the source, and we hence include only upper limits based on the average fluxes measured in each of the three \textit{Herschel} bands within apertures centered on the BCG position and with size equal to the respective beams. We consider a spectral model given by the sum of a power law and a modified blackbody spectrum, with the latter parametrized as \citep{Hildebrand1983}
\begin{equation}
  F_{\mathrm{dust}}(\nu) = (1+z)\,d^{-2}_{\textsc{l}}(z)\, M_{\mathrm{dust}} \, \kappa(\nu_{\textsc{rf}}) \, B_{\nu_{\textsc{rf}}}(T_{\mathrm{dust}}).
\end{equation}
Here, $d_{\textsc{l}}(z)$ is the luminosity distance for a source at redshift $z$, and the parameter $M_{\mathrm{dust}}$ corresponds the mass of the dust component. The terms $k_{\nu_{\textsc{rf}}}$ and $B_{\nu_{\textsc{rf}}}(T_{\mathrm{dust}})$ are the dust emissivity and the blackbody spectrum for a dust temperature $T_{\mathrm{dust}}$ measured at the rest-frame frequency $\nu_{\textsc{rf}}=(1+z)\,\nu$. For the dust opacity function, we consider a power-law form, $k_{\nu_{\textsc{rf}}}=k_{0}[(1+z)\nu/\nu_{0}]^{\beta}$, where the dust emission efficiency index is fixed to $\beta=1.5$ \citep{PlanckXLIII2016,Erler2018}. We set the reference frequency to $\nu_0=c/350~\mathrm{\mu m}$, hence considering the typical value for the mass absorption coefficient at $350~\mathrm{\mu m}$ of $\kappa_{0}=0.192~\mathrm{m^{2}~kg^{-1}}$ \citep{Draine2003}.

The inferred spectrum is shown in Fig.~\ref{fig:indexc}. We observe a general agreement with the modified power law model, except for a discrepancy over ALMA's Band~3 spectral window. First, this can be related to the obvious difference between the two models employed in our analysis, with the dust-like component being more physically motivated than the simple modification to power law spectrum introduced before. Second, the improved sensitivity of ALMA Band~4 observations compared to the Band~3 data (Sect.~\ref{sec:data:alma}) is making the latter subdominant with respect to the former in driving the fit at $\sim100~\mathrm{GHz}$.

As for the reconstructed parameters, we measure values for the dust mass and temperature of  $3.5\substack{+2.3\\-1.9}\cdot10^8~\mathrm{M_{\odot}}$ and $19.8\substack{+7.0\\-6.6}~\mathrm{K}$, respectively, consistent with the masses and temperatures observed in other star forming BCGs (e.g., \citealt{Edge2010}, \citealt{McDonald2015}, \citealt{Fogarty2017}, \citealt{Castignani2020}). 
Despite this, the large uncertainties on the best fit parameters of the modified blackbody spectrum model express the heavy degeneracy affecting the posterior probability distribution of the reconstructed mass and temperature. The proper characterization of the submillimeter properties of the central galaxy will thus require better observational constraints on the high frequency end of its spectral energy distribution.

\subsubsection{Multicomponent model}\label{sec:res:sz:twosz}
\begin{figure*}
    \centering
    \includegraphics[clip,trim=0 1.75cm 0 2.75cm,width=\textwidth]{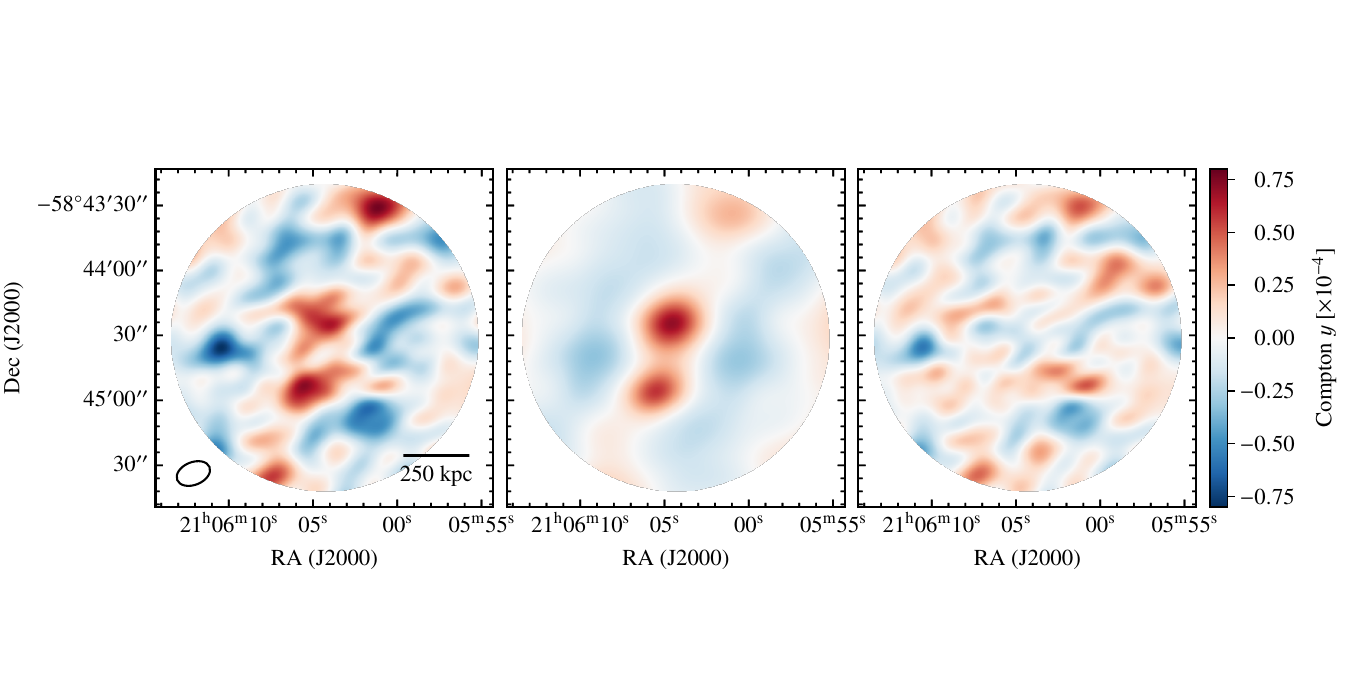}
    \caption{Combined dirty images of the raw (left), model (center), and residual (right) using Band 3 and Band 4 ACA interferometric data. We note that no primary beam correction is applied to the images. We note that the signal is heavily attenuated by antenna pattern as one moves to the edges of the field of view. Hence, despite the noise appears roughly uniform across the entire map, the signal-to-noise ratio decreases abruptly away from the image center.}
    \label{fig:szres}
\end{figure*}

Given the robustness in our detection of the two SZ modes, we repeat the analysis this time including two separate SZ components along with the point-like feature with a modified power-law spectrum describing the central radio source. Again, we assume a universal profile \citep{Arnaud2010} with free mass parameter and inner slope for each of the SZ model components. We consider the same uniform priors as in Sect.~\ref{sec:res:sz:onesz} for each of the SZ components. However, in order to enforce the identifiability of the separate modes, we introduce an ordering condition on their declinations as detailed in Appendix A2 of \citet{Handley2015}.

\begin{table*}
    \centering
    \caption{Parameters of the model employed in the analysis of the Band 3 and Band 4 ALMA+ACA data.}
    \begin{tabular}{lrlc}
    \hline\hline\noalign{\smallskip}
    Parameter & \multicolumn{2}{>{\centering}p{4.00cm}}{Prior distribution} & Best-fit value \\ \noalign{\smallskip}
    \hline
    \noalign{\medskip}
    \multicolumn{4}{l}{\textit{Northern gNFW component}} \\ \noalign{\smallskip}
    RA           & uniform      & $\mathrm{min}=\ra{21;05;56.00}$, $\mathrm{max}=\ra{21;06;13.00}$ &  \ra{21;06;04.57}$\substack{+1.7^{\prime\prime}\\-1.8^{\prime\prime}}$    \\ \noalign{\smallskip}
    Dec          & ordered uniform      & $\mathrm{min}=-\dec{58;45;30.00}$, $\mathrm{max}=-\dec{58;43;30}$ & \dec{-58;44;25.44}$\substack{+0.80^{\prime\prime}\\-0.78^{\prime\prime}}$ \\ \noalign{\smallskip}
    $M_{500}$\tablefootmark{a}    & log-uniform  & $\mathrm{min}=10^{12}~\mathrm{M_{\odot}}$, $\mathrm{max}=10^{16}~\mathrm{M_{\odot}}$ & $2.60\substack{+0.57\\-0.54}\cdot 10^{14}~\mathrm{M_{\odot}}$ \\ \noalign{\smallskip}
    Inner slope  & uniform      & $\mathrm{min}=0.00$, $\mathrm{max}=2.00$ & $0.47\substack{+0.11\\-0.12}$ \\\noalign{\smallskip}
    \noalign{\medskip}
    \multicolumn{4}{l}{\textit{Southern gNFW component}} \\
    \noalign{\smallskip}
    RA           & uniform      & $\mathrm{min}=\ra{21;05;56.00}$, $\mathrm{max}=\ra{21;06;13.00}$ &  \ra{21;06;05.49}$\substack{+1.5^{\prime\prime}\\-1.3^{\prime\prime}}$    \\ \noalign{\smallskip}
    Dec          & ordered uniform      & $\mathrm{min}=-\dec{58;45;30.00}$, $\mathrm{max}=-\dec{58;43;30}$ & \dec{-58;44;57.81}$\substack{+0.72^{\prime\prime}\\-0.78^{\prime\prime}}$ \\ \noalign{\smallskip}
    $M_{500}$\tablefootmark{a}    & log-uniform  & $\mathrm{min}=10^{12}~\mathrm{M_{\odot}}$, $\mathrm{max}=10^{16}~\mathrm{M_{\odot}}$ & $1.34 \substack{+0.52\\-0.52}\cdot 10^{14}~\mathrm{M_{\odot}}$ \\ \noalign{\smallskip}
    Inner slope  & uniform      & $\mathrm{min}=0.00$, $\mathrm{max}=2.00$ & $0.86\substack{+0.25\\-0.24}$\\\noalign{\smallskip}
    \noalign{\medskip}
    \multicolumn{4}{l}{\textit{Compact source}\tablefootmark{b}} \\
    \noalign{\smallskip}
    RA                 & uniform & $\mathrm{min}=\ra{21;05;56.00}$, $\mathrm{max}=\ra{21;06;13.00}$                      &   \ra{21;06;04.60}$\substack{+0.31^{\prime\prime}\\-0.31^{\prime\prime}}$ \\\noalign{\smallskip} 
    Dec                & uniform & $\mathrm{min}=-\dec{58;45;30.00}$, $\mathrm{max}=-\dec{58;43;30}$                      & \dec{-58;44;28.78}$\substack{+0.27^{\prime\prime}\\-0.26^{\prime\prime}}$ \\\noalign{\smallskip}
    Flux at 5 GHz      & uniform  & $\mathrm{min}=0.00~\mathrm{mJy}$, $\mathrm{max}=5.00~\mathrm{mJy}$ & $0.690\substack{+0.037\\-0.037}~\mathrm{mJy}$ \\\noalign{\smallskip}
    Spectral index     & uniform  & $\mathrm{min}=-3.00$, $\mathrm{max}=5.00$                            & $-1.10\substack{+0.05\\-0.05}$   \\\noalign{\smallskip}
    Spectral curvature & uniform & $\mathrm{min}=-1.00$, $\mathrm{max}=1.00$             & $0.123\substack{+0.022\\-0.022}$ \\
    \noalign{\smallskip}
    \hline
    \end{tabular}
    \tablefoot{
        \tablefoottext{a}{$M_{500}$ corresponds to the mass of the galaxy cluster within the radius $r_{500}$.}
        \tablefoottext{b}{We are assuming a spectral dependence of the form $F(\nu) = F_{5\mathrm{GHz}}\,(\nu/5~\mathrm{GHz})^{\alpha+\beta\log(\nu/5~\mathrm{GHz})}$}, with $\alpha$ and $\beta$ corresponding to the spectral index and curvature of the source spectrum.
    }
    \label{tab:sz}
\end{table*}

A summary of the best-fitting parameters is provided in Table~\ref{tab:sz}.
In terms of statistical significance of the SZ detection, the inclusion of the two-component model for the thermal SZ effect from SPT-CL~J2106-5844 is found to improve the Bayesian log-evidence by $\Delta\log\mathcal{Z}=120.72\pm0.16$ with respect to the model without any SZ components, roughly corresponding to a $15.6\sigma$ detection in comparison to the null model. Also, two components are found to be preferred over one SZ model at the $\sim9.9\sigma$ level.\footnote{The conversion between the Bayesian log-evidence difference $\Delta\mathcal{Z}$ and the classical significance level is performed under the assumption of a Gaussian posterior distribution and neglecting any change in the prior volume due to the difference in the parameter sets between the models with and without the thermal SZ components.} We also note that the centroid of the SZ effect determined by SPT lies midway between the two SZ substructures, consistent with the signals from the individual components being blended into a single SZ feature at the SPT angular resolution.

There are a few important caveats associated with the SZ mass estimates provided in this work. It is important to consider that the presence of $M_{500}$ in our parameter set follows from the specific parametrization adopted by \citet{Arnaud2010} to get a self-similar pressure profile under the assumption of hydrostatic equilibrium. Given the disturbed nature of SPT-CL~J2106-5844, we may instead expect the intracluster gas not to be hydrostatically supported and to be characterized by a significant nonthermal contribution to its pressure \citep{Shi2015, Biffi2016}. This would result in a non-negligible bias in the mass inferred from the analysis of the thermal SZ effect \citep{Battaglia2012, Nelson2012, Angelinelli2020, Ansarifard2020}. In addition, the strong degeneracy between the mass and inner slope parameters may have an impact on the accuracy of the mass estimate, as well as inducing the large uncertainties we measure on such parameters. For reference, when fixing all the gNFW slopes to the values for the universal pressure profile in \citet{Arnaud2010}, we obtain masses of $3.10\substack{+0.10\\-0.10} \cdot 10^{14}~\mathrm{M_{\odot}}$ and $3.26\substack{+0.25\\-0.25} \cdot 10^{14}~\mathrm{M_{\odot}}$ for the northern and southern SZ components, respectively.

For reference, we show in Fig.~\ref{fig:szres} the reconstructed images of the raw, best-fit model, and corresponding residual interferometric data obtained by combining the Band 3 and Band 4 ACA observations na\"ively converted to Compton $\vary$ units (i.e., assuming the signal after removing the central AGN is dominated by the classical thermal SZ effect alone). We do not include any images generated from the ALMA data as they show no prominent structures apart from the point-like central radio source. The reported images are produced using the \texttt{CASA} package version 5.6.1. However, we emphasize again that all results based on the analysis of the ACA and ALMA observations of SPT-CL~J2106-5844 are derived by working directly on the raw interferometric data. The ACA maps are presented only for display purposes and not used for any quantitative study.

\subsubsection{Constraining the kinetic SZ signal}\label{sec:res:sz:kinsz}
For all analyses presented in the previous sections, we did not include in our models any contribution from the kinetic SZ effect \citep{Sunyaev1980}, proportional to line-of-sight component of the peculiar velocity of a given ICM structure. Given the highly disturbed state of the system, we could however expect large velocity components and, hence, potential non-negligible biases introduced by ignoring such an effect. It is however worth noting that, even in the extreme case of MACS~J0717.5+3745 (e.g., \citealt{Mroczkowski2012}, \citealt{Sayers2013}, \citealt{Adam2017}, \citealt{Sayers2019}), the kinetic SZ effect has been observed to contribute  $\lesssim10\%$ the total SZ signal from a cluster.

\begin{figure}
    \centering
    \includegraphics[clip,trim=0 1.25cm 0.75cm 0,width=\columnwidth]{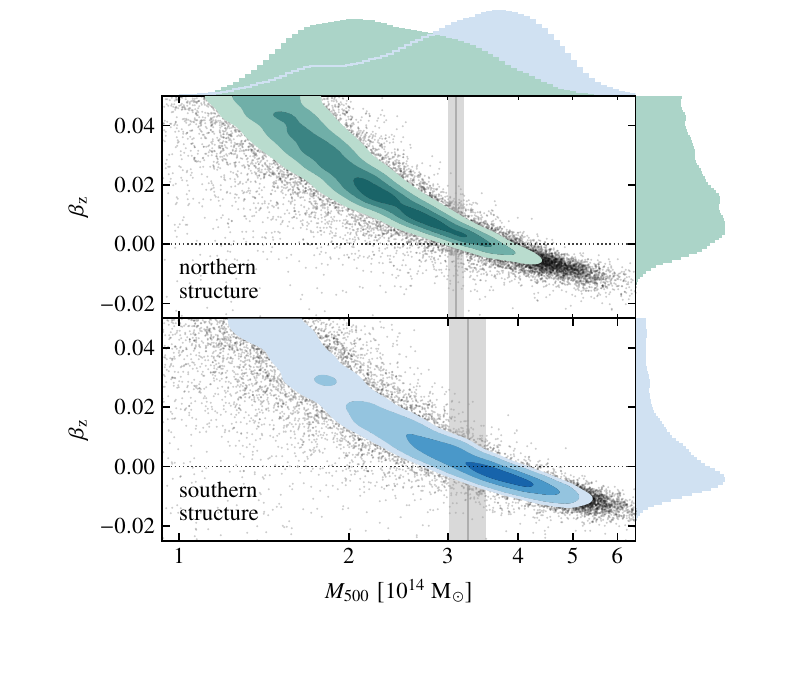}
    \caption{Marginalized posterior for the parameters corresponding to masses and line-of-sight peculiar velocities of the two SZ features (Sect.~\ref{sec:res:sz:twosz}). The contours correspond to the same significance levels used in Fig.~\ref{fig:bimodal}. Due to the limited spectral coverage and sensitivity of the ALMA+ACA data, it was not possible to derive tight constraints on the potential line-of-sight motion of the ICM features identified through the SZ analysis. The gray lines and shaded bands correspond to the mass estimates obtained in Sect.~\ref{sec:res:sz:twosz} under the assumption of a universal pressure profile \citep{Arnaud2010}.}
    \label{fig:ksz}
\end{figure}

The main limitation for constraining any kinetic SZ signature from the available ALMA+ACA data is their limited spectral coverage, insufficient for performing a full separation of the different SZ components in a single system. In fact, the lack of measurements at frequencies corresponding to the null or to the increment part of the thermal SZ spectrum makes the thermal and kinetic SZ signal strongly degenerate within the measurement uncertainties. We present in Fig.~\ref{fig:ksz} the marginalized posterior for the main parameters --- the mass $M_{500}$ and the line-of-sight component of the peculiar velocity in units of the speed of light $\beta_{\mathrm{z}}$ --- of a multicomponent model analogous to the one introduced in the previous section but including a kinetic SZ term for both the subclusters. Under the same assumption adopted in the previous section of an isothermal model, the introduction of a kinetic contribution is equivalent to changing the SZ spectral scaling to \citep[e.g.,][]{Mroczkowski2019}
\begin{equation}
    g(x) = \frac{x^4 e^x}{(e^x-1)^2}\left[\left(x\frac{e^x+1}{e^x-1}-4\right)-\beta_{\mathrm{z}}\frac{m_{e} c^2}{k_{\textsc{b}} T_{\mathrm{e}}}+\delta_{\mathrm{e}}(x,T_{\mathrm{e}},\beta_{\mathrm{z}})\right]
    \label{eq:ksz}
\end{equation}
for a dimensionless frequency  $x=h\nu/k_{\textsc{b} T_{\textsc{cmb}}}$. Here, $h$ and $k_{\textsc{b}}$ are the Planck and Boltzmann constants, $m_{\mathrm{e}}$ the electron mass, and $T_{\mathrm{e}}$ and $T_{\textsc{cmb}}$ the temperatures of the intracluster electrons and the cosmic microwave background (CMB), respectively. The last term in the parentheses $\delta_{\mathrm{e}}(x,T_{\mathrm{e}},\beta_{\mathrm{z}})$ accounts for the corrections to both the thermal and kinetic SZ terms due to the relativistic velocities of the intracluster electrons, and is computed according to the approximation provided by \citet{Nozawa2006}. To facilitate interpretation, we restricted the analysis to the case with all the slopes of the gNFW pressure distributions fixed to the values for the universal pressure model \citep{Arnaud2010}. Further, we introduced a prior on the integrated SZ flux in order to break (or, at least, reduce) the degeneracy between the thermal and kinetic spectral components of the SZ signal. In particular, we considered the estimate of the cylindrically integrated SZ flux within $r_{500}$ --- the radius within which the average cluster density equals $\times500$ the critical density of the Universe at the redshift of the cluster. --- based on the latest AdvACT measurements of the cluster SZ signal \citep{Hilton2020}.

We measure values for $\beta_{\mathrm{z}}$ of $1.7\substack{+2.0\\-1.5}\cdot10^{-2}$ and $0.2\substack{+2.0\\-0.8}\cdot10^{-2}$ for the northern and southern SZ features, respectively, equivalent to line-of-sight velocities of $5088\substack{+6092\\-4435}~\mathrm{km~s^{-1}}$ and $463\substack{+6027\\-2538}~\mathrm{km~s^{-1}}$. Although the velocity of the northern component is nominally different from zero at the $\sim1.1\sigma$ level, the corresponding one-dimensional histogram in Fig.~\ref{fig:ksz} shows that the determination of the best-fit value of the parameter is driven by the high $\beta_{\mathrm{z}}$ tail of the posterior distribution. In a minor way, a similar behavior is observed for the southern component. This, together with the large uncertainties affecting the velocity parameters, suggests that the limitations in the available ALMA+ACA data leave the kinetic SZ contribution practically unconstrained. 

Related to the amplitude of the thermal SZ effect from the two ICM features, the inclusion of the kinetic SZ term causes a variation in the masses to $2.26\substack{+0.83\\-0.72}\cdot10^{14}~\mathrm{M_{\odot}}$ and $3.22\substack{+0.93\\-1.46}$, systematically lower but consistent with the estimates reported in the previous section considering the same universal profile slopes.

We note that in our analysis we are completely neglecting any contribution from primary CMB anisotropies on small scales, spectroscopically indistinguishable from the kinetic SZ signal. However, their amplitude is expected to be $\lesssim2~\mathrm{\mu Jy}$ on the angular scales probed by the ALMA+ACA observations \citep{DiMascolo2019}. As reported in Sect.~\ref{sec:data:alma} and according to the results in Fig.~\ref{fig:ksz}, one can thus immediately infer that the statistical uncertainties in the ALMA+ACA data are much larger than any signal from either a plausible kinetic SZ term or primary CMB anisotropies.

%-----------------------------------------------------------------------
\section{Discussion}\label{sec:discuss}
\begin{figure*}
    \centering
    \includegraphics[clip,trim=12.15cm 2.00cm 0.25cm 2.68cm,height=0.646\columnwidth]{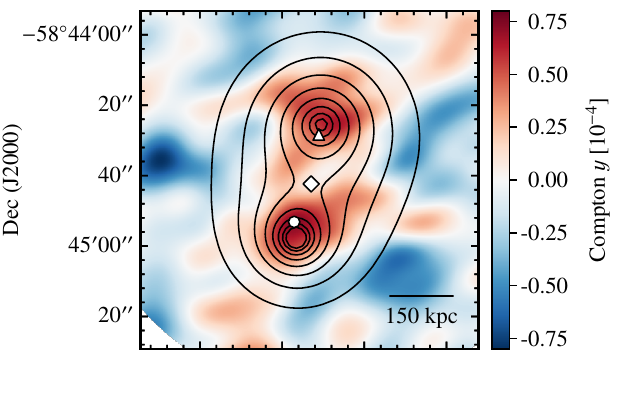}
    \includegraphics[clip,trim=13.45cm 2.00cm 0.25cm 2.68cm,height=0.646\columnwidth]{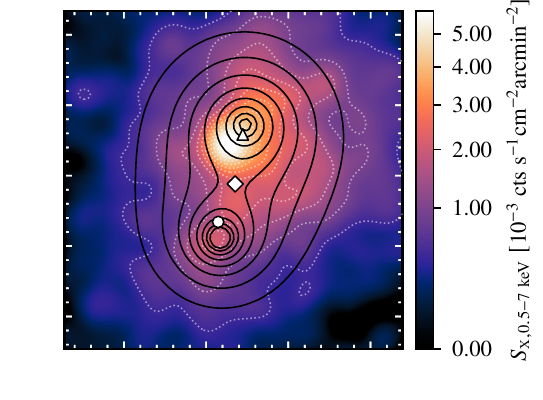} \\
    \includegraphics[clip,trim=12.15cm 2.00cm 0.25cm 2.68cm,height=0.646\columnwidth]{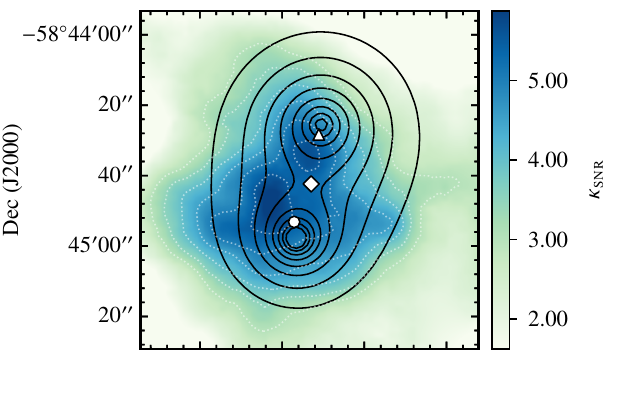}
    \includegraphics[clip,trim=13.45cm 2.00cm 0.25cm 2.68cm,height=0.646\columnwidth]{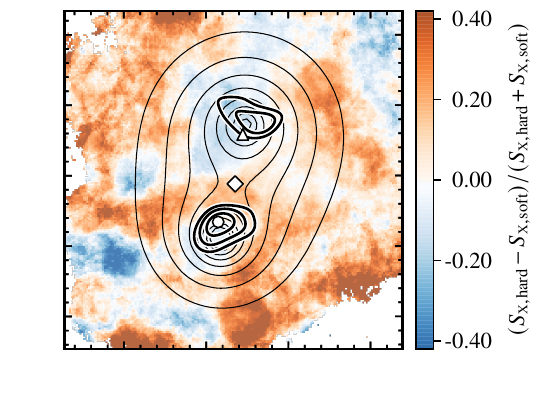} \\
    \includegraphics[clip,trim=12.15cm 1.80cm 0.25cm 2.68cm,height=0.665\columnwidth]{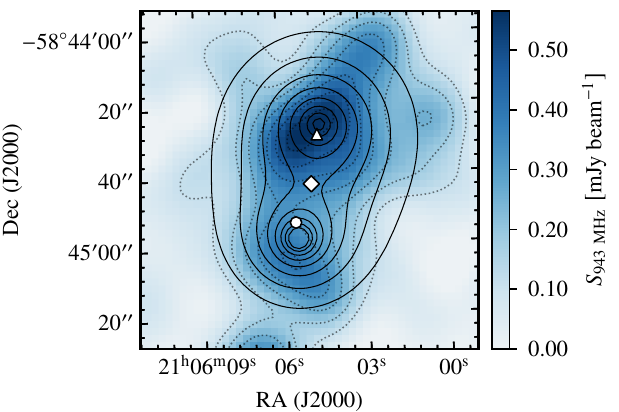}
    \includegraphics[clip,trim=13.45cm 1.80cm 0.25cm 2.68cm,height=0.665\columnwidth]{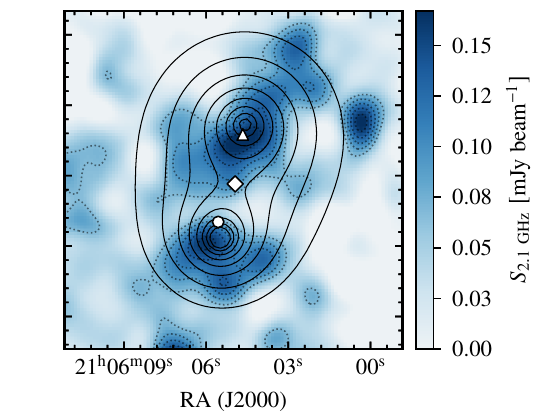}
    \caption{Comparison of the SZ effect (upper left), smoothed X-ray surface brightness (upper right), signal-to-noise map of the weak-lensing convergence (middle left; \citealt{Kim2019}),  X-ray hardness ratio (middle right) maps, and EMU (lower left) and ATCA (lower right) images for SPT-CL~J2106-5844. The black contours appearing in all panels represent the unfiltered SZ model derived in Section~\ref{sec:res:sz:twosz}. To facilitate comparison between the main cluster features, we overlay to the \textit{Chandra}, EMU, and ATCA the same contours as in Fig.~\ref{fig:xsurf}, \ref{fig:emu}, and \ref{fig:atca}, respectively, while the ones for the mass distribution reproduce the contours in Fig.~6 of \citet{Kim2019}. The symbols in all the maps denote the positions of the BCG \citep[triangle;][]{Song2012}, the peak in the galaxy number density \citep[circle;][]{Kim2019}, and the SZ centroid as measured by SPT \citep[diamond;][]{Bleem2015}. The two SZ features can be seen to correspond to two weak-lensing mass peaks, as well as mildly soft components of the X-ray emission, in turn corresponding to the X-ray peak and the southern extension. We note that, interestingly}, the curved morphology of the southern feature in the SZ dirty image (thick black contours in the middle right panel) appears to trace an arc-shaped patch of hard X-ray emission.
    \label{fig:totsz}
\end{figure*}

Figure~\ref{fig:totsz} provides a comparative view of the SZ signature of SPT-CL~J2106-5844 (both model and dirty image) with the weak lensing and X-ray surface brightness distribution in the cluster. Overall, the SZ effect from SPT-CL~J2106-5844 is found to be aligned with the mass distribution measured by \citet{Kim2019}, with the two subclusters roughly corresponding to the separate clumps in the main mass concentration within the cluster.
We observe a moderate offset ($>50~\mathrm{kpc}$) between the southern SZ peak and the corresponding weak-lensing component, suggesting that the southern gas may have already started dissociating from its galaxy population. However, the available weak-lensing map has a positional precision of $\sim5\arcsec$, which, combined with the uncertainties on the recovered SZ centroids, limits our confidence about the observed separation.

Given the clear identification of two individual SZ components with comparable strength, we interpret SPT-CL~J2106-5844 as a major merger.  While the exact merger scenario is difficult to ascertain with high certainty, we posit two possible scenarios that we consider likely:
\begin{enumerate}
    \item We are observing the merger before first core passage. This scenario is supported by the existence of two distinct SZ peaks that are well fit by spherical gNFW models, and the X-ray bright, well-formed core in northern component. \\ 
    \item The merger is observed after the first core passage and has a relatively large impact parameter (i.e., the merger is off-axis). In this scenario, the lack of a prominent X-ray  core in the southern component may be explained by the disruption of the southern subcluster during the first core-core interaction.
\end{enumerate}
It is important to note that, despite the dominant contribution to the X-ray emission by the bright cool core in the northern part of the cluster, it is still possible to identify a moderate enhancement of the X-ray signal south of the northern cool core (Fig.~\ref{fig:xsurf} and upper right panel in Fig.~\ref{fig:totsz}) and coinciding with southern SZ feature. In the double gNFW scenario used to the fit the ALMA+ACA SZ data, one can predict the X-ray luminosities of the two substructures using the scaling relation from \citet{Mantz2016}. When considering the masses estimated under the assumption of a universal pressure model, this leads to the total X-ray luminosity in the $0.1-2.4~\mathrm{keV}$ band of $11.4\substack{+2.1\\-1.8}\cdot10^{44}~\mathrm{erg~s^{-1}}$, lower than the corresponding value of $20.5\substack{+1.0\\-1.0}\cdot10^{44}~\mathrm{erg~s^{-1}}$ measured from the core-excised X-ray observations (Fig.~\ref{fig:power}). It is worth noting that a low X-ray luminosity of $13.0\substack{+3.6\\-3.0}\cdot10^{44}~\mathrm{erg~s^{-1}}$ is found also when applying the X-ray scaling to the total mass estimate reported in AdvACT catalog ($6.00\substack{+0.91\\-0.79}\cdot 10^{14}~\mathrm{M_{\odot}}$; \citealt{Hilton2020}). This is however not surprising, given both the impact of the interferometric filtering on the SZ model (see Sect.~\ref{sec:discuss:dyn} below) and the disturbed nature of the system . In interacting clusters, both the X-ray luminosity and the thermal SZ effect can be boosted, with the magnitude of the boost being dependent on the characteristic Mach number $\mathcal{M}$ of the shocks in the interacting region \citep[see, e.g.,][]{Churazov2020}. For low $\mathcal{M}$ values, the X-ray brightness might be more strongly affected than the thermal SZ signal, while the opposite is true for very high $\mathcal{M}$. In addition, the X-ray luminosity (and the appearance of the X-ray image) can be affected by the presence or absence of the cool core, which is the case for the northern region of SPT-CL~J2106-5844. It is clear that much better X-ray and SZ data are needed to cleanly distinguish these two proposed scenarios. At present, the data appears consistent with both cases.

\subsection{The origin of the extended radio structure}
The morphology of the radio emission around the BCG suggests that it may be associated with a radio jet from the central AGN with mildly bent arms \citep{Sakelliou2000}. Such sources are not uncommon in disturbed galaxy clusters, as the motion of a radio galaxy with respect to the intracluster gas can induce a significant bending of the radio lobes (see, e.g., \citealt{PaternoMahler2017}, \citealt{Garon2019}, \citealt{GoldenMarx2019}, \citealt{Moravec2020}). On the other hand, the southern diffuse emission C3 (Fig.~\ref{fig:compact}) appears cospatial with a collection of elliptical galaxies identified in the HST image, with a few of them manifesting a mild correspondence with bright spots in the extended radio feature. It is hence possible that the southern radio structure does not belong to the radio galaxy system associated with the BCG, but arises as a superposition of relatively extended emission due to past or ongoing AGN activity within any of the members of the observed galaxy overdensity. 

In addition, it is worth considering that the combination of the spatial filtering with the limited sensitivity of the ATCA data may result in images highlighting only the most compact and bright component of the total radio emission, obscuring a more extended component beneath the narrower structure. Thus, independently of the above scenarios, the  possibility that a fraction of the extended emission is associated with an diffuse radio halo cannot be entirely ruled out. 

In the following subsections, we explore a few key aspects and implications of these possibilities for the origin of the extended radio emission.

\subsubsection{Connection with the central AGN}\label{sec:discuss:agn}
The EMU and ATCA data do not allow us to narrow down the hypotheses regarding the nature of the diffuse radio emission and, in particular, whether the extended feature belongs in its entirety to the radio galaxy coinciding with the BCG, or if it is a blend of several extended radio sources and radio-loud galaxies.
In the former case, the whole radio structure should remain associated with the environment of only the northern ICM substructure. If the system has not yet begun to interact, no opportunity for mixing has occurred, but even if the merging subclusters have experienced first contact or undergone an off-axis (i.e., high impact parameter) core passage, a contact discontinuity separating the ICM components of the two subclusters may persist for some time.

With only $\sim800$ source counts in the \textit{Chandra} imaging, we find no conclusive evidence in the X-ray hardness ratio map (Fig.~\ref{fig:xhard}) for X-ray cavities or dredging by the radio jet of cold material from the core region up to scales larger than the extent of the radio structure (i.e., $\sim 200~\mathrm{kpc}$ from the compact core) that would conclusively support the interpretation of the diffuse radio structure as only due to AGN activity from the central galaxy. Further, each of the dominant features in the extended radio plasma (i.e., the core region of the radio galaxy and C3; Sect.~\ref{sec:res:radio:ext} and Fig.~\ref{fig:compact}) spatially correspond to one of the distinct SZ subcomponents, consistent with the potential interpretation for the radio structures to be generated by separate populations of radio sources. Independently of the specific origin of the observed radio signal, the cluster merger may be the primary responsible for making the elongated radio feature visible, as shocks and adiabatic perturbations may induce reacceleration or compression of fossil electrons \citep{vanWeeren2019}. In fact, in the case of the diffuse radio emission originating in its entirety from the central AGN, a process able to rejuvenate the radio plasma would be required to boost the emission to the levels seen at ATCA frequencies. In particular, a characteristic lifetime for synchrotron-emitting electrons due to synchrotron and, especially, inverse Compton energy losses is less than $10~\mathrm{Myr}$ at the cluster redshift for any value of the magnetic field. This is roughly between one and two orders of magnitude shorter than the time required to advect relativistic electrons across 200 kpc with a velocity equal to the sound speed in the ICM. 

We note that, while the presence of a radio-loud AGN in the center of a cluster at low redshift is known to correlate with the presence of a cool core \citep[e.g.,][]{Pasini2020},
SPT-CL~J2106-5844 is not unique in being a merger hosting a radio-loud AGN. 
In general, \cite{Birzan2017} found that for an SZ-selected sample of clusters at $0.75 < z < 1.2$, there was no correlation between the presence of an AGN and whether the cluster had a cool core or not.
For more specific examples, \cite{Savini2019} report the disturbed, non-cool-core cluster RXCJ0142.0+2131 hosts a similarly bright AGN, while \cite{Moravec2020} and \cite{Ruppin2020} report on the merging clusters MOO~J1506+5137 and MOO~J1142+1527, both of which host radio bright AGNs and are located at a similar redshift as SPT-CL~J2106-5844. 
As another example, RXJ1347.5-1145 hosts a radio luminous AGN in the center of its intact cool core, and is thought to be a dramatic late-stage, off-axis merger \citep{Ueda2018}. In contrast with the double-peaked SPT-CL~J2106-5844, \citet{DiMascolo2019} found that the SZ signal in RXJ1347.5-1145 can be described by a singular ellipsoidal gNFW pressure distribution on scales from approximately 4\arcsec to beyond 10\arcmin, supporting the post- or late-stage merger interpretation.

\subsubsection{Upper limit on the radio halo contribution}
The ATCA data clearly show that the jet-like feature provides a dominant contribution to the radio signal on modestly large scales ($\sim400~\mathrm{kpc}$; Fig.~\ref{fig:atca}). Despite the multiple pieces of evidence presented in the previous section in support of the AGN origin of the radio structure, though, the similarity in distribution between the EMU diffuse radio structure and the ICM within SPT-CL~J2106-5844 (lower panel of Fig.~\ref{fig:emu}) makes the extended radio emission in principle consistent with the general trend in other radio haloes \citep{vanWeeren2019}.

In fact, mini-halos are sometimes observed around BCG galaxies residing in cool cores \citep[e.g.,][]{Richard-Laferriere2020}, and giant radio halos might be expected here, as well, given the mass of the system and its merger status. Further, at the characteristic luminosities expected for such structures (Fig.~\ref{fig:indexe}), they would not be visible given the very strong radio emission from the proposed radio galaxy associated with the BCG in SPT-CL~J2106-5844. 
Due to the increased inverse Compton cooling efficiency at high redshift, we might expect high-redshift radio haloes to be less powerful than their low-redshift counterparts that have been used to construct the mass- and X-ray luminosity-halo luminosity scaling relationships (e.g., \citealt{Cassano2013}, \citealt{vanWeeren2019}, \citealt{Cuciti2021b}). However, \citet{DiGennaro2020} recently reported the detection of luminous radio haloes for a sample of galaxy clusters with redshift up to $z\simeq0.90$, which unexpectedly closely follow the low-redshift correlations. They interpret this as being due to a combination of the increased availability of electrons due to higher star-forming and AGN activity plus an unexpectedly high magnetic field at these early epochs.  If similar conditions prevail at $z=1.132$, we could expect a halo associated with SPT-CL~J2106-5844 to also lie on the low-redshift scaling relations. In particular, when accounting for both the increase with the redshift in the inverse Compton efficiency and in the turbulence injection rate, one finds that, at $z=1.132$, a magnetic field with strength $\sim2\%$ lower than for clusters at $z=0.7$ is sufficient to generate a halo with the same radio luminosity as measured at lower redshift \citep{DiGennaro2020}. To date, though, no information is available on the population of radio haloes in galaxy clusters in the same redshift range of SPT-CL~J2106-5844, limiting the possibility of robustly predicting the radio properties of the cluster.

To explore the implications of possible halo-related radio emission, we compute the radio power for SPT-CL~J2106-5844 and compare this with the typical observable-radio power scaling relations for radio haloes. We compute the radio power by integrating the radio signal measured in the naturally-weighted 2.1~GHz ATCA map within the region with a significance $>4\sigma$ (i.e., inside the second contour in the middle panel of Fig.~\ref{fig:atca}). Since ATCA and EMU provide measurements of fluxes consistent with the same power-law spectrum (Fig.~\ref{fig:indexe}), they can provide equivalent constraints on the halo power. We however choose to use the ATCA data as they allow for a better discrimination against the contamination from more compact emission. The contribution from the radio source coincident with the BCG is expected to be minimal, as a model for its signal has been preliminary subtracted from the ATCA data (Sect.~\ref{sec:data:atca}). At the same time, the selected region directly excludes the sources C1 and C2 (Fig.~\ref{fig:compact}). The measured flux density is then scaled to 1.4~GHz using a spectral index of -0.87 (Sect.~\ref{sec:res:radio:ext}) giving a 1.4~GHz rest-frame radio luminosity of $1.2\cdot10^{25}~\mathrm{W~Hz^{-1}}$. Because much or all of the diffuse emission may be related to clumps of galaxies or AGN emission, the calculated values should be considered only as upper limits to the radio halo.

\begin{figure}
    \centering
    \includegraphics[clip,trim=0 0.15cm 0 0.45cm,width=\columnwidth]{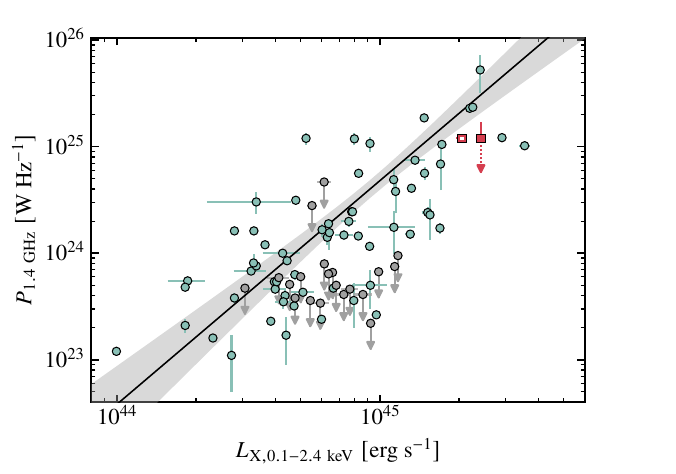}\\\vspace{10pt}
    \includegraphics[clip,trim=0 0.15cm 0 0.45cm,width=\columnwidth]{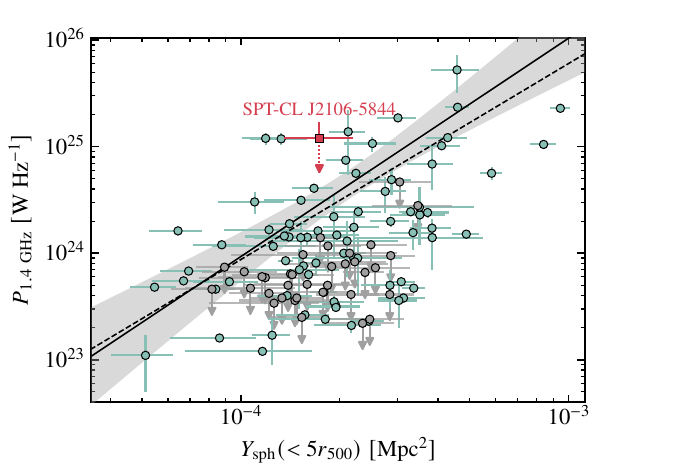}
    \caption{Radio power versus X-ray luminosity (top) and spherically integrated SZ signal (bottom). The solid lines and shaded regions in both panels correspond to the scaling relations and given 95\% confidence intervals reported by \citet{Cassano2013}, respectively. For comparison, we report as a dashed line the scaling relation derived by \citet{Basu2012} directly on the integrated SZ values $Y_{\mathrm{sph}}(<5 r_{500})$.
    The data points are from \citet{Birzan2019}, \citet{DiGennaro2020}, \citet{Cuciti2021a}, and references therein, and correspond to galaxy clusters with clear identification of radio haloes (green circles) or non detections (gray circles). We note that many systems are not included in the top panel as there are no published values for their X-ray luminosities. For the $Y_{\mathrm{sph}}(<5 r_{500})$ estimates, we consider the values from \textit{Planck} SZ catalog \citep{PlanckXXVII2016} except for SPT-CL~J2106-5844, for which we use the integrated SZ flux computed from the AdvACT measurement of its mass \citep{Hilton2020}. The solid and open red markers correspond to the X-ray luminosities measured including or excising the X-ray bright core, respectively. Although the X-ray luminosity of SPT-CL~J2106-5844 (red square) is consistent with the expectation for the measured radio power, the cluster shows a large offset with respect to the radio power-integrated SZ effect relation. We caution that the constraint on the radio power from SPT-CL~J2106-5844 does not imply a non-detection, but an upper limit to the contribution from a potential radio halo.}
    \label{fig:power}
\end{figure}

Fig.~\ref{fig:power} shows that the upper limit to the radio power is lower than the average trend with X-ray luminosity, but consistent with the overall scatter seen in that quantity. 
This lower than average ratio could be due to a temporary large enhancement of the X-ray luminosity, akin to that seen in merger simulations (e.g., \citealt{Ricker2001}, \citealt{Randall2002}, \citealt{Poole2007}, \citealt{Wik2008}). The radio halo power can also be lower than the average trend during pre-core passage \citep{Donnert2013}.  We also considered the possibility that an X-ray bright cool core in the north could increase the observed X-ray luminosity; however, excising it resulted in only a minor change. In summary, given the variations expected in both radio and X-ray luminosities, the observations of SPT-CL~J2106-5844 appear consistent with the current established relationship. 

Figure ~\ref{fig:power} also shows that the upper limit on the radio halo luminosity for SPT-CL~J2106-5844 is slightly larger than the scaling relations by \citet{Basu2012} and \citet{Cassano2013} for the spherically integrated SZ signal. It is among the highest radio values for clusters with comparable SZ properties. Given that our measurement is an upper limit only, it is therefore consistent with the SZ results. However, we note that if the actual radio power were an order of magnitude below our upper limit, it would be comparable to the radio luminosities expected from the SZ scaling, but would be part of the low-radio-power population based on the X-ray luminosity. It is not clear exactly what to expect for SPT-CL~J2106-5844, though, since the merger status would lead to expectations for a high radio halo power, whereas clusters with bright radio BCGs and cool cores only rarely show large radio halos \citep[see, e.g.,][]{Kale2019}. Nevertheless, given the multiple radio structures and range of scale sizes present in the cluster, it is not clear that a robust measurement of the halo power is practical.

\subsection{The dynamics of the ICM substructures}\label{sec:discuss:dyn}
The northern SZ component is observed to be located around the position of the BCG and the brightest region of the X-ray emission. We measure an offset of around $24~\mathrm{kpc}$ and $69~\mathrm{kpc}$ between the reconstructed SZ centroid and, respectively, the BCG and X-ray peak. Such a situation is common in galaxy clusters whose central regions are experiencing gas sloshing \citep{Ascasibar2006, ZuHone2011a, ZuHone2013, Bykov2019, Simionescu2019}. In such cases, minor or off-axis merger events induce oscillatory motions of the gas within the gravitational potential well of clusters. The net result is the generation of anti-correlated perturbations in the cluster density and temperature fields spiraling around the BCG. Since the sloshing motion happens in the subsonic or transonic regime \citep[e.g.,][]{Ueda2018,Ueda2019}, the SZ effect should remain relatively unperturbed. For sloshing taking place in the plane of sky, the measured X-ray surface brightness would be dominated by the emission from the colder and denser gas, which does not necessarily coincide with the peak of the line-of-sight-integrated electron pressure. As a result, the X-ray peak may be displaced with respect to the SZ centroid. Such a scenario is consistent with the X-ray surface brightness and hardness ratio measured for the northern component. The brightest region coincides with a relatively soft region, presumably belonging to a cool core disrupted by the gas sloshing, while the harder region west of the BCG may be tracing the outer edge of a sloshing cold front. Independently of the specific merger scenario, a strong sloshing activity in the northern region of the cluster could further explain the the relative faintness of the southern X-ray emission with respect to the main X-ray peak.

A similar correspondence with a mildly soft X-ray emitting region is found for the southern SZ features, indicating that the observed gas may belong to the remnant core of the merging subcluster. As mentioned before, this structure is observed to correspond to an extension toward the southeast in the X-ray surface brightness and peaking around the centroid of the southern SZ component. Although it is not possible to draw any firm conclusion on the specific dynamics of the ICM components, the comparison of the hardness ratio map with the dirty image of the ALMA+ACA data further shows that the shape of the southern SZ structure seems to follow an arc-like X-ray-hard feature observed in the hardness ratio map north of the southern SZ component. An exciting possibility is that this feature is a merger-driven shock. We emphasize again, though, that the significance of the sole hardness measurement for the aforementioned ICM structures is only marginally larger than $2\sigma$ (bottom panel of Fig.~\ref{fig:xhard}), and further observations, as well as a joint SZ and X-ray analysis, will be required to confirm (or reject) them.

\begin{figure}
    \centering
    \includegraphics[clip,trim=12.3cm 1.85cm 0.4cm 2.78cm,width=\columnwidth]{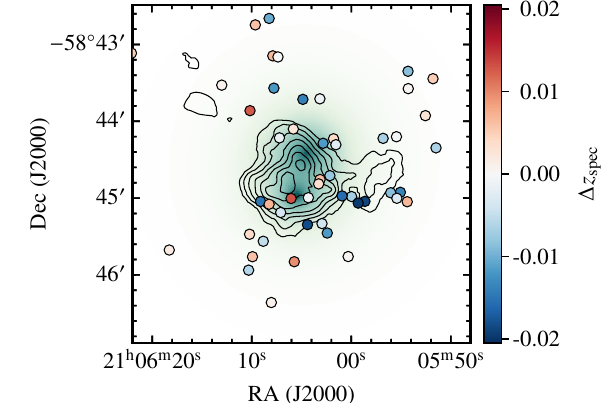}
    \caption{Difference between the spectroscopic redshifts of the cluster members from the GOGREEN survey \citep{Balogh2017,Balogh2020} and cluster nominal redshift $z=1.132$. The cluster galaxy distribution shows no clear asymmetry in the velocity distribution on the plane of sky, but highlights that the cluster members are organized in a filamentary structure around SPT-CL~J2106-5844. The SZ model in units of Compton $\vary$ is shown on the background. The black contours correspond the weak-lensing convergence map as in \citet{Kim2019}.}
    \label{fig:redshifts}
\end{figure}

In Figure \ref{fig:redshifts}, we show the redshifts for the galaxies measured by the Gemini Observations of Galaxies in Rich Early ENvironments\footnote{\url{http://gogreensurvey.ca/data-releases/data-packages/}} (GOGREEN; \citealt{Balogh2017}, \citealt{Old2020}, \citealt{vanderBurg2020}, and \citealt{Balogh2020}) survey. Similarly to what was inferred from the kinetic SZ analysis (Sect.~\ref{sec:res:sz:kinsz}), we find no obvious velocity offset between the two main components that would indicate the major merger is oriented predominantly along the line of sight. We do however see evidence for the western extension identified in the weak-lensing study by \citet{Kim2019} to be blue-shifted. The structure seems to be associated with an interloper or filament, $\sim500~\mathrm{kpc}$ in length, infalling from the northwest into the cluster with a line-of-sight component that may be responsible, in the pre-merger hypothesis, for setting in place the putative sloshing in the northern region of SPT-CL~J2106-5844. Alternatively, if we are observing SPT-CL~J2106-5844 in an early post-merger phase, the sloshing motion within the cluster core may have been induced by the interaction of the two subclusters during the off-axis core passage. 

The lack of symmetry in the amplitude of the X-ray peaks coinciding with the two distinct SZ features is unusual and possibly suggests a merger configuration more complicated than a clean, textbook-case head-on major merger happening on the plane of the sky. A binary thermodynamic structure is observed, also, in the merging galaxy cluster Abell~115 \citep{Hallman2018,Lee2020a} or Abell~141 \citep{Caglar2018}. On the other hand, the off-axis merger Abell~754 also manifests a clear asymmetry in its X-ray signal, with a low surface-brightness tail extending behind an X-ray bright clump, but the X-ray inferred map for the ICM pressure is measured to flatten toward the cluster core \citep{Henry2004}. We note that the X-ray pseudo-pressure map inferred in \cite{Henry2004} may differ systematically from a true Compton-$\vary$ map, due to the differing line-of-sight dependences of X-ray emission and the thermal SZ effect. However, if the underlying pressure distribution in SPT-CL~J2106-5844 is similar to that inferred for Abell~754, it may happen that the bimodal structure in the SZ signal is an artifact of the interferometric large-scale filtering, as most of the signal from the broad shallow core would not be sampled by ALMA+ACA. It is in fact worth considering that we measure an integrated SZ flux density\footnote{The reported SZ flux density is computed as the cylindrically integrated SZ signal within an aperture of $1.69\arcmin$, equivalent to the AdvACT estimate for SPT-CL~J2106-5844's $r_{500}$.} for our SZ model of $2.17\substack{+0.26\\-0.18}\cdot10^{-4}~\mathrm{arcmin^2}$ ($3.01\substack{+0.74\\-0.77}\cdot10^{-4}~\mathrm{arcmin^2}$ in the case of universal pressure model), lower than the measurement $4.7\substack{+1.1\\-1.3}\cdot 10^{-4}~\mathrm{arcmin^2}$ inferred from the AdvACT catalog \citep{Hilton2020}. The lower value obtained from our ALMA+ACA modeling is symptomatic of how the interferometric filtering acts on the model degeneracies by favoring a more compact and low-mass pressure distribution than the universal average (\citealt{Arnaud2010}; Sect.~\ref{sec:res:sz:twosz}), hence limiting the possibility of a successful extrapolation of the reconstructed SZ model to large scales.

Finally, if the merger axis were in the plane of the sky and the two subclusters were separated by some 300 kpc, a significant amount of shocked or compressed gas should lie between them (see, e.g., the 1.2~Gyr slice in the 1:1 merger with 500~kpc impact parameter;\footnote{\url{http://gcmc.hub.yt/fiducial/1to1_b0.5/index.html}} \citealt{ZuHone2011b}).  However, a combination of sensitivity and filtering prevents us from reaching an unambiguous conclusion about the presence of such shocked gas using the current data. In particular, spatial filtering can remove much of the large-scale structure associated with the shock, while the cores of the two subclusters would remain largely unaffected. This could explain the success of our double gNFW model for this system. On the other hand, the observed cluster properties may be the result of projection effects, with the observed signal from the pre-merging components being produced well before their first contact. A dual feature in the weak lensing distribution analogous to the one observed in SPT-CL~J2106-5844 was in fact found by \citet{Frye2019} for PLCK~G165.7+67.0, whose lack of significant SZ and X-ray signatures was interpreted as due to the line-of-sight configuration of pre-merging subclusters. Overall, the combination of a non-negligible impact parameter, line-of-sight projection effects, and the limitations of the available X-ray and SZ data restrict the possibility of fully disentangling the intrinsic structure of the ICM within SPT-CL~J2106--5844.

Important information on the true dynamical state of the cluster could be revealed by a strong-lensing analysis of the cluster field. In fact, tighter constraints on the mass distribution would help in understanding whether this is still being traced by the ICM and, hence, discriminating between the various merger scenarios. Unfortunately, we were not able to observe any strong-lensing feature in the available HST and GOGREEN images. We do however report the identification of a putative arc-like feature in the F140W HST image, centered at (\ra{21;06;05.54},\dec{-58;44;41.53}) and stretching across $7\arcsec$ in the northeast-southwest direction (i.e., perpendicular to the main cluster axis and in between the two mass components). Nevertheless, although the geometry is consistent with the expectation for the observed mass distribution, we note that it is not possible to draw any firm conclusion on the source being real or just a noise artifact solely based on the available observations.

\begin{figure*}
    \centering
    \includegraphics[clip,trim=0 0 0 11cm,width=\textwidth]{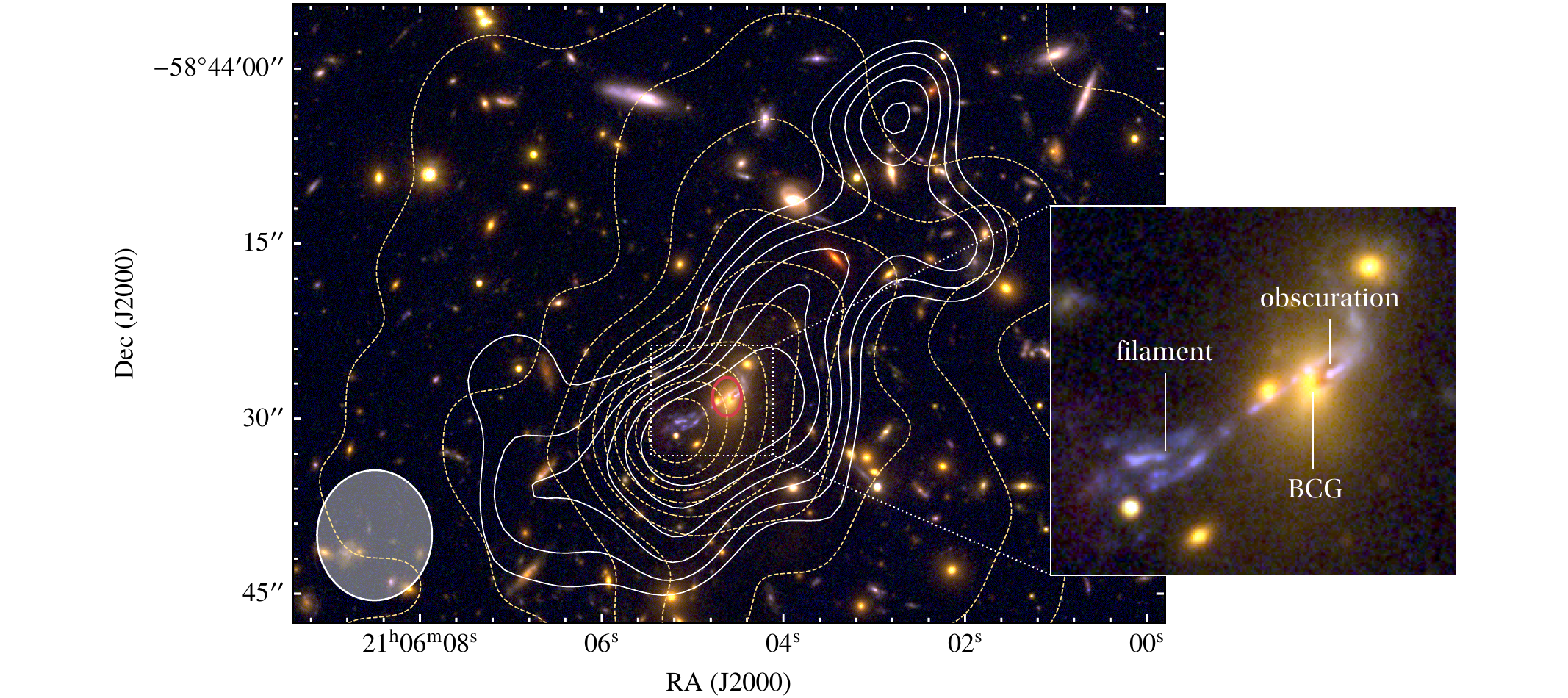}
    \caption{Composite HST F140W (red), F105W (green), and F606W (blue) images with contours from the smoothed \textit{Chandra} surface brightness map (yellow dashed; upper right panel in Fig.~\ref{fig:totsz}) and from the 2.1~GHz ATCA data after subtraction of the core radio emission, and imaged using Briggs weighting with robust parameter equal to 0.00 and a $u\varv$-tapering at $10~\mathrm{k\lambda}$ (white solid, covering the $3\sigma-5.5\sigma$ range in steps of $0.5\sigma$; upper panel of Fig.~\ref{fig:atca}). The resulting synthesized beam is shown in the bottom right corner. We denote the position of the subtracted radio core model with a red ellipse, whose axes and position angle correspond to the synthesized beam of the uniformly weighted 2.1~GHz ATCA image used to subtract the core emission. The morphology of the central region of the extended radio emission, consistent with a wide-angle tailed radio galaxy (Sect.~\ref{sec:discuss:agn}; \citealt{Sakelliou2000}), appears to align in the same direction as the optical filament.}
    \label{fig:filament}
\end{figure*}

\subsection{A cold star-forming filament surrounding the BCG}
The GOGREEN optical-to-near-infrared data further confirm the presence of a strongly star forming structure surrounding the BCG, as previously identified by \citet{McDonald2016} in HST observations (Fig.~\ref{fig:filament}). The origin of such a feature is however unclear. If associated with the central radio galaxy, it could be due to the cool gas uplifted from the central cluster region by the AGN jet or the gas already falling back inward after being cooled \citep{Churazov2001,Tremblay2018,Qiu2020}.
Some support for this hypothesis comes from the apparent correspondence of the orientations of the optical filament and the extended radio structure surrounding the BCG (Fig.~\ref{fig:filament}), which is consistent with cooling gas in the relatively undisturbed northern subcluster core (c.f. the soft X-ray emission seen in Fig.~\ref{fig:xhard}). This would provide an immediate explanation for the observed disbalance in the distribution of the X-ray emission from SPT-CL~J2106-5844. Analogous to results reported by \citet{Diego2020} for ACT-CL~J0102-4915 (``El Gordo''), it would in fact be possible to ascribe the X-ray peak to cooler, denser gas experiencing a massive cooling flow. 

On the other hand, consistent with the scenario of galaxy-galaxy interactions being the dominant mechanism for triggering star formation within the cores of clusters at high redshift \citep{Webb2015,McDonald2016,Bonaventura2017}, the filamentary structure seen in the HST F606W data could consist of the remnant gas of a tidally disrupted galaxy merging with the BCG. The darkened shell surrounding the star-forming clump that is observed to sit right in front of the BCG could hence be due to the absorption by an optically thick medium (e.g., dust) within the infalling galaxy (labeled ``obscuration'' in the inset panel in Fig.~\ref{fig:filament}). We note that, despite of the specific scenario invoked to explain the observed filamentary structure, the synchrotron emission from the star-forming clumps could contribute to the overall radio signal in the vicinity of the BCG. Unfortunately, the resolution and sensitivity of the current radio data do not allow one to probe properly the environment around the central galaxy. We aim at examining in a future work the dynamics of the cluster galaxies as well as the physical properties of the optical filamentary structure around the BCG, complementing the available measurements with new radio observations at subarcsecond resolution.

%-----------------------------------------------------------------------

\section{Conclusions and future work}\label{sec:conc}
In this paper, we have used archival observations from ALMA and ACA to detect and resolve the gas associated with the two main subclusters comprising the dominant components of the massive $z=1.132$ merging cluster SPT-CL~J2106-5844. These observations exploit several of the main strengths of ALMA+ACA measurements of the signal due to the thermal SZ effect --- high spatial (i.e., subarcminute) resolution, redshift-independent surface brightness, a large collecting area that yields $\sim 10-100 \mu$Jy-level sensitivity, and the spatial dynamic range required to model contamination by compact sources --- to reveal two unambiguous SZ components of approximately equal amplitude.

We also present archival, multiband radio observations from ATCA, along with new data from the EMU Pilot Survey performed by ASKAP. Removal of the compact radio sources within the field reveals a diffuse radio structure on scales of $\sim400~\mathrm{kpc}$. The overall morphology suggests that the extended radio signal could be associated with large-scale jets from a radio-loud AGN within the BCG and, plausibly, with a star-forming filament previously  identified in optical data. However, there is ample evidence to support the possibility for the southernmost radio extension to be due to a complex of radio-loud sources independent of the central radio galaxy, although the available radio data do not provide sufficient constraints on the origin of such structure.

When combined with evidence from weak lensing \citep{Kim2019} and X-ray observations, our results indicate the system could be just after first core passage, but only if the merger has a high impact parameter (i.e., did not undergo a direct collision in its first pass), or could be in a pre-merger phase. The latter would allow for the possibility that the cluster is just before first core passage if a low impact parameter merger, or that the merger geometry includes a large line-of-sight component. However, the former --- that this is an off-axis merger post first core passage --- has the advantage that it more readily explains the asymmetry observed in the X-ray surface brightness. Further, it would naturally be compatible with the possibility for the large-scale, radio-emitting plasma to be produced solely by the AGN associated with the BCG and to permeate a portion of the ICM associated with the southern subcluster component. An exciting possibility, compatible with both merger hypotheses, is that the arced morphology of the southern SZ peak is due to a merger-driven shock. An arc-shaped band of hard X-ray emission is tentatively detected at $\sim3\sigma$ ahead of this SZ feature (see middle-right panel of Fig.~\ref{fig:szres}). The location of the southeastern lensing peak (Fig.~\ref{fig:szres}, middle-left), ostensibly propagating ahead of the feature, would support this interpretation.

Future deep radio follow-up campaigns using, for example, MeerKAT \citep{Jonas2009,Jonas2016} at higher resolution will be required to elucidate the nature of the extended radio emission and determine if the correspondence between the optical filaments and radio structure are in fact reliable. Such observations can also be used to discern if the extended filamentary structure near the BCG is due to radio jets or synchrotron from star formation, and if a knotty substructure contributes significantly to these emission seen in the current ATCA and ASKAP observations.
One possible explanation for some of the extended radio features not associated with compact sources is that the motion of the ICM induced by the merger event has distorted and reaccelerated electrons from past radio outbursts \citep{Zuhone2020}. Deeper and higher-resolution low-frequency radio data could also be used to search for evidence of reacceleration due to merger-induced shocks. In addition, better polarization constraints could be used to search for potential relic complexes, thus helping to shed light on the merger dynamics.

In addition, deeper ALMA+ACA observations could be used to confirm (or dismiss) the putative shock ahead of the southern SZ component as well as to search for other merger shocks and SZ substructures \citep[e.g.,][]{Basu2016,DiMascolo2019b}.  Such observations will benefit from the improved sensitivity to SZ brightness and larger scales expected for ALMA Bands 1 and 2 \citep{DiFrancesco2013,Huang2016,Fuller2016,Yagoubov2020}. 
And beyond this, future large single-dish (sub)millimeter-wave observatories like the Atacama Large Aperture Submillimeter Telescope (AtLAST; \citealt{Klaassen2019,Klaassen2020}) will provide a more complete picture of the SZ effect in the clusters it discovers and targets on scales of $\sim 80$~kpc to beyond the virial radius.

Apart from SZ observations, further high-resolution ALMA measurements of the CO lines could be used to determine if there is a cold molecular component to the filaments surrounding the BCG. Deep spectroscopy of these lines would enable dynamical analyses to yield information on whether they contain multiple velocity components, which would support the hypothesis that we are observing star formation induced by a galaxy-galaxy merger \citep[e.g.,][]{Greve2005}.
Such observations could be included when performing deeper SZ observations in ALMA Bands 3 and 4, but were excluded from the archival data analyzed in this work.

%-----------------------------------------------------------------------

\section*{Acknowledgments}
We are grateful to the referee for the stimulating suggestions and comments. These provided valuable indications that helped us improve this manuscript.

This paper makes use of the following ALMA data: ADS/JAO.ALMA\#2016.1.01175.S and ADS/JAO.ALMA\#2017.1.01649.S. ALMA is a partnership of ESO (representing its member states), NSF (USA) and NINS (Japan), together with NRC (Canada), MOST and ASIAA (Taiwan), and KASI (Republic of Korea), in cooperation with the Republic of Chile. The Joint ALMA Observatory (JAO) is operated by ESO, AUI/NRAO and NAOJ. 
The National Radio Astronomy Observatory is a facility of the National Science Foundation operated under cooperative agreement by Associated Universities, Inc.

The Australian SKA Pathfinder is part of the Australia Telescope National Facility which is managed by CSIRO. Operation of ASKAP is funded by the Australian Government with support from the National Collaborative Research Infrastructure Strategy. ASKAP uses the resources of the Pawsey Supercomputing Centre. Establishment of ASKAP, the Murchison Radio-astronomy Observatory and the Pawsey Supercomputing Centre are initiatives of the Australian Government, with support from the Government of Western Australia and the Science and Industry Endowment Fund. We acknowledge the Wajarri Yamatji people as the traditional owners of the Observatory site. The POSSUM project has been made possible through funding from the Australian Research Council, the Natural Sciences and Engineering Research Council of Canada, the Canada Research Chairs Program, and the Canada Foundation for Innovation. The Dunlap Institute is funded through an endowment established by the David Dunlap family and the University of Toronto. 

Partial support for LR comes from U.S. National Science Foundation grant AST 17-14205 to the University of Minnesota.

MJJ acknowledges support from the National Research Foundation of Korea under the program nos. 2017R1A2B2004644 and 2017R1A4A1015178.

JLW acknowledges the support of the Natural Sciences and Engineering Research Council of Canada (NSERC) through grant RGPIN-2015-05948, and of the Canada Research Chairs program. 

LDM is supported by the ERC-StG `ClustersXCosmo' grant agreement 716762. He would like to personally thank Alexandro Saro and Andrea Biviano for the constructive suggestions, and Gabriella Di Gennaro for helping me getting my mind around radio haloes.

The authors thank Ettore Carretti and Michał Michałowski for the useful comments, and the GOGREEN collaboration for fruitful discussions and for providing the spectroscopic redshifts for the cluster members. We additionally thank the Additional Representative Images for Legacy (ARI-L) project for making us aware of the rich ALMA and ACA archival data sets.

The results presented in this paper were produced using the following \texttt{python} packages: \texttt{NumPy} \citep{Harris2020}, \texttt{SciPy} \citep{Virtanen2020}, \texttt{Matplotlib} \citep{Hunter2007}, \texttt{dynesty} \citep{Speagle2020}, \texttt{Astropy} \citep{Astropy2018}, \texttt{APLpy} \citep{aplpy2012,aplpy2019}.

\bibliographystyle{aa}
\bibliography{spt2106}

\begin{thebibliography}{133}
\expandafter\ifx\csname natexlab\endcsname\relax\def\natexlab#1{#1}\fi

\bibitem[{{Adam} {et~al.}(2017){Adam}, {Bartalucci}, {Pratt}, {Ade},
  {Andr{\'e}}, {Arnaud}, {Beelen}, {Beno{\^\i}t}, {Bideaud}, {Billot},
  {Bourdin}, {Bourrion}, {Calvo}, {Catalano}, {Coiffard}, {Comis}, {D'Addabbo},
  {De Petris}, {D{\'e}mocl{\`e}s}, {D{\'e}sert}, {Doyle}, {Egami}, {Ferrari},
  {Goupy}, {Kramer}, {Lagache}, {Leclercq}, {Mac{\'\i}as-P{\'e}rez},
  {Maurogordato}, {Mauskopf}, {Mayet}, {Monfardini}, {Mroczkowski}, {Pajot},
  {Pascale}, {Perotto}, {Pisano}, {Pointecouteau}, {Ponthieu}, {Rev{\'e}ret},
  {Ritacco}, {Rodriguez}, {Romero}, {Ruppin}, {Schuster}, {Sievers},
  {Triqueneaux}, {Tucker}, {Zemcov}, \& {Zylka}}]{Adam2017}
{Adam}, R., {Bartalucci}, I., {Pratt}, G.~W., {et~al.} 2017, \aap, 598, A115

\bibitem[{{Amodeo} {et~al.}(2016){Amodeo}, {Ettori}, {Capasso}, \&
  {Sereno}}]{Amodeo2016}
{Amodeo}, S., {Ettori}, S., {Capasso}, R., \& {Sereno}, M. 2016, \aap, 590,
  A126

\bibitem[{{Angelinelli} {et~al.}(2020){Angelinelli}, {Vazza}, {Giocoli},
  {Ettori}, {Jones}, {Brunetti}, {Br{\"u}ggen}, \& {Eckert}}]{Angelinelli2020}
{Angelinelli}, M., {Vazza}, F., {Giocoli}, C., {et~al.} 2020, \mnras, 495, 864

\bibitem[{{Ansarifard} {et~al.}(2020){Ansarifard}, {Rasia}, {Biffi}, {Borgani},
  {Cui}, {De Petris}, {Dolag}, {Ettori}, {Movahed}, {Murante}, \&
  {Yepes}}]{Ansarifard2020}
{Ansarifard}, S., {Rasia}, E., {Biffi}, V., {et~al.} 2020, \aap, 634, A113

\bibitem[{{Arnaud} {et~al.}(2010)}]{Arnaud2010}
{Arnaud}, M. {et~al.} 2010, \aap, 517, A92

\bibitem[{{Ascasibar} \& {Markevitch}(2006)}]{Ascasibar2006}
{Ascasibar}, Y. \& {Markevitch}, M. 2006, \apj, 650, 102

\bibitem[{{Balogh} {et~al.}(2017){Balogh}, {Gilbank}, {Muzzin}, {Rudnick},
  {Cooper}, {Lidman}, {Biviano}, {Demarco}, {McGee}, {Nantais}, {Noble}, {Old},
  {Wilson}, {Yee}, {Bellhouse}, {Cerulo}, {Chan}, {Pintos-Castro}, {Simpson},
  {van der Burg}, {Zaritsky}, {Ziparo}, {Alonso}, {Bower}, {De Lucia},
  {Finoguenov}, {Lambas}, {Muriel}, {Parker}, {Rettura}, {Valotto}, \&
  {Wetzel}}]{Balogh2017}
{Balogh}, M.~L., {Gilbank}, D.~G., {Muzzin}, A., {et~al.} 2017, \mnras, 470,
  4168

\bibitem[{{Balogh} {et~al.}(2020){Balogh}, {van der Burg}, {Muzzin}, {Rudnick},
  {Wilson}, {Webb}, {Biviano}, {Boak}, {Cerulo}, {Chan}, {Cooper}, {Gilbank},
  {Gwyn}, {Lidman}, {Matharu}, {McGee}, {Old}, {Pintos-Castro}, {Reeves},
  {Shipley}, {Vulcani}, {Yee}, {Alonso}, {Bellhouse}, {Cooke}, {Davidson}, {De
  Lucia}, {Demarco}, {Drakos}, {Fillingham}, {Finoguenov}, {Forrest},
  {Golledge}, {Jablonka}, {Lambas Garcia}, {McNab}, {Muriel}, {Nantais},
  {Noble}, {Parker}, {Petter}, {Poggianti}, {Townsend}, {Valotto}, {Webb}, \&
  {Zaritsky}}]{Balogh2020}
{Balogh}, M.~L., {van der Burg}, R. F.~J., {Muzzin}, A., {et~al.} 2020, \mnras,
  500, 358

\bibitem[{{Bartalucci} {et~al.}(2017){Bartalucci}, {Arnaud}, {Pratt},
  {D{\'e}mocl{\`e}s}, {van der Burg}, \& {Mazzotta}}]{Bartalucci2017}
{Bartalucci}, I., {Arnaud}, M., {Pratt}, G.~W., {et~al.} 2017, \aap, 598, A61

\bibitem[{{Basu}(2012)}]{Basu2012}
{Basu}, K. 2012, \mnras, 421, L112

\bibitem[{{Basu} {et~al.}(2016){Basu}, {Sommer}, {Erler}, {Eckert}, {Vazza},
  {Magnelli}, {Bertoldi}, \& {Tozzi}}]{Basu2016}
{Basu}, K., {Sommer}, M., {Erler}, J., {et~al.} 2016, \apjl, 829, L23

\bibitem[{{Battaglia} {et~al.}(2012){Battaglia}, {Bond}, {Pfrommer}, \&
  {Sievers}}]{Battaglia2012}
{Battaglia}, N., {Bond}, J.~R., {Pfrommer}, C., \& {Sievers}, J.~L. 2012, \apj,
  758, 74

\bibitem[{{Biffi} {et~al.}(2016){Biffi}, {Borgani}, {Murante}, {Rasia},
  {Planelles}, {Granato}, {Ragone-Figueroa}, {Beck}, {Gaspari}, \&
  {Dolag}}]{Biffi2016}
{Biffi}, V., {Borgani}, S., {Murante}, G., {et~al.} 2016, \apj, 827, 112

\bibitem[{{B{\^\i}rzan} {et~al.}(2017){B{\^\i}rzan}, {Rafferty}, {Br{\"u}ggen},
  \& {Intema}}]{Birzan2017}
{B{\^\i}rzan}, L., {Rafferty}, D.~A., {Br{\"u}ggen}, M., \& {Intema}, H.~T.
  2017, \mnras, 471, 1766

\bibitem[{{B{\^\i}rzan} {et~al.}(2019){B{\^\i}rzan}, {Rafferty}, {Cassano},
  {Brunetti}, {van Weeren}, {Br{\"u}ggen}, {Intema}, {de Gasperin},
  {Andrade-Santos}, {Botteon}, {R{\"o}ttgering}, \& {Shimwell}}]{Birzan2019}
{B{\^\i}rzan}, L., {Rafferty}, D.~A., {Cassano}, R., {et~al.} 2019, \mnras,
  487, 4775

\bibitem[{{Bleem} {et~al.}(2020){Bleem}, {Bocquet}, {Stalder}, {Gladders},
  {Ade}, {Allen}, {Anderson}, {Annis}, {Ashby}, {Austermann}, {Avila}, {Avva},
  {Bayliss}, {Beall}, {Bechtol}, {Bender}, {Benson}, {Bertin}, {Bianchini},
  {Blake}, {Brodwin}, {Brooks}, {Buckley-Geer}, {Burke}, {Carlstrom}, {Rosell},
  {Carrasco Kind}, {Carretero}, {Chang}, {Chiang}, {Citron}, {Moran},
  {Costanzi}, {Crawford}, {Crites}, {da Costa}, {de Haan}, {De Vicente},
  {Desai}, {Diehl}, {Dietrich}, {Dobbs}, {Eifler}, {Everett}, {Flaugher},
  {Floyd}, {Frieman}, {Gallicchio}, {Garc{\'\i}a-Bellido}, {George}, {Gerdes},
  {Gilbert}, {Gruen}, {Gruendl}, {Gschwend}, {Gupta}, {Gutierrez}, {Halverson},
  {Harrington}, {Henning}, {Heymans}, {Holder}, {Hollowood}, {Holzapfel},
  {Honscheid}, {Hrubes}, {Huang}, {Hubmayr}, {Irwin}, {James}, {Jeltema},
  {Joudaki}, {Khullar}, {Klein}, {Knox}, {Kuropatkin}, {Lee}, {Li}, {Lidman},
  {Lowitz}, {MacCrann}, {Mahler}, {Maia}, {Marshall}, {McDonald}, {McMahon},
  {Melchior}, {Menanteau}, {Meyer}, {Miquel}, {Mocanu}, {Mohr}, {Montgomery},
  {Nadolski}, {Natoli}, {Nibarger}, {Noble}, {Novosad}, {Padin}, {Palmese},
  {Parkinson}, {Patil}, {Paz-Chinch{\'o}n}, {Plazas}, {Pryke}, {Ramachandra},
  {Reichardt}, {Remolina Gonz{\'a}lez}, {Romer}, {Roodman}, {Ruhl}, {Rykoff},
  {Saliwanchik}, {Sanchez}, {Saro}, {Sayre}, {Schaffer}, {Schrabback},
  {Serrano}, {Sharon}, {Sievers}, {Smecher}, {Smith}, {Soares-Santos}, {Stark},
  {Story}, {Suchyta}, {Tarle}, {Tucker}, {Vanderlinde}, {Veach}, {Vieira},
  {Wang}, {Weller}, {Whitehorn}, {Wu}, {Yefremenko}, \& {Zhang}}]{Bleem2020}
{Bleem}, L.~E., {Bocquet}, S., {Stalder}, B., {et~al.} 2020, \apjs, 247, 25

\bibitem[{{Bleem} {et~al.}(2015){Bleem}, {Stalder}, {de Haan}, {Aird}, {Allen},
  {Applegate}, {Ashby}, {Bautz}, {Bayliss}, {Benson}, {Bocquet}, {Brodwin},
  {Carlstrom}, {Chang}, {Chiu}, {Cho}, {Clocchiatti}, {Crawford}, {Crites},
  {Desai}, {Dietrich}, {Dobbs}, {Foley}, {Forman}, {George}, {Gladders},
  {Gonzalez}, {Halverson}, {Hennig}, {Hoekstra}, {Holder}, {Holzapfel},
  {Hrubes}, {Jones}, {Keisler}, {Knox}, {Lee}, {Leitch}, {Liu}, {Lueker},
  {Luong-Van}, {Mantz}, {Marrone}, {McDonald}, {McMahon}, {Meyer}, {Mocanu},
  {Mohr}, {Murray}, {Padin}, {Pryke}, {Reichardt}, {Rest}, {Ruel}, {Ruhl},
  {Saliwanchik}, {Saro}, {Sayre}, {Schaffer}, {Schrabback}, {Shirokoff},
  {Song}, {Spieler}, {Stanford}, {Staniszewski}, {Stark}, {Story}, {Stubbs},
  {Vand erlinde}, {Vieira}, {Vikhlinin}, {Williamson}, {Zahn}, \&
  {Zenteno}}]{Bleem2015}
{Bleem}, L.~E., {Stalder}, B., {de Haan}, T., {et~al.} 2015, \apjs, 216, 27

\bibitem[{{Bonaventura} {et~al.}(2017){Bonaventura}, {Webb}, {Muzzin}, {Noble},
  {Lidman}, {Wilson}, {Yee}, {Geach}, {Hezaveh}, {Shupe}, \&
  {Surace}}]{Bonaventura2017}
{Bonaventura}, N.~R., {Webb}, T.~M.~A., {Muzzin}, A., {et~al.} 2017, \mnras,
  469, 1259

\bibitem[{{Bulbul} {et~al.}(2019){Bulbul}, {Chiu}, {Mohr}, {McDonald},
  {Benson}, {Bautz}, {Bayliss}, {Bleem}, {Brodwin}, {Bocquet}, {Capasso},
  {Dietrich}, {Forman}, {Hlavacek-Larrondo}, {Holzapfel}, {Khullar}, {Klein},
  {Kraft}, {Miller}, {Reichardt}, {Saro}, {Sharon}, {Stalder}, {Schrabback}, \&
  {Stanford}}]{Bulbul2019}
{Bulbul}, E., {Chiu}, I.~N., {Mohr}, J.~J., {et~al.} 2019, \apj, 871, 50

\bibitem[{{Bykov} {et~al.}(2019){Bykov}, {Vazza}, {Kropotina}, {Levenfish}, \&
  {Paerels}}]{Bykov2019}
{Bykov}, A.~M., {Vazza}, F., {Kropotina}, J.~A., {Levenfish}, K.~P., \&
  {Paerels}, F.~B.~S. 2019, \ssr, 215, 14

\bibitem[{{Caglar}(2018)}]{Caglar2018}
{Caglar}, T. 2018, \mnras, 475, 2870

\bibitem[{{Carlstrom} {et~al.}(2011){Carlstrom}, {Ade}, {Aird}, {Benson},
  {Bleem}, {Busetti}, {Chang}, {Chauvin}, {Cho}, {Crawford}, {Crites}, {Dobbs},
  {Halverson}, {Heimsath}, {Holzapfel}, {Hrubes}, {Joy}, {Keisler}, {Lanting},
  {Lee}, {Leitch}, {Leong}, {Lu}, {Lueker}, {Luong-Van}, {McMahon}, {Mehl},
  {Meyer}, {Mohr}, {Montroy}, {Padin}, {Plagge}, {Pryke}, {Ruhl}, {Schaffer},
  {Schwan}, {Shirokoff}, {Spieler}, {Staniszewski}, {Stark}, {Tucker}, {Vand
  erlinde}, {Vieira}, \& {Williamson}}]{Carlstrom2011}
{Carlstrom}, J.~E., {Ade}, P.~A.~R., {Aird}, K.~A., {et~al.} 2011, \pasp, 123,
  568

\bibitem[{{Cassano} {et~al.}(2013){Cassano}, {Ettori}, {Brunetti},
  {Giacintucci}, {Pratt}, {Venturi}, {Kale}, {Dolag}, \&
  {Markevitch}}]{Cassano2013}
{Cassano}, R., {Ettori}, S., {Brunetti}, G., {et~al.} 2013, \apj, 777, 141

\bibitem[{{Castignani} {et~al.}(2020){Castignani}, {Combes}, \&
  {Salom{\'e}}}]{Castignani2020}
{Castignani}, G., {Combes}, F., \& {Salom{\'e}}, P. 2020, \aap, 635, L10

\bibitem[{{Churazov} {et~al.}(2001){Churazov}, {Br{\"u}ggen}, {Kaiser},
  {B{\"o}hringer}, \& {Forman}}]{Churazov2001}
{Churazov}, E., {Br{\"u}ggen}, M., {Kaiser}, C.~R., {B{\"o}hringer}, H., \&
  {Forman}, W. 2001, \apj, 554, 261

\bibitem[{{Churazov} {et~al.}(2020){Churazov}, {Khabibullin}, {Lyskova},
  {Sunyaev}, \& {Bykov}}]{Churazov2020}
{Churazov}, E., {Khabibullin}, I., {Lyskova}, N., {Sunyaev}, R., \& {Bykov},
  A.~M. 2020, arXiv e-prints, arXiv:2012.11627

\bibitem[{{Cuciti} {et~al.}(2021{\natexlab{a}}){Cuciti}, {Cassano}, {Brunetti},
  {Dallacasa}, {de Gasperin}, {Ettori}, {Giacintucci}, {Kale}, {Pratt}, {van
  Weeren}, \& {Venturi}}]{Cuciti2021b}
{Cuciti}, V., {Cassano}, R., {Brunetti}, G., {et~al.} 2021{\natexlab{a}}, \aap,
  647, A51

\bibitem[{{Cuciti} {et~al.}(2021{\natexlab{b}}){Cuciti}, {Cassano}, {Brunetti},
  {Dallacasa}, {van Weeren}, {Giacintucci}, {Bonafede}, {de Gasperin},
  {Ettori}, {Kale}, {Pratt}, \& {Venturi}}]{Cuciti2021a}
{Cuciti}, V., {Cassano}, R., {Brunetti}, G., {et~al.} 2021{\natexlab{b}}, \aap,
  647, A50

\bibitem[{{Di Francesco} {et~al.}(2013){Di Francesco}, {Johnstone}, {Matthews},
  {Bartel}, {Bronfman}, {Casassus}, {Chitsazzadeh}, {Chou}, {Cunningham},
  {Duchene}, {Geisbuesch}, {Hales}, {Ho}, {Houde}, {Iono}, {Kemper}, {Kepley},
  {Koch}, {Kohno}, {Kothes}, {Lai}, {Lin}, {Liu}, {Mason}, {Maccarone},
  {Mizuno}, {Morata}, {Schieven}, {Scaife}, {Scott}, {Shang}, {Shimojo}, {Su},
  {Takakuwa}, {Wagg}, {Wootten}, \& {Yusef-Zadeh}}]{DiFrancesco2013}
{Di Francesco}, J., {Johnstone}, D., {Matthews}, B.~C., {et~al.} 2013, arXiv
  e-prints, arXiv:1310.1604

\bibitem[{{Di Gennaro} {et~al.}(2020){Di Gennaro}, {van Weeren}, {Brunetti},
  {Cassano}, {Br{\"u}ggen}, {Hoeft}, {Shimwell}, {R{\"o}ttgering}, {Bonafede},
  {Botteon}, {Cuciti}, {Dallacasa}, {de Gasperin},
  {Dom{\'\i}nguez-Fern{\'a}ndez}, {En{\ss}lin}, {Gastaldello}, {Mandal},
  {Rossetti}, \& {Simionescu}}]{DiGennaro2020}
{Di Gennaro}, G., {van Weeren}, R.~J., {Brunetti}, G., {et~al.} 2020, Nature
  Astronomy [\eprint[arXiv]{2011.01628}]

\bibitem[{{Di Mascolo} {et~al.}(2019{\natexlab{a}}){Di Mascolo}, {Churazov}, \&
  {Mroczkowski}}]{DiMascolo2019}
{Di Mascolo}, L., {Churazov}, E., \& {Mroczkowski}, T. 2019{\natexlab{a}},
  \mnras, 487, 4037

\bibitem[{{Di Mascolo} {et~al.}(2019{\natexlab{b}}){Di Mascolo}, {Mroczkowski},
  {Churazov}, {Markevitch}, {Basu}, {Clarke}, {Devlin}, {Mason}, {Randall},
  {Reese}, {Sunyaev}, \& {Wik}}]{DiMascolo2019b}
{Di Mascolo}, L., {Mroczkowski}, T., {Churazov}, E., {et~al.}
  2019{\natexlab{b}}, \aap, 628, A100

\bibitem[{{Di Mascolo} {et~al.}(2020){Di Mascolo}, {Mroczkowski}, {Churazov},
  {Moravec}, {Brodwin}, {Gonzalez}, {Decker}, {Eisenhardt}, {Stanford},
  {Stern}, {Sunyaev}, \& {Wylezalek}}]{DiMascolo2020}
{Di Mascolo}, L., {Mroczkowski}, T., {Churazov}, E., {et~al.} 2020, \aap, 638,
  A70

\bibitem[{{Dicker} {et~al.}(2020){Dicker}, {Romero}, {Di Mascolo},
  {Mroczkowski}, {Sievers}, {Moravec}, {Bhandarkar}, {Brodwin}, {Connor},
  {Decker}, {Devlin}, {Gonzalez}, {Lowe}, {Mason}, {Sarazin}, {Stanford},
  {Stern}, {Thongkham}, {Wylezalek}, \& {Zago}}]{Dicker2020}
{Dicker}, S.~R., {Romero}, C.~E., {Di Mascolo}, L., {et~al.} 2020, \apj, 902,
  144

\bibitem[{{Diego} {et~al.}(2020){Diego}, {Molnar}, {Cerny}, {Broadhurst},
  {Windhorst}, {Zitrin}, {Bouwens}, {Coe}, {Conselice}, \&
  {Sharon}}]{Diego2020}
{Diego}, J.~M., {Molnar}, S.~M., {Cerny}, C., {et~al.} 2020, \apj, 904, 106

\bibitem[{{Donnert} {et~al.}(2013){Donnert}, {Dolag}, {Brunetti}, \&
  {Cassano}}]{Donnert2013}
{Donnert}, J., {Dolag}, K., {Brunetti}, G., \& {Cassano}, R. 2013, \mnras, 429,
  3564

\bibitem[{{Draine}(2003)}]{Draine2003}
{Draine}, B.~T. 2003, \araa, 41, 241

\bibitem[{{Edge} {et~al.}(2010){Edge}, {Oonk}, {Mittal}, {Allen}, {Baum},
  {B{\"o}hringer}, {Bregman}, {Bremer}, {Combes}, {Crawford}, {Donahue},
  {Egami}, {Fabian}, {Ferland}, {Hamer}, {Hatch}, {Jaffe}, {Johnstone},
  {McNamara}, {O'Dea}, {Popesso}, {Quillen}, {Salom{\'e}}, {Sarazin}, {Voit},
  {Wilman}, \& {Wise}}]{Edge2010}
{Edge}, A.~C., {Oonk}, J.~B.~R., {Mittal}, R., {et~al.} 2010, \aap, 518, L47

\bibitem[{{Egami} {et~al.}(2010){Egami}, {Rex}, {Rawle},
  {P{\'e}rez-Gonz{\'a}lez}, {Richard}, {Kneib}, {Schaerer}, {Altieri},
  {Valtchanov}, {Blain}, {Fadda}, {Zemcov}, {Bock}, {Boone}, {Bridge},
  {Clement}, {Combes}, {Dessauges-Zavadsky}, {Dowell}, {Ilbert}, {Ivison},
  {Jauzac}, {Lutz}, {Metcalfe}, {Omont}, {Pell{\'o}}, {Pereira}, {Rieke},
  {Rodighiero}, {Smail}, {Smith}, {Tramoy}, {Walth}, {van der Werf}, \&
  {Werner}}]{Egami2010}
{Egami}, E., {Rex}, M., {Rawle}, T.~D., {et~al.} 2010, \aap, 518, L12

\bibitem[{{Erler} {et~al.}(2018){Erler}, {Basu}, {Chluba}, \&
  {Bertoldi}}]{Erler2018}
{Erler}, J., {Basu}, K., {Chluba}, J., \& {Bertoldi}, F. 2018, \mnras, 476,
  3360

\bibitem[{{Fogarty} {et~al.}(2017){Fogarty}, {Postman}, {Larson}, {Donahue}, \&
  {Moustakas}}]{Fogarty2017}
{Fogarty}, K., {Postman}, M., {Larson}, R., {Donahue}, M., \& {Moustakas}, J.
  2017, \apj, 846, 103

\bibitem[{{Fogarty} {et~al.}(2019){Fogarty}, {Postman}, {Li}, {Dannerbauer},
  {Liu}, {Donahue}, {Ziegler}, {Koekemoer}, \& {Frye}}]{Fogarty2019}
{Fogarty}, K., {Postman}, M., {Li}, Y., {et~al.} 2019, \apj, 879, 103

\bibitem[{{Foley} {et~al.}(2011){Foley}, {Andersson}, {Bazin}, {de Haan},
  {Ruel}, {Ade}, {Aird}, {Armstrong}, {Ashby}, {Bautz}, {Benson}, {Bleem},
  {Bonamente}, {Brodwin}, {Carlstrom}, {Chang}, {Clocchiatti}, {Crawford},
  {Crites}, {Desai}, {Dobbs}, {Dudley}, {Fazio}, {Forman}, {Garmire}, {George},
  {Gladders}, {Gonzalez}, {Halverson}, {High}, {Holder}, {Holzapfel}, {Hoover},
  {Hrubes}, {Jones}, {Joy}, {Keisler}, {Knox}, {Lee}, {Leitch}, {Lueker},
  {Luong-Van}, {Marrone}, {McMahon}, {Mehl}, {Meyer}, {Mohr}, {Montroy},
  {Murray}, {Padin}, {Plagge}, {Pryke}, {Reichardt}, {Rest}, {Ruhl},
  {Saliwanchik}, {Saro}, {Schaffer}, {Shaw}, {Shirokoff}, {Song}, {Spieler},
  {Stalder}, {Stanford}, {Staniszewski}, {Stark}, {Story}, {Stubbs}, {Vand
  erlinde}, {Vieira}, {Vikhlinin}, {Williamson}, \& {Zenteno}}]{Foley2011}
{Foley}, R.~J., {Andersson}, K., {Bazin}, G., {et~al.} 2011, \apj, 731, 86

\bibitem[{{Frye} {et~al.}(2019){Frye}, {Pascale}, {Qin}, {Zitrin}, {Diego},
  {Walth}, {Yan}, {Conselice}, {Alpaslan}, {Bauer}, {Busoni}, {Coe}, {Cohen},
  {Dole}, {Donahue}, {Georgiev}, {Jansen}, {Limousin}, {Livermore}, {Norman},
  {Rabien}, \& {Windhorst}}]{Frye2019}
{Frye}, B.~L., {Pascale}, M., {Qin}, Y., {et~al.} 2019, \apj, 871, 51

\bibitem[{{Fuller} {et~al.}(2016){Fuller}, {Avison}, {Beltran}, {Casasola},
  {Caselli}, {Cicone}, {Costagliola}, {De Breuck}, {Hunt}, {Jimenez-Serra},
  {Laing}, {Longmore}, {Massardi}, {Mroczkowski}, {Paladino}, {Ramstedt},
  {Richards}, {Testi}, {Vergani}, {Viti}, \& {Wagg}}]{Fuller2016}
{Fuller}, G.~A., {Avison}, A., {Beltran}, M., {et~al.} 2016, arXiv e-prints,
  arXiv:1602.02414

\bibitem[{{Gaensler} {et~al.}(2010){Gaensler}, {Landecker}, {Taylor}, \&
  {POSSUM Collaboration}}]{Gaensler2010}
{Gaensler}, B.~M., {Landecker}, T.~L., {Taylor}, A.~R., \& {POSSUM
  Collaboration}. 2010, in American Astronomical Society Meeting Abstracts,
  Vol. 215, American Astronomical Society Meeting Abstracts \#215, 470.13

\bibitem[{{Garon} {et~al.}(2019){Garon}, {Rudnick}, {Wong}, {Jones}, {Kim},
  {Andernach}, {Shabala}, {Kapi{\'n}ska}, {Norris}, {de Gasperin}, {Tate}, \&
  {Tang}}]{Garon2019}
{Garon}, A.~F., {Rudnick}, L., {Wong}, O.~I., {et~al.} 2019, \aj, 157, 126

\bibitem[{{Golden-Marx} {et~al.}(2019){Golden-Marx}, {Blanton},
  {Paterno-Mahler}, {Brodwin}, {Ashby}, {Lemaux}, {Lubin}, {Gal}, \&
  {Tomczak}}]{GoldenMarx2019}
{Golden-Marx}, E., {Blanton}, E.~L., {Paterno-Mahler}, R., {et~al.} 2019, \apj,
  887, 50

\bibitem[{{Gonzalez} {et~al.}(2015){Gonzalez}, {Decker}, {Brodwin},
  {Eisenhardt}, {Marrone}, {Stanford}, {Stern}, {Wylezalek}, {Aldering},
  {Abdulla}, {Boone}, {Carlstrom}, {Fagrelius}, {Gettings}, {Greer}, {Hayden},
  {Leitch}, {Lin}, {Mantz}, {Muchovej}, {Perlmutter}, \&
  {Zeimann}}]{Gonzalez2015}
{Gonzalez}, A.~H., {Decker}, B., {Brodwin}, M., {et~al.} 2015, \apjl, 812, L40

\bibitem[{{Gonzalez} {et~al.}(2019){Gonzalez}, {Gettings}, {Brodwin},
  {Eisenhardt}, {Stanford}, {Wylezalek}, {Decker}, {Marrone}, {Moravec},
  {O'Donnell}, {Stalder}, {Stern}, {Abdulla}, {Brown}, {Carlstrom}, {Chambers},
  {Hayden}, {Lin}, {Magnier}, {Masci}, {Mantz}, {McDonald}, {Mo}, {Perlmutter},
  {Wright}, \& {Zeimann}}]{Gonzalez2019}
{Gonzalez}, A.~H., {Gettings}, D.~P., {Brodwin}, M., {et~al.} 2019, \apjs, 240,
  33

\bibitem[{{Greve} {et~al.}(2005){Greve}, {Bertoldi}, {Smail}, {Neri},
  {Chapman}, {Blain}, {Ivison}, {Genzel}, {Omont}, {Cox}, {Tacconi}, \&
  {Kneib}}]{Greve2005}
{Greve}, T.~R., {Bertoldi}, F., {Smail}, I., {et~al.} 2005, \mnras, 359, 1165

\bibitem[{{Griffin} {et~al.}(2010){Griffin}, {Abergel}, {Abreu}, {Ade},
  {Andr{\'e}}, {Augueres}, {Babbedge}, {Bae}, {Baillie}, {Baluteau}, {Barlow},
  {Bendo}, {Benielli}, {Bock}, {Bonhomme}, {Brisbin}, {Brockley-Blatt},
  {Caldwell}, {Cara}, {Castro-Rodriguez}, {Cerulli}, {Chanial}, {Chen},
  {Clark}, {Clements}, {Clerc}, {Coker}, {Communal}, {Conversi}, {Cox},
  {Crumb}, {Cunningham}, {Daly}, {Davis}, {de Antoni}, {Delderfield}, {Devin},
  {di Giorgio}, {Didschuns}, {Dohlen}, {Donati}, {Dowell}, {Dowell}, {Duband},
  {Dumaye}, {Emery}, {Ferlet}, {Ferrand}, {Fontignie}, {Fox}, {Franceschini},
  {Frerking}, {Fulton}, {Garcia}, {Gastaud}, {Gear}, {Glenn}, {Goizel},
  {Griffin}, {Grundy}, {Guest}, {Guillemet}, {Hargrave}, {Harwit}, {Hastings},
  {Hatziminaoglou}, {Herman}, {Hinde}, {Hristov}, {Huang}, {Imhof}, {Isaak},
  {Israelsson}, {Ivison}, {Jennings}, {Kiernan}, {King}, {Lange}, {Latter},
  {Laurent}, {Laurent}, {Leeks}, {Lellouch}, {Levenson}, {Li}, {Li},
  {Lilienthal}, {Lim}, {Liu}, {Lu}, {Madden}, {Mainetti}, {Marliani}, {McKay},
  {Mercier}, {Molinari}, {Morris}, {Moseley}, {Mulder}, {Mur}, {Naylor},
  {Nguyen}, {O'Halloran}, {Oliver}, {Olofsson}, {Olofsson}, {Orfei}, {Page},
  {Pain}, {Panuzzo}, {Papageorgiou}, {Parks}, {Parr-Burman}, {Pearce},
  {Pearson}, {P{\'e}rez-Fournon}, {Pinsard}, {Pisano}, {Podosek}, {Pohlen},
  {Polehampton}, {Pouliquen}, {Rigopoulou}, {Rizzo}, {Roseboom}, {Roussel},
  {Rowan-Robinson}, {Rownd}, {Saraceno}, {Sauvage}, {Savage}, {Savini},
  {Sawyer}, {Scharmberg}, {Schmitt}, {Schneider}, {Schulz}, {Schwartz},
  {Shafer}, {Shupe}, {Sibthorpe}, {Sidher}, {Smith}, {Smith}, {Smith},
  {Spencer}, {Stobie}, {Sudiwala}, {Sukhatme}, {Surace}, {Stevens}, {Swinyard},
  {Trichas}, {Tourette}, {Triou}, {Tseng}, {Tucker}, {Turner}, {Vaccari},
  {Valtchanov}, {Vigroux}, {Virique}, {Voellmer}, {Walker}, {Ward}, {Waskett},
  {Weilert}, {Wesson}, {White}, {Whitehouse}, {Wilson}, {Winter}, {Woodcraft},
  {Wright}, {Xu}, {Zavagno}, {Zemcov}, {Zhang}, \& {Zonca}}]{Griffin2010}
{Griffin}, M.~J., {Abergel}, A., {Abreu}, A., {et~al.} 2010, \aap, 518, L3

\bibitem[{{Hallman} {et~al.}(2018){Hallman}, {Alden}, {Rapetti}, {Datta}, \&
  {Burns}}]{Hallman2018}
{Hallman}, E.~J., {Alden}, B., {Rapetti}, D., {Datta}, A., \& {Burns}, J.~O.
  2018, \apj, 859, 44

\bibitem[{{Handley} {et~al.}(2015){Handley}, {Hobson}, \&
  {Lasenby}}]{Handley2015}
{Handley}, W.~J., {Hobson}, M.~P., \& {Lasenby}, A.~N. 2015, \mnras, 453, 4384

\bibitem[{Harris {et~al.}(2020)Harris, Millman, van~der Walt, Gommers,
  Virtanen, Cournapeau, Wieser, Taylor, Berg, Smith, Kern, Picus, Hoyer, van
  Kerkwijk, Brett, Haldane, del R{'{\i}}o, Wiebe, Peterson,
  G{'{e}}rard-Marchant, Sheppard, Reddy, Weckesser, Abbasi, Gohlke, \&
  Oliphant}]{Harris2020}
Harris, C.~R., Millman, K.~J., van~der Walt, S.~J., {et~al.} 2020, Nature, 585,
  357

\bibitem[{{Hatziminaoglou} {et~al.}(2015){Hatziminaoglou}, {Zwaan}, {Andreani},
  {Barta}, {Bertoldi}, {Brand}, {Gueth}, {Hogerheijde}, {Maercker}, {Massardi},
  {Muehle}, {Muxlow}, {Richards}, {Schilke}, {Tilanus}, {Vlemmings}, {Afonso},
  \& {Messias}}]{Hatziminaoglou2015}
{Hatziminaoglou}, E., {Zwaan}, M., {Andreani}, P., {et~al.} 2015, The
  Messenger, 162, 24

\bibitem[{{Henderson} {et~al.}(2016){Henderson}, {Allison}, {Austermann},
  {Baildon}, {Battaglia}, {Beall}, {Becker}, {De Bernardis}, {Bond},
  {Calabrese}, {Choi}, {Coughlin}, {Crowley}, {Datta}, {Devlin}, {Duff},
  {Dunkley}, {D{\"u}nner}, {van Engelen}, {Gallardo}, {Grace}, {Hasselfield},
  {Hills}, {Hilton}, {Hincks}, {Hloẑek}, {Ho}, {Hubmayr}, {Huffenberger},
  {Hughes}, {Irwin}, {Koopman}, {Kosowsky}, {Li}, {McMahon}, {Munson}, {Nati},
  {Newburgh}, {Niemack}, {Niraula}, {Page}, {Pappas}, {Salatino}, {Schillaci},
  {Schmitt}, {Sehgal}, {Sherwin}, {Sievers}, {Simon}, {Spergel}, {Staggs},
  {Stevens}, {Thornton}, {Van Lanen}, {Vavagiakis}, {Ward}, \&
  {Wollack}}]{Henderson2016}
{Henderson}, S.~W., {Allison}, R., {Austermann}, J., {et~al.} 2016, Journal of
  Low Temperature Physics, 184, 772

\bibitem[{{Henry} {et~al.}(2004){Henry}, {Finoguenov}, \& {Briel}}]{Henry2004}
{Henry}, J.~P., {Finoguenov}, A., \& {Briel}, U.~G. 2004, \apj, 615, 181

\bibitem[{{Hildebrand}(1983)}]{Hildebrand1983}
{Hildebrand}, R.~H. 1983, \qjras, 24, 267

\bibitem[{{Hilton} {et~al.}(2021){Hilton}, {Sif{\'o}n}, {Naess},
  {Madhavacheril}, {Oguri}, {Rozo}, {Rykoff}, {Abbott}, {Adhikari}, {Aguena},
  {Aiola}, {Allam}, {Amodeo}, {Amon}, {Annis}, {Ansarinejad}, {Aros-Bunster},
  {Austermann}, {Avila}, {Bacon}, {Battaglia}, {Beall}, {Becker}, {Bernstein},
  {Bertin}, {Bhandarkar}, {Bhargava}, {Bond}, {Brooks}, {Burke}, {Calabrese},
  {Carrasco Kind}, {Carretero}, {Choi}, {Choi}, {Conselice}, {da Costa},
  {Costanzi}, {Crichton}, {Crowley}, {D{\"u}nner}, {Denison}, {Devlin},
  {Dicker}, {Diehl}, {Dietrich}, {Doel}, {Duff}, {Duivenvoorden}, {Dunkley},
  {Everett}, {Ferraro}, {Ferrero}, {Fert{\'e}}, {Flaugher}, {Frieman},
  {Gallardo}, {Garc{\'\i}a-Bellido}, {Gaztanaga}, {Gerdes}, {Giles}, {Golec},
  {Gralla}, {Grandis}, {Gruen}, {Gruendl}, {Gschwend}, {Gutierrez}, {Han},
  {Hartley}, {Hasselfield}, {Hill}, {Hilton}, {Hincks}, {Hinton}, {Ho},
  {Honscheid}, {Hoyle}, {Hubmayr}, {Huffenberger}, {Hughes}, {Jaelani}, {Jain},
  {James}, {Jeltema}, {Kent}, {Knowles}, {Koopman}, {Kuehn}, {Lahav}, {Lima},
  {Lin}, {Lokken}, {Loubser}, {MacCrann}, {Maia}, {Marriage}, {Martin},
  {McMahon}, {Melchior}, {Menanteau}, {Miquel}, {Miyatake}, {Moodley},
  {Morgan}, {Mroczkowski}, {Nati}, {Newburgh}, {Niemack}, {Nishizawa},
  {Ogando}, {Orlowski-Scherer}, {Page}, {Palmese}, {Partridge},
  {Paz-Chinch{\'o}n}, {Phakathi}, {Plazas}, {Robertson}, {Romer}, {Carnero
  Rosell}, {Salatino}, {Sanchez}, {Schaan}, {Schillaci}, {Sehgal}, {Serrano},
  {Shin}, {Simon}, {Smith}, {Soares-Santos}, {Spergel}, {Staggs}, {Storer},
  {Suchyta}, {Swanson}, {Tarle}, {Thomas}, {To}, {Trac}, {Ullom}, {Vale}, {Van
  Lanen}, {Vavagiakis}, {De Vicente}, {Wilkinson}, {Wollack}, {Xu}, \&
  {Zhang}}]{Hilton2020}
{Hilton}, M., {Sif{\'o}n}, C., {Naess}, S., {et~al.} 2021, \apjs, 253, 3

\bibitem[{{Hlavacek-Larrondo} {et~al.}(2015){Hlavacek-Larrondo}, {McDonald},
  {Benson}, {Forman}, {Allen}, {Bleem}, {Ashby}, {Bocquet}, {Brodwin},
  {Dietrich}, {Jones}, {Liu}, {Reichardt}, {Saliwanchik}, {Saro}, {Schrabback},
  {Song}, {Stalder}, {Vikhlinin}, \& {Zenteno}}]{Hlavacek2015}
{Hlavacek-Larrondo}, J., {McDonald}, M., {Benson}, B.~A., {et~al.} 2015, \apj,
  805, 35

\bibitem[{{Huang} {et~al.}(2016){Huang}, {Morata}, {Koch}, {Kemper}, {Hwang},
  {Chiong}, {Ho}, {Chu}, {Huang}, {Liu}, {Hsieh}, {Tseng}, {Weng}, {Ho},
  {Chiang}, {Wu}, {Chang}, {Jian}, {Lee}, {Lee}, {Iguchi}, {Asayama}, {Iono},
  {Gonzalez}, {Effland}, {Saini}, {Pospieszalski}, {Henke}, {Yeung}, {Finger},
  {Tapia}, \& {Reyes}}]{Huang2016}
{Huang}, Y. D.~T., {Morata}, O., {Koch}, P.~M., {et~al.} 2016, in Society of
  Photo-Optical Instrumentation Engineers (SPIE) Conference Series, Vol. 9911,
  Modeling, Systems Engineering, and Project Management for Astronomy VI, ed.
  G.~Z. {Angeli} \& P.~{Dierickx}, 99111V

\bibitem[{Hunter(2007)}]{Hunter2007}
Hunter, J.~D. 2007, Computing in Science \& Engineering, 9, 90

\bibitem[{{Iguchi} {et~al.}(2009){Iguchi}, {Morita}, {Sugimoto}, {Vilar{\'o}},
  {Saito}, {Hasegawa}, {Kawabe}, {Tatematsu}, {Sakamoto}, {Kiuchi}, {Okumura},
  {Kosugi}, {Inatani}, {Takakuwa}, {Iono}, {Kamazaki}, {Ogasawara}, \&
  {Ishiguro}}]{Iguchi2009}
{Iguchi}, S., {Morita}, K.-I., {Sugimoto}, M., {et~al.} 2009, \pasj, 61, 1

\bibitem[{{Itoh} \& {Nozawa}(2004)}]{Itoh2004}
{Itoh}, N. \& {Nozawa}, S. 2004, \aap, 417, 827

\bibitem[{{Johnston} {et~al.}(2007){Johnston}, {Bailes}, {Bartel}, {Baugh},
  {Bietenholz}, {Blake}, {Braun}, {Brown}, {Chatterjee}, {Darling}, {Deller},
  {Dodson}, {Edwards}, {Ekers}, {Ellingsen}, {Feain}, {Gaensler}, {Haverkorn},
  {Hobbs}, {Hopkins}, {Jackson}, {James}, {Joncas}, {Kaspi}, {Kilborn},
  {Koribalski}, {Kothes}, {Landecker}, {Lenc}, {Lovell}, {Macquart},
  {Manchester}, {Matthews}, {McClure-Griffiths}, {Norris}, {Pen}, {Phillips},
  {Power}, {Protheroe}, {Sadler}, {Schmidt}, {Stairs}, {Staveley-Smith},
  {Stil}, {Taylor}, {Tingay}, {Tzioumis}, {Walker}, {Wall}, \&
  {Wolleben}}]{Johnston2007}
{Johnston}, S., {Bailes}, M., {Bartel}, N., {et~al.} 2007, \pasa, 24, 174

\bibitem[{{Jonas} \& {MeerKAT Team}(2016)}]{Jonas2016}
{Jonas}, J. \& {MeerKAT Team}. 2016, in MeerKAT Science: On the Pathway to the
  SKA, 1

\bibitem[{{Jonas}(2009)}]{Jonas2009}
{Jonas}, J.~L. 2009, IEEE Proceedings, 97, 1522

\bibitem[{{Kale} {et~al.}(2019){Kale}, {Shende}, \& {Parekh}}]{Kale2019}
{Kale}, R., {Shende}, K.~M., \& {Parekh}, V. 2019, \mnras, 486, L80

\bibitem[{{Kim} {et~al.}(2019){Kim}, {Jee}, {Perlmutter}, {Hayden}, {Rubin},
  {Huang}, {Aldering}, \& {Ko}}]{Kim2019}
{Kim}, J., {Jee}, M.~J., {Perlmutter}, S., {et~al.} 2019, \apj, 887, 76

\bibitem[{{Klaassen} {et~al.}(2019){Klaassen}, {Mroczkowski}, {Bryan},
  {Groppi}, {Basu}, {Cicone}, {Dannerbauer}, {De Breuck}, {Fischer}, {Geach},
  {Hatziminaoglou}, {Holland}, {Kawabe}, {Sehgal}, {Stanke}, \& {van
  Kampen}}]{Klaassen2019}
{Klaassen}, P., {Mroczkowski}, T., {Bryan}, S., {et~al.} 2019, in Bulletin of
  the American Astronomical Society, Vol.~51, 58

\bibitem[{{Klaassen} {et~al.}(2020){Klaassen}, {Mroczkowski}, {Cicone},
  {Hatziminaoglou}, {Sartori}, {De Breuck}, {Bryan}, {Dicker}, {Duran},
  {Groppi}, {Kaercher}, {Kawabe}, {Kohno}, \& {Geach}}]{Klaassen2020}
{Klaassen}, P.~D., {Mroczkowski}, T.~K., {Cicone}, C., {et~al.} 2020, in
  Society of Photo-Optical Instrumentation Engineers (SPIE) Conference Series,
  Vol. 11445, Society of Photo-Optical Instrumentation Engineers (SPIE)
  Conference Series, 114452F

\bibitem[{{Lee} {et~al.}(2020){Lee}, {Jee}, {Kang}, {Ryu}, {Kimm}, \&
  {Br{\"u}ggen}}]{Lee2020a}
{Lee}, W., {Jee}, M.~J., {Kang}, H., {et~al.} 2020, \apj, 894, 60

\bibitem[{{Mantz} {et~al.}(2016){Mantz}, {Allen}, {Morris}, {von der Linden},
  {Applegate}, {Kelly}, {Burke}, {Donovan}, \& {Ebeling}}]{Mantz2016}
{Mantz}, A.~B., {Allen}, S.~W., {Morris}, R.~G., {et~al.} 2016, \mnras, 463,
  3582

\bibitem[{{McDonald} {et~al.}(2013){McDonald}, {Benson}, {Vikhlinin},
  {Stalder}, {Bleem}, {de Haan}, {Lin}, {Aird}, {Ashby}, {Bautz}, {Bayliss},
  {Bocquet}, {Brodwin}, {Carlstrom}, {Chang}, {Cho}, {Clocchiatti}, {Crawford},
  {Crites}, {Desai}, {Dobbs}, {Dudley}, {Foley}, {Forman}, {George},
  {Gettings}, {Gladders}, {Gonzalez}, {Halverson}, {High}, {Holder},
  {Holzapfel}, {Hoover}, {Hrubes}, {Jones}, {Joy}, {Keisler}, {Knox}, {Lee},
  {Leitch}, {Liu}, {Lueker}, {Luong-Van}, {Mantz}, {Marrone}, {McMahon},
  {Mehl}, {Meyer}, {Miller}, {Mocanu}, {Mohr}, {Montroy}, {Murray},
  {Nurgaliev}, {Padin}, {Plagge}, {Pryke}, {Reichardt}, {Rest}, {Ruel}, {Ruhl},
  {Saliwanchik}, {Saro}, {Sayre}, {Schaffer}, {Shirokoff}, {Song},
  {{\v{S}}uhada}, {Spieler}, {Stanford}, {Staniszewski}, {Stark}, {Story}, {van
  Engelen}, {Vanderlinde}, {Vieira}, {Williamson}, {Zahn}, \&
  {Zenteno}}]{McDonald2013}
{McDonald}, M., {Benson}, B.~A., {Vikhlinin}, A., {et~al.} 2013, \apj, 774, 23

\bibitem[{{McDonald} {et~al.}(2016){McDonald}, {Stalder}, {Bayliss}, {Allen},
  {Applegate}, {Ashby}, {Bautz}, {Benson}, {Bleem}, {Brodwin}, {Carlstrom},
  {Chiu}, {Desai}, {Gonzalez}, {Hlavacek-Larrondo}, {Holzapfel}, {Marrone},
  {Miller}, {Reichardt}, {Saliwanchik}, {Saro}, {Schrabback}, {Stanford},
  {Stark}, {Vieira}, \& {Zenteno}}]{McDonald2016}
{McDonald}, M., {Stalder}, B., {Bayliss}, M., {et~al.} 2016, \apj, 817, 86

\bibitem[{{McDonald} {et~al.}(2015){McDonald}, {Werner}, {Oonk}, \&
  {Veilleux}}]{McDonald2015}
{McDonald}, M., {Werner}, N., {Oonk}, J.~B.~R., \& {Veilleux}, S. 2015, \apj,
  804, 16

\bibitem[{{McMullin} {et~al.}(2007){McMullin}, {Waters}, {Schiebel}, {Young},
  \& {Golap}}]{McMullin2007}
{McMullin}, J.~P., {Waters}, B., {Schiebel}, D., {Young}, W., \& {Golap}, K.
  2007, in Astronomical Society of the Pacific Conference Series, Vol. 376,
  Astronomical Data Analysis Software and Systems XVI, ed. R.~A. {Shaw},
  F.~{Hill}, \& D.~J. {Bell}, 127

\bibitem[{{Moravec} {et~al.}(2020){Moravec}, {Gonzalez}, {Dicker}, {Alberts},
  {Brodwin}, {Clarke}, {Connor}, {Decker}, {Devlin}, {Eisenhardt}, {Mason},
  {Mo}, {Mroczkowski}, {Pope}, {Romero}, {Sarazin}, {Sievers}, {Stanford},
  {Stern}, {Wylezalek}, \& {Zago}}]{Moravec2020}
{Moravec}, E., {Gonzalez}, A., {Dicker}, S., {et~al.} 2020, \apj, 898, 145

\bibitem[{{Mroczkowski} {et~al.}(2009){Mroczkowski}, {Bonamente}, {Carlstrom},
  {Culverhouse}, {Greer}, {Hawkins}, {Hennessy}, {Joy}, {Lamb}, {Leitch},
  {Loh}, {Maughan}, {Marrone}, {Miller}, {Muchovej}, {Nagai}, {Pryke}, {Sharp},
  \& {Woody}}]{Mroczkowski2009}
{Mroczkowski}, T., {Bonamente}, M., {Carlstrom}, J.~E., {et~al.} 2009, \apj,
  694, 1034

\bibitem[{{Mroczkowski} {et~al.}(2012){Mroczkowski}, {Dicker}, {Sayers},
  {Reese}, {Mason}, {Czakon}, {Romero}, {Young}, {Devlin}, {Golwala},
  {Korngut}, {Sarazin}, {Bock}, {Koch}, {Lin}, {Molnar}, {Pierpaoli}, {Umetsu},
  \& {Zemcov}}]{Mroczkowski2012}
{Mroczkowski}, T., {Dicker}, S., {Sayers}, J., {et~al.} 2012, \apj, 761, 47

\bibitem[{{Mroczkowski} {et~al.}(2019){Mroczkowski}, {Nagai}, {Basu}, {Chluba},
  {Sayers}, {Adam}, {Churazov}, {Crites}, {Di Mascolo}, {Eckert},
  {Macias-Perez}, {Mayet}, {Perotto}, {Pointecouteau}, {Romero}, {Ruppin},
  {Scannapieco}, \& {ZuHone}}]{Mroczkowski2019}
{Mroczkowski}, T., {Nagai}, D., {Basu}, K., {et~al.} 2019, \ssr, 215, 17

\bibitem[{{Nagai} {et~al.}(2007){Nagai}, {Kravtsov}, \&
  {Vikhlinin}}]{Nagai2007}
{Nagai}, D., {Kravtsov}, A.~V., \& {Vikhlinin}, A. 2007, \apj, 668, 1

\bibitem[{{Nelson} {et~al.}(2012){Nelson}, {Rudd}, {Shaw}, \&
  {Nagai}}]{Nelson2012}
{Nelson}, K., {Rudd}, D.~H., {Shaw}, L., \& {Nagai}, D. 2012, \apj, 751, 121

\bibitem[{{Norris} {et~al.}(2011){Norris}, {Hopkins}, {Afonso}, {Brown},
  {Condon}, {Dunne}, {Feain}, {Hollow}, {Jarvis}, {Johnston-Hollitt}, {Lenc},
  {Middelberg}, {Padovani}, {Prandoni}, {Rudnick}, {Seymour}, {Umana},
  {Andernach}, {Alexander}, {Appleton}, {Bacon}, {Banfield}, {Becker}, {Brown},
  {Ciliegi}, {Jackson}, {Eales}, {Edge}, {Gaensler}, {Giovannini}, {Hales},
  {Hancock}, {Huynh}, {Ibar}, {Ivison}, {Kennicutt}, {Kimball}, {Koekemoer},
  {Koribalski}, {L{\'o}pez-S{\'a}nchez}, {Mao}, {Murphy}, {Messias},
  {Pimbblet}, {Raccanelli}, {Randall}, {Reiprich}, {Roseboom},
  {R{\"o}ttgering}, {Saikia}, {Sharp}, {Slee}, {Smail}, {Thompson}, {Urquhart},
  {Wall}, \& {Zhao}}]{Norris2011}
{Norris}, R.~P., {Hopkins}, A.~M., {Afonso}, J., {et~al.} 2011, \pasa, 28, 215

\bibitem[{{Nozawa} {et~al.}(2006){Nozawa}, {Itoh}, {Suda}, \&
  {Ohhata}}]{Nozawa2006}
{Nozawa}, S., {Itoh}, N., {Suda}, Y., \& {Ohhata}, Y. 2006, Nuovo Cimento B
  Serie, 121, 487

\bibitem[{{Old} {et~al.}(2020){Old}, {Balogh}, {van der Burg}, {Biviano},
  {Yee}, {Pintos-Castro}, {Webb}, {Muzzin}, {Rudnick}, {Vulcani}, {Poggianti},
  {Cooper}, {Zaritsky}, {Cerulo}, {Wilson}, {Chan}, {Lidman}, {McGee},
  {Demarco}, {Forrest}, {De Lucia}, {Gilbank}, {Kukstas}, {McCarthy},
  {Jablonka}, {Nantais}, {Noble}, {Reeves}, \& {Shipley}}]{Old2020}
{Old}, L.~J., {Balogh}, M.~L., {van der Burg}, R. F.~J., {et~al.} 2020, \mnras,
  493, 5987

\bibitem[{{Pasini} {et~al.}(2020){Pasini}, {Br{\"u}ggen}, {de Gasperin},
  {B{\^\i}rzan}, {O'Sullivan}, {Finoguenov}, {Jarvis}, {Gitti}, {Brighenti},
  {Whittam}, {Collier}, {Heywood}, \& {Gozaliasl}}]{Pasini2020}
{Pasini}, T., {Br{\"u}ggen}, M., {de Gasperin}, F., {et~al.} 2020, \mnras, 497,
  2163

\bibitem[{{Paterno-Mahler} {et~al.}(2017){Paterno-Mahler}, {Blanton},
  {Brodwin}, {Ashby}, {Golden-Marx}, {Decker}, {Wing}, \&
  {Anand}}]{PaternoMahler2017}
{Paterno-Mahler}, R., {Blanton}, E.~L., {Brodwin}, M., {et~al.} 2017, \apj,
  844, 78

\bibitem[{{Pilbratt} {et~al.}(2010){Pilbratt}, {Riedinger}, {Passvogel},
  {Crone}, {Doyle}, {Gageur}, {Heras}, {Jewell}, {Metcalfe}, {Ott}, \&
  {Schmidt}}]{Pilbratt2010}
{Pilbratt}, G.~L., {Riedinger}, J.~R., {Passvogel}, T., {et~al.} 2010, \aap,
  518, L1

\bibitem[{{Planck Collaboration} {et~al.}(2016{\natexlab{a}}){Planck
  Collaboration}, {Adam}, {Ade}, {Aghanim}, {Ashdown}, {Aumont}, {Baccigalupi},
  {Banday}, {Barreiro}, {Bartolo}, {Battaner}, {Benabed}, {Benoit-L{\'e}vy},
  {Bersanelli}, {Bielewicz}, {Bikmaev}, {Bonaldi}, {Bond}, {Borrill},
  {Bouchet}, {Burenin}, {Burigana}, {Calabrese}, {Cardoso}, {Catalano},
  {Chiang}, {Christensen}, {Churazov}, {Colombo}, {Combet}, {Comis}, {Couchot},
  {Crill}, {Curto}, {Cuttaia}, {Danese}, {Davis}, {de Bernardis}, {de Rosa},
  {de Zotti}, {Delabrouille}, {D{\'e}sert}, {Diego}, {Dole}, {Dor{\'e}},
  {Douspis}, {Ducout}, {Dupac}, {Elsner}, {En{\ss}lin}, {Finelli}, {Forni},
  {Frailis}, {Fraisse}, {Franceschi}, {Galeotta}, {Ganga}, {G{\'e}nova-Santos},
  {Giard}, {Giraud-H{\'e}raud}, {Gjerl{\o}w}, {Gonz{\'a}lez-Nuevo},
  {G{\'o}rski}, {Gregorio}, {Gruppuso}, {Gudmundsson}, {Hansen}, {Harrison},
  {Hern{\'a}ndez-Monteagudo}, {Herranz}, {Hildebrandt}, {Hivon}, {Hobson},
  {Hornstrup}, {Hovest}, {Hurier}, {Jaffe}, {Jaffe}, {Jones}, {Keih{\"a}nen},
  {Keskitalo}, {Khamitov}, {Kisner}, {Kneissl}, {Knoche}, {Kunz},
  {Kurki-Suonio}, {Lagache}, {L{\"a}hteenm{\"a}ki}, {Lamarre}, {Lasenby},
  {Lattanzi}, {Lawrence}, {Leonardi}, {Levrier}, {Liguori}, {Lilje},
  {Linden-V{\o}rnle}, {L{\'o}pez-Caniego}, {Mac{\'\i}as-P{\'e}rez}, {Maffei},
  {Maggio}, {Mandolesi}, {Mangilli}, {Maris}, {Martin},
  {Mart{\'\i}nez-Gonz{\'a}lez}, {Masi}, {Matarrese}, {Melchiorri}, {Mennella},
  {Migliaccio}, {Miville-Desch{\^e}nes}, {Moneti}, {Montier}, {Morgante},
  {Mortlock}, {Munshi}, {Murphy}, {Naselsky}, {Nati}, {Natoli},
  {N{\o}rgaard-Nielsen}, {Novikov}, {Novikov}, {Oxborrow}, {Pagano}, {Pajot},
  {Paoletti}, {Pasian}, {Perdereau}, {Perotto}, {Pettorino}, {Piacentini},
  {Piat}, {Plaszczynski}, {Pointecouteau}, {Polenta}, {Ponthieu}, {Pratt},
  {Prunet}, {Puget}, {Rachen}, {Rebolo}, {Reinecke}, {Remazeilles}, {Renault},
  {Renzi}, {Ristorcelli}, {Rocha}, {Rosset}, {Rossetti}, {Roudier},
  {Rubi{\~n}o-Mart{\'\i}n}, {Rusholme}, {Santos}, {Savelainen}, {Savini},
  {Scott}, {Stolyarov}, {Stompor}, {Sudiwala}, {Sunyaev}, {Sutton},
  {Suur-Uski}, {Sygnet}, {Tauber}, {Terenzi}, {Toffolatti}, {Tomasi},
  {Tristram}, {Tucci}, {Valenziano}, {Valiviita}, {Van Tent}, {Vielva},
  {Villa}, {Wade}, {Wehus}, {Yvon}, {Zacchei}, \& {Zonca}}]{PlanckXLIII2016}
{Planck Collaboration}, {Adam}, R., {Ade}, P.~A.~R., {et~al.}
  2016{\natexlab{a}}, \aap, 596, A104

\bibitem[{{Planck Collaboration} {et~al.}(2016{\natexlab{b}}){Planck
  Collaboration}, {Ade}, {Aghanim}, {Arnaud}, {Ashdown}, {Aumont},
  {Baccigalupi}, {Banday}, {Barreiro}, {Barrena}, {Bartlett}, {Bartolo},
  {Battaner}, {Battye}, {Benabed}, {Beno{\^\i}t}, {Benoit-L{\'e}vy}, {Bernard},
  {Bersanelli}, {Bielewicz}, {Bikmaev}, {B{\"o}hringer}, {Bonaldi}, {Bonavera},
  {Bond}, {Borrill}, {Bouchet}, {Bucher}, {Burenin}, {Burigana}, {Butler},
  {Calabrese}, {Cardoso}, {Carvalho}, {Catalano}, {Challinor}, {Chamballu},
  {Chary}, {Chiang}, {Chon}, {Christensen}, {Clements}, {Colombi}, {Colombo},
  {Combet}, {Comis}, {Couchot}, {Coulais}, {Crill}, {Curto}, {Cuttaia},
  {Dahle}, {Danese}, {Davies}, {Davis}, {de Bernardis}, {de Rosa}, {de Zotti},
  {Delabrouille}, {D{\'e}sert}, {Dickinson}, {Diego}, {Dolag}, {Dole},
  {Donzelli}, {Dor{\'e}}, {Douspis}, {Ducout}, {Dupac}, {Efstathiou},
  {Eisenhardt}, {Elsner}, {En{\ss}lin}, {Eriksen}, {Falgarone}, {Fergusson},
  {Feroz}, {Ferragamo}, {Finelli}, {Forni}, {Frailis}, {Fraisse}, {Franceschi},
  {Frejsel}, {Galeotta}, {Galli}, {Ganga}, {G{\'e}nova-Santos}, {Giard},
  {Giraud-H{\'e}raud}, {Gjerl{\o}w}, {Gonz{\'a}lez-Nuevo}, {G{\'o}rski},
  {Grainge}, {Gratton}, {Gregorio}, {Gruppuso}, {Gudmundsson}, {Hansen},
  {Hanson}, {Harrison}, {Hempel}, {Henrot-Versill{\'e}},
  {Hern{\'a}ndez-Monteagudo}, {Herranz}, {Hildebrandt}, {Hivon}, {Hobson},
  {Holmes}, {Hornstrup}, {Hovest}, {Huffenberger}, {Hurier}, {Jaffe}, {Jaffe},
  {Jin}, {Jones}, {Juvela}, {Keih{\"a}nen}, {Keskitalo}, {Khamitov}, {Kisner},
  {Kneissl}, {Knoche}, {Kunz}, {Kurki-Suonio}, {Lagache}, {Lamarre}, {Lasenby},
  {Lattanzi}, {Lawrence}, {Leonardi}, {Lesgourgues}, {Levrier}, {Liguori},
  {Lilje}, {Linden-V{\o}rnle}, {L{\'o}pez-Caniego}, {Lubin},
  {Mac{\'\i}as-P{\'e}rez}, {Maggio}, {Maino}, {Mak}, {Mandolesi}, {Mangilli},
  {Martin}, {Mart{\'\i}nez-Gonz{\'a}lez}, {Masi}, {Matarrese}, {Mazzotta},
  {McGehee}, {Mei}, {Melchiorri}, {Melin}, {Mendes}, {Mennella}, {Migliaccio},
  {Mitra}, {Miville-Desch{\^e}nes}, {Moneti}, {Montier}, {Morgante},
  {Mortlock}, {Moss}, {Munshi}, {Murphy}, {Naselsky}, {Nastasi}, {Nati},
  {Natoli}, {Netterfield}, {N{\o}rgaard-Nielsen}, {Noviello}, {Novikov},
  {Novikov}, {Olamaie}, {Oxborrow}, {Paci}, {Pagano}, {Pajot}, {Paoletti},
  {Pasian}, {Patanchon}, {Pearson}, {Perdereau}, {Perotto}, {Perrott},
  {Perrotta}, {Pettorino}, {Piacentini}, {Piat}, {Pierpaoli}, {Pietrobon},
  {Plaszczynski}, {Pointecouteau}, {Polenta}, {Pratt}, {Pr{\'e}zeau}, {Prunet},
  {Puget}, {Rachen}, {Reach}, {Rebolo}, {Reinecke}, {Remazeilles}, {Renault},
  {Renzi}, {Ristorcelli}, {Rocha}, {Rosset}, {Rossetti}, {Roudier}, {Rozo},
  {Rubi{\~n}o-Mart{\'\i}n}, {Rumsey}, {Rusholme}, {Rykoff}, {Sandri}, {Santos},
  {Saunders}, {Savelainen}, {Savini}, {Schammel}, {Scott}, {Seiffert},
  {Shellard}, {Shimwell}, {Spencer}, {Stanford}, {Stern}, {Stolyarov},
  {Stompor}, {Streblyanska}, {Sudiwala}, {Sunyaev}, {Sutton}, {Suur-Uski},
  {Sygnet}, {Tauber}, {Terenzi}, {Toffolatti}, {Tomasi}, {Tramonte},
  {Tristram}, {Tucci}, {Tuovinen}, {Umana}, {Valenziano}, {Valiviita}, {Van
  Tent}, {Vielva}, {Villa}, {Wade}, {Wandelt}, {Wehus}, {White}, {Wright},
  {Yvon}, {Zacchei}, \& {Zonca}}]{PlanckXXVII2016}
{Planck Collaboration}, {Ade}, P.~A.~R., {Aghanim}, N., {et~al.}
  2016{\natexlab{b}}, \aap, 594, A27

\bibitem[{{Poole} {et~al.}(2007){Poole}, {Babul}, {McCarthy}, {Fardal},
  {Bildfell}, {Quinn}, \& {Mahdavi}}]{Poole2007}
{Poole}, G.~B., {Babul}, A., {McCarthy}, I.~G., {et~al.} 2007, \mnras, 380, 437

\bibitem[{{Qiu} {et~al.}(2020){Qiu}, {Bogdanovi{\'c}}, {Li}, {McDonald}, \&
  {McNamara}}]{Qiu2020}
{Qiu}, Y., {Bogdanovi{\'c}}, T., {Li}, Y., {McDonald}, M., \& {McNamara}, B.~R.
  2020, Nature Astronomy, 4, 900

\bibitem[{{Rajpurohit} {et~al.}(2021){Rajpurohit}, {Wittor}, {van Weeren},
  {Vazza}, {Hoeft}, {Rudnick}, {Locatelli}, {Eilek}, {Forman}, {Bonafede},
  {Bonnassieux}, {Riseley}, {Brienza}, {Brunetti}, {Br{\"u}ggen}, {Loi},
  {Rajpurohit}, {R{\"o}ttgering}, {Botteon}, {Clarke}, {Drabent},
  {Dom{\'\i}nguez-Fern{\'a}ndez}, {Di Gennaro}, \&
  {Gastaldello}}]{Rajpurohit2020}
{Rajpurohit}, K., {Wittor}, D., {van Weeren}, R.~J., {et~al.} 2021, \aap, 646,
  A56

\bibitem[{{Randall} {et~al.}(2002){Randall}, {Sarazin}, \&
  {Ricker}}]{Randall2002}
{Randall}, S.~W., {Sarazin}, C.~L., \& {Ricker}, P.~M. 2002, in American
  Astronomical Society Meeting Abstracts, Vol. 201, American Astronomical
  Society Meeting Abstracts, 67.06

\bibitem[{{Rau} \& {Cornwell}(2011)}]{Rau2011}
{Rau}, U. \& {Cornwell}, T.~J. 2011, \aap, 532, A71

\bibitem[{Reynolds(1994)}]{Reynolds1994}
Reynolds, J. 1994, ATNF Internal

\bibitem[{{Richard-Laferri{\`e}re} {et~al.}(2020){Richard-Laferri{\`e}re},
  {Hlavacek-Larrondo}, {Nemmen}, {Rhea}, {Taylor}, {Prasow-{\'E}mond},
  {Gendron-Marsolais}, {Latulippe}, {Edge}, {Fabian}, {Sanders}, {Hogan}, \&
  {Demontigny}}]{Richard-Laferriere2020}
{Richard-Laferri{\`e}re}, A., {Hlavacek-Larrondo}, J., {Nemmen}, R.~S.,
  {et~al.} 2020, \mnras, 499, 2934

\bibitem[{{Ricker} \& {Sarazin}(2001)}]{Ricker2001}
{Ricker}, P.~M. \& {Sarazin}, C.~L. 2001, \apj, 561, 621

\bibitem[{Robitaille(2019)}]{aplpy2019}
Robitaille, T. 2019, {APLpy v2.0: The Astronomical Plotting Library in Python}

\bibitem[{{Robitaille} \& {Bressert}(2012)}]{aplpy2012}
{Robitaille}, T. \& {Bressert}, E. 2012, {APLpy: Astronomical Plotting Library
  in Python}, Astrophysics Source Code Library

\bibitem[{{Romero} {et~al.}(2020){Romero}, {Sievers}, {Ghirardini}, {Dicker},
  {Giacintucci}, {Mroczkowski}, {Mason}, {Sarazin}, {Devlin}, {Gaspari},
  {Battaglia}, {Hilton}, {Bulbul}, {Lowe}, \& {Stanchfield}}]{Romero2020}
{Romero}, C.~E., {Sievers}, J., {Ghirardini}, V., {et~al.} 2020, \apj, 891, 90

\bibitem[{{Ruppin} {et~al.}(2020){Ruppin}, {Adam}, {Ade}, {Andr{\'e}},
  {Andrianasolo}, {Arnaud}, {Aussel}, {Bartalucci}, {Bautz}, {Beelen},
  {Beno{\^\i}t}, {Bideaud}, {Bourrion}, {Brodwin}, {Calvo}, {Catalano},
  {Comis}, {Decker}, {De Petris}, {D{\'e}sert}, {Doyle}, {Driessen},
  {Eisenhardt}, {Gomez}, {Gonzalez}, {Goupy}, {K{\'e}ruzor{\'e}}, {Kramer},
  {Ladjelate}, {Lagache}, {Leclercq}, {Lestrade}, {Mac{\'\i}as-P{\'e}rez},
  {Mauskopf}, {Mayet}, {McDonald}, {Monfardini}, {Moravec}, {Perotto},
  {Pisano}, {Pointecouteau}, {Ponthieu}, {Pratt}, {Rev{\'e}ret}, {Ritacco},
  {Romero}, {Roussel}, {Schuster}, {Shu}, {Sievers}, {Stanford}, {Stern},
  {Tucker}, \& {Zylka}}]{Ruppin2020}
{Ruppin}, F., {Adam}, R., {Ade}, P., {et~al.} 2020, in European Physical
  Journal Web of Conferences, Vol. 228, European Physical Journal Web of
  Conferences, 00026

\bibitem[{{Sakelliou} \& {Merrifield}(2000)}]{Sakelliou2000}
{Sakelliou}, I. \& {Merrifield}, M.~R. 2000, \mnras, 311, 649

\bibitem[{{Savini} {et~al.}(2019){Savini}, {Bonafede}, {Br{\"u}ggen},
  {Rafferty}, {Shimwell}, {Botteon}, {Brunetti}, {Intema}, {Wilber}, {Cassano},
  {Vazza}, {van Weeren}, {Cuciti}, {De Gasperin}, {R{\"o}ttgering}, {Sommer},
  {B{\^\i}rzan}, \& {Drabent}}]{Savini2019}
{Savini}, F., {Bonafede}, A., {Br{\"u}ggen}, M., {et~al.} 2019, \aap, 622, A24

\bibitem[{{Sayers} {et~al.}(2019){Sayers}, {Monta{\~n}a}, {Mroczkowski},
  {Wilson}, {Zemcov}, {Zitrin}, {Cibirka}, {Golwala}, {Hughes}, {Nagai},
  {Reese}, {S{\'a}nchez}, \& {Zuhone}}]{Sayers2019}
{Sayers}, J., {Monta{\~n}a}, A., {Mroczkowski}, T., {et~al.} 2019, \apj, 880,
  45

\bibitem[{{Sayers} {et~al.}(2013){Sayers}, {Mroczkowski}, {Zemcov}, {Korngut},
  {Bock}, {Bulbul}, {Czakon}, {Egami}, {Golwala}, {Koch}, {Lin}, {Mantz},
  {Molnar}, {Moustakas}, {Pierpaoli}, {Rawle}, {Reese}, {Rex}, {Shitanishi},
  {Siegel}, \& {Umetsu}}]{Sayers2013}
{Sayers}, J., {Mroczkowski}, T., {Zemcov}, M., {et~al.} 2013, \apj, 778, 52

\bibitem[{{Schrabback} {et~al.}(2018){Schrabback}, {Applegate}, {Dietrich},
  {Hoekstra}, {Bocquet}, {Gonzalez}, {von der Linden}, {McDonald}, {Morrison},
  {Raihan}, {Allen}, {Bayliss}, {Benson}, {Bleem}, {Chiu}, {Desai}, {Foley},
  {de Haan}, {High}, {Hilbert}, {Mantz}, {Massey}, {Mohr}, {Reichardt}, {Saro},
  {Simon}, {Stern}, {Stubbs}, \& {Zenteno}}]{Schrabback2018}
{Schrabback}, T., {Applegate}, D., {Dietrich}, J.~P., {et~al.} 2018, \mnras,
  474, 2635

\bibitem[{{Shi} {et~al.}(2015){Shi}, {Komatsu}, {Nelson}, \& {Nagai}}]{Shi2015}
{Shi}, X., {Komatsu}, E., {Nelson}, K., \& {Nagai}, D. 2015, \mnras, 448, 1020

\bibitem[{{Simionescu} {et~al.}(2019){Simionescu}, {ZuHone}, {Zhuravleva},
  {Churazov}, {Gaspari}, {Nagai}, {Werner}, {Roediger}, {Canning}, {Eckert},
  {Gu}, \& {Paerels}}]{Simionescu2019}
{Simionescu}, A., {ZuHone}, J., {Zhuravleva}, I., {et~al.} 2019, \ssr, 215, 24

\bibitem[{{Song} {et~al.}(2012){Song}, {Zenteno}, {Stalder}, {Desai}, {Bleem},
  {Aird}, {Armstrong}, {Ashby}, {Bayliss}, {Bazin}, {Benson}, {Bertin},
  {Brodwin}, {Carlstrom}, {Chang}, {Cho}, {Clocchiatti}, {Crawford}, {Crites},
  {de Haan}, {Dobbs}, {Dudley}, {Foley}, {George}, {Gettings}, {Gladders},
  {Gonzalez}, {Halverson}, {Harrington}, {High}, {Holder}, {Holzapfel},
  {Hoover}, {Hrubes}, {Joy}, {Keisler}, {Knox}, {Lee}, {Leitch}, {Liu},
  {Lueker}, {Luong-Van}, {Marrone}, {McDonald}, {McMahon}, {Mehl}, {Meyer},
  {Mocanu}, {Mohr}, {Montroy}, {Natoli}, {Nurgaliev}, {Padin}, {Plagge},
  {Pryke}, {Reichardt}, {Rest}, {Ruel}, {Ruhl}, {Saliwanchik}, {Saro}, {Sayre},
  {Schaffer}, {Shaw}, {Shirokoff}, {{\v{S}}uhada}, {Spieler}, {Stanford},
  {Staniszewski}, {Stark}, {Story}, {Stubbs}, {van Engelen}, {Vanderlinde},
  {Vieira}, {Williamson}, \& {Zahn}}]{Song2012}
{Song}, J., {Zenteno}, A., {Stalder}, B., {et~al.} 2012, \apj, 761, 22

\bibitem[{{Speagle}(2020)}]{Speagle2020}
{Speagle}, J.~S. 2020, \mnras, 493, 3132

\bibitem[{{Strickland} {et~al.}(2002){Strickland}, {Heckman}, {Weaver},
  {Hoopes}, \& {Dahlem}}]{Strickland2002}
{Strickland}, D.~K., {Heckman}, T.~M., {Weaver}, K.~A., {Hoopes}, C.~G., \&
  {Dahlem}, M. 2002, \apj, 568, 689

\bibitem[{{Sunyaev} \& {Zeldovich}(1972)}]{Sunyaev1972}
{Sunyaev}, R.~A. \& {Zeldovich}, Y.~B. 1972, Comments on Astrophysics and Space
  Physics, 4, 173

\bibitem[{{Sunyaev} \& {Zeldovich}(1980)}]{Sunyaev1980}
{Sunyaev}, R.~A. \& {Zeldovich}, Y.~B. 1980, \mnras, 190, 413

\bibitem[{{The Astropy Collaboration} {et~al.}(2018){The Astropy
  Collaboration}, {Price-Whelan}, {Sip{\\H o}cz}, \t{G{\\"u}nther}, {Lim},
  {Crawford}, \t{Conseil}, {Shupe}, {Craig}, {Dencheva}, \t{Ginsburg},
  {VanderPlas}, {Bradley}, \t{P{\\\'e}rez-Su{\\\'a}rez}, {de Val-Borro}, {Paper
  Contributors}, \t{Aldcroft}, {Cruz}, {Robitaille}, \t{Tollerud},
  {Coordination Committee}, {Ardelean}, \t{Babej}, {Bach}, {Bachetti},
  {Bakanov}, \t{Bamford}, {Barentsen}, {Barmby}, {Baumbach}, \t{Berry},
  {Biscani}, {Boquien}, {Bostroem}, \t{Bouma}, {Brammer}, {Bray},
  {Breytenbach}, \t{Buddelmeijer}, {Burke}, {Calderone}, \t{Cano
  Rodr{\\\'{\\i}}guez}, {Cara}, {Cardoso}, \t{Cheedella}, {Copin}, {Corrales},
  {Crichton}, \t{D{\\rsquo}Avella}, {Deil}, {Depagne}, \t{Dietrich}, {Donath},
  {Droettboom}, \t{Earl}, {Erben}, {Fabbro}, {Ferreira}, \t{Finethy}, {Fox},
  {Garrison}, {Gibbons}, \t{Goldstein}, {Gommers}, {Greco}, \t{Greenfield},
  {Groener}, {Grollier}, \t{Hagen}, {Hirst}, {Homeier}, {Horton},
  \t{Hosseinzadeh}, {Hu}, {Hunkeler}, {Ivezi{\\\'c}}, \t{Jain}, {Jenness},
  {Kanarek}, {Kendrew}, \t{Kern}, {Kerzendorf}, {Khvalko}, \t{King}, {Kirkby},
  {Kulkarni}, {Kumar}, \t{Lee}, {Lenz}, {Littlefair}, {Ma}, \t{Macleod},
  {Mastropietro}, {McCully}, \t{Montagnac}, {Morris}, {Mueller}, {Mumford},
  \t{Muna}, {Murphy}, {Nelson}, {Nguyen}, \t{Ninan}, {N{\\"o}the}, {Ogaz},
  {Oh}, \t{Parejko}, {Parley}, {Pascual}, {Patil}, \t{Patil}, {Plunkett},
  {Prochaska}, \t{Rastogi}, {Reddy Janga}, {Sabater}, {Sakurikar}, \t{Seifert},
  {Sherbert}, {Sherwood-Taylor}, \t{Shih}, {Sick}, {Silbiger}, {Singanamalla},
  \t{Singer}, {Sladen}, {Sooley}, \t{Sornarajah}, {Streicher}, {Teuben},
  {Thomas}, \t{Tremblay}, {Turner}, {Terr{\\\'o}n}, \t{van Kerkwijk}, {de la
  Vega}, {Watkins}, \t{Weaver}, {Whitmore}, {Woillez}, \t{Zabalza}, \&
  {Contributors}}]{Astropy2018}
{The Astropy Collaboration}, {Price-Whelan}, A.~M., {Sip{\\H o}cz}, B.~M.,
  {et~al.} 2018, \\aj, 156, 123

\bibitem[{{Tremblay} {et~al.}(2018){Tremblay}, {Combes}, {Oonk}, {Russell},
  {McDonald}, {Gaspari}, {Husemann}, {Nulsen}, {McNamara}, {Hamer}, {O'Dea},
  {Baum}, {Davis}, {Donahue}, {Voit}, {Edge}, {Blanton}, {Bremer}, {Bulbul},
  {Clarke}, {David}, {Edwards}, {Eggerman}, {Fabian}, {Forman}, {Jones},
  {Kerman}, {Kraft}, {Li}, {Powell}, {Randall}, {Salom{\'e}}, {Simionescu},
  {Su}, {Sun}, {Urry}, {Vantyghem}, {Wilkes}, \& {ZuHone}}]{Tremblay2018}
{Tremblay}, G.~R., {Combes}, F., {Oonk}, J.~B.~R., {et~al.} 2018, \apj, 865, 13

\bibitem[{{Ueda} {et~al.}(2019){Ueda}, {Ichinohe}, {Kitayama}, \&
  {Umetsu}}]{Ueda2019}
{Ueda}, S., {Ichinohe}, Y., {Kitayama}, T., \& {Umetsu}, K. 2019, \apj, 871,
  207

\bibitem[{{Ueda} {et~al.}(2018){Ueda}, {Kitayama}, {Oguri}, {Komatsu},
  {Akahori}, {Iono}, {Izumi}, {Kawabe}, {Kohno}, {Matsuo}, {Ota}, {Suto},
  {Takakuwa}, {Takizawa}, {Tsutsumi}, \& {Yoshikawa}}]{Ueda2018}
{Ueda}, S., {Kitayama}, T., {Oguri}, M., {et~al.} 2018, \apj, 866, 48

\bibitem[{{van der Burg} {et~al.}(2020){van der Burg}, {Rudnick}, {Balogh},
  {Muzzin}, {Lidman}, {Old}, {Shipley}, {Gilbank}, {McGee}, {Biviano},
  {Cerulo}, {Chan}, {Cooper}, {De Lucia}, {Demarco}, {Forrest}, {Gwyn},
  {Jablonka}, {Kukstas}, {Marchesini}, {Nantais}, {Noble}, {Pintos-Castro},
  {Poggianti}, {Reeves}, {Stefanon}, {Vulcani}, {Webb}, {Wilson}, {Yee}, \&
  {Zaritsky}}]{vanderBurg2020}
{van der Burg}, R. F.~J., {Rudnick}, G., {Balogh}, M.~L., {et~al.} 2020, \aap,
  638, A112

\bibitem[{{van Weeren} {et~al.}(2019){van Weeren}, {de Gasperin}, {Akamatsu},
  {Br{\"u}ggen}, {Feretti}, {Kang}, {Stroe}, \& {Zandanel}}]{vanWeeren2019}
{van Weeren}, R.~J., {de Gasperin}, F., {Akamatsu}, H., {et~al.} 2019, \ssr,
  215, 16

\bibitem[{Virtanen {et~al.}(2020)Virtanen, Gommers, Oliphant, Haberland, Reddy,
  Cournapeau, Burovski, Peterson, Weckesser, Bright, {van der Walt}, Brett,
  Wilson, Millman, Mayorov, Nelson, Jones, Kern, Larson, Carey, Polat, Feng,
  Moore, {VanderPlas}, Laxalde, Perktold, Cimrman, Henriksen, Quintero, Harris,
  Archibald, Ribeiro, Pedregosa, {van Mulbregt}, \& {SciPy 1.0
  Contributors}}]{Virtanen2020}
Virtanen, P., Gommers, R., Oliphant, T.~E., {et~al.} 2020, Nature Methods, 17,
  261

\bibitem[{{Webb} {et~al.}(2015){Webb}, {Noble}, {DeGroot}, {Wilson}, {Muzzin},
  {Bonaventura}, {Cooper}, {Delahaye}, {Foltz}, {Lidman}, {Surace}, {Yee},
  {Chapman}, {Dunne}, {Geach}, {Hayden}, {Hildebrandt}, {Huang}, {Pope},
  {Smith}, {Perlmutter}, \& {Tudorica}}]{Webb2015}
{Webb}, T., {Noble}, A., {DeGroot}, A., {et~al.} 2015, \apj, 809, 173

\bibitem[{{Wik} {et~al.}(2008){Wik}, {Sarazin}, {Ricker}, \&
  {Randall}}]{Wik2008}
{Wik}, D.~R., {Sarazin}, C.~L., {Ricker}, P.~M., \& {Randall}, S.~W. 2008,
  \apj, 680, 17

\bibitem[{{Williamson} {et~al.}(2011){Williamson}, {Benson}, {High}, {Vand
  erlinde}, {Ade}, {Aird}, {Andersson}, {Armstrong}, {Ashby}, {Bautz}, {Bazin},
  {Bertin}, {Bleem}, {Bonamente}, {Brodwin}, {Carlstrom}, {Chang}, {Chapman},
  {Clocchiatti}, {Crawford}, {Crites}, {de Haan}, {Desai}, {Dobbs}, {Dudley},
  {Fazio}, {Foley}, {Forman}, {Garmire}, {George}, {Gladders}, {Gonzalez},
  {Halverson}, {Holder}, {Holzapfel}, {Hoover}, {Hrubes}, {Jones}, {Joy},
  {Keisler}, {Knox}, {Lee}, {Leitch}, {Lueker}, {Luong-Van}, {Marrone},
  {McMahon}, {Mehl}, {Meyer}, {Mohr}, {Montroy}, {Murray}, {Padin}, {Plagge},
  {Pryke}, {Reichardt}, {Rest}, {Ruel}, {Ruhl}, {Saliwanchik}, {Saro},
  {Schaffer}, {Shaw}, {Shirokoff}, {Song}, {Spieler}, {Stalder}, {Stanford},
  {Staniszewski}, {Stark}, {Story}, {Stubbs}, {Vieira}, {Vikhlinin}, \&
  {Zenteno}}]{Williamson2011}
{Williamson}, R., {Benson}, B.~A., {High}, F.~W., {et~al.} 2011, \apj, 738, 139

\bibitem[{{Wilson} {et~al.}(2011){Wilson}, {Ferris}, {Axtens}, {Brown},
  {Davis}, {Hampson}, {Leach}, {Roberts}, {Saunders}, {Koribalski}, {Caswell},
  {Lenc}, {Stevens}, {Voronkov}, {Wieringa}, {Brooks}, {Edwards}, {Ekers},
  {Emonts}, {Hindson}, {Johnston}, {Maddison}, {Mahony}, {Malu}, {Massardi},
  {Mao}, {McConnell}, {Norris}, {Schnitzeler}, {Subrahmanyan}, {Urquhart},
  {Thompson}, \& {Wark}}]{Wilson2011}
{Wilson}, W.~E., {Ferris}, R.~H., {Axtens}, P., {et~al.} 2011, \mnras, 416, 832

\bibitem[{{Wootten} \& {Thompson}(2009)}]{Wootten2009}
{Wootten}, A. \& {Thompson}, A.~R. 2009, IEEE Proceedings, 97, 1463

\bibitem[{{Yagoubov} {et~al.}(2020){Yagoubov}, {Mroczkowski}, {Belitsky},
  {Cuadrado-Calle}, {Cuttaia}, {Fuller}, {Gallego}, {Gonzalez}, {Kaneko},
  {Mena}, {Molina}, {Nesti}, {Tapia}, {Villa}, {Beltr{\'a}n}, {Cavaliere},
  {Ceru}, {Chesmore}, {Coughlin}, {De Breuck}, {Fredrixon}, {George}, {Gibson},
  {Golec}, {Josaitis}, {Kemper}, {Kotiranta}, {Lapkin},
  {L{\'o}pez-Fern{\'a}ndez}, {Marconi}, {Mariotti}, {McGenn}, {McMahon},
  {Murk}, {Pezzotta}, {Phillips}, {Reyes}, {Ricciardi}, {Sandri}, {Strandberg},
  {Terenzi}, {Testi}, {Thomas}, {Uzawa}, {Vigan{\`o}}, \&
  {Wadefalk}}]{Yagoubov2020}
{Yagoubov}, P., {Mroczkowski}, T., {Belitsky}, V., {et~al.} 2020, \aap, 634,
  A46

\bibitem[{{ZuHone}(2011)}]{ZuHone2011b}
{ZuHone}, J.~A. 2011, \apj, 728, 54

\bibitem[{{ZuHone} {et~al.}(2011){ZuHone}, {Markevitch}, \&
  {Lee}}]{ZuHone2011a}
{ZuHone}, J.~A., {Markevitch}, M., \& {Lee}, D. 2011, \apj, 743, 16

\bibitem[{{ZuHone} {et~al.}(2013){ZuHone}, {Markevitch}, {Ruszkowski}, \&
  {Lee}}]{ZuHone2013}
{ZuHone}, J.~A., {Markevitch}, M., {Ruszkowski}, M., \& {Lee}, D. 2013, \apj,
  762, 69

\bibitem[{{ZuHone} {et~al.}(2020){ZuHone}, {Markevitch}, {Weinberger},
  {Nulsen}, \& {Ehlert}}]{Zuhone2020}
{ZuHone}, J.~A., {Markevitch}, M., {Weinberger}, R., {Nulsen}, P., \& {Ehlert},
  K. 2020, arXiv e-prints, arXiv:2012.02001

\end{thebibliography}

\end{document}